# Mic-hackathon 2024: Hackathon on Machine Learning for Electron and Scanning Probe Microscopy

Utkarsh Pratiush[1], Austin Houston[1], Kamyar Barakati[1], Aditya Raghavan[1], Dasol Yoon[10], Harikrishnan KP[11], Zhaslan Baraissov[11], Desheng Ma[11], Samuel S. Welborn[13], Mikolaj Jakowski[16], Shawn-Patrick Barhorst[16], Alexander J. Pattison[14], Panayotis Manganaris[38], Sita Sirisha Madugula[2], Sai Venkata Gayathri Ayyagari[15], Vishal Kennedy[16], Ralph Bulanadi[17], Michelle Wang[18], Kieran J. Pang[19], Ian Addison-Smith[20,21b], Willy Menacho[21,21b], Horacio V. Guzman[21,21b], Alexander Kiefer[16], Nicholas Furth[16], Nikola L. Kolev[34], Mikhail Petrov[37], Viktoriia Liu[39], Sergey Ilyev[35], Srikar Rairao[16], Tommaso Rodani[36], Ivan Pinto-Huguet[22], Xuli Chen[22], Josep Cruañes[22], Marta Torrens[22], Jovan Pomar[22], Fanzhi Su[22], Pawan Vedanti[23], Zhiheng Lyu[24], Xingzhi Wang[24], Lehan Yao[3], Amir Taqieddin[25], Forrest Laskowski[25], Xiangyu Yin[26], Yu-Tsun Shao[27], Benjamin Fein-Ashley[27a], Yi Jiang[26], Vineet Kumar[28], Himanshu Mishra[28], Yogesh Paul[29], Adib Bazgir[31], Rama chandra Praneeth Madugula[30], Yuwen Zhang[31], Pravan Omprakash[32], Jian Huang[32], Eric Montufar-Morales[32], Vivek Chawla[1], Harshit Sethi[33], Jie Huang[33], Lauri Kurki[33], Grace Guinan[4], Addison Salvador[4,5], Arman Ter-Petrosyan[6], Madeline Van Winkle[4], Steven R. Spurgeon[4,7,8], Ganesh Narasimha[2], Zijie Wu[2], Richard Liu[1], Yongtao Liu[2], Boris Slautin[9], Andrew R Lupini[2], Rama Vasudevan[2], Gerd Duscher[1], and Sergei V. Kalinin[1,3]

***Affiliation***

*[1] Department of Materials Science and Engineering, University of Tennessee, Knoxville, USA*
*[2] Center for Nanophase Materials Sciences, Oak Ridge National Laboratory, Oak Ridge, TN 37831, USA*
*[3] Pacific Northwest National Laboratory, Richland, WA 99354*
*[4] National Renewable Energy Laboratory, Golden, CO, 80401*
*[5] University of Cincinnati, Cincinnati, OH, 45221*
*[6] University of California Irvine, Irvine, CA, 92697*
*[7] Metallurgical and Materials Engineering Department, Colorado School of Mines, Golden, CO 80401*
*[8] Renewable and Sustainable Energy Institute (RASEI), University of Colorado, Boulder, Boulder, CO 80309*
*[9] Institute for Materials Science and Center for Nanointegration, Duisburg-Essen (CENIDE), University of Duisburg-Essen, Essen, 45141, Germany*
*[10] Department of Materials Science and Engineering, Cornell University, Ithaca, NY, USA*
*[11] School of Applied and Engineering Physics, Cornell University, Ithaca, NY, USA*
*[12] Tickle College of Engineering, University of Tennessee - Knoxville, Knoxville, TN 37996, USA*
*[13]National Energy Research Scientific Computing Center, Lawrence Berkeley National Laboratory (LBNL), Berkeley, CA*
*[14] Molecular Foundry, Lawrence Berkeley National Laboratory, Berkeley, CA 94720, USA*
*[15] Department of Materials Science and Engineering, The Pennsylvania State University, University Park, Pennsylvania, USA*
*[16] University of Tennessee, Knoxville - TN, USA*
*[17] Department of Quantum Matter Physics, University of Geneva, 1211 Geneva, Switzerland*
*[18] Department of Electrical and Photonics Engineering, Technical University of Denmark, 2800 Kongens Lyngby, Denmark*
*[19] Department of Experimental Psychology, Justus Liebig University Giessen, 35394 Giessen, Germany*
*[20] Department of Mechanical Engineering, Universidad de Chile, Beauchef 851, Santiago, Chile*
*[21]Institut de Ciència de Materials de Barcelona, CSIC, 08193 Barcelona, Spain*

[21b] *Biophysics and Intelligent Matter Lab, E-08193 Barcelona, Spain*
[22] *Department of Materials Science and Metallurgy, University of Cambridge, 27 Charles Babbage Road, Cambridge, CB3 0FS, UK*
[23] *Department of Materials Science and Engineering, University of Pennsylvania, Philadelphia, PA 19104, USA*
[24] *Department of Materials Science and Engineering, University of Illinois at Urbana−Champaign, Urbana ,Illinois 61801, United States*
[25] *Technology Integration – Materials Informatics and Modeling Department, Solid Power Operating Inc, Louisville, CO 80027*
[26] *Advanced Photon Source, Argonne National Laboratory, 60439, IL, USA*
[27] *Mork Family Department of Chemical Engineering and Materials Science, University of Southern California, Los Angeles, CA 90089, USA*
[27a] *Ming Hsieh Department of Electrical and Computer Engineering, University of Southern California, Los Angeles, CA 90089, USA*
[28] *Dept. of Surface and Plasma Science, Faculty of Mathematics and Physics, Charles University, 18000 Prague 8, Czech Republic*
[29] *Institute for Neuromodulation and Neurotechnology, University Hospital and University of Tuebingen, Tübingen, Germany*
[30] *Department of Mechanical Engineering, New York University, New York, NY 10012, USA*
[31] *Department of Mechanical and Aerospace Engineering, University of Missouri-Columbia, Columbia, MO 65211, USA*
[32] *Institute of Material Science and Engineering, Washington University in St. Louis, St. Louis, MO 63130, USA*
[33] *Department of Applied Physics, Aalto University, Helsinki, FI-02150, Finland*
[34] *University College London*
[35] *Moscow Institute of Physics and Technology*
[36] *AREA Science Park, Università degli Studi di Trieste*
[37] *Tufts University*
[38] *Department of Nuclear Engineering, North Carolina State University, Raleigh, NC*
[39] *Aspiring Scholars Directed Research Program, Fremont, California, USA.*

∗Corresponding author: upratius@vols.utk.edu, sergei2@utk.edu

Microscopy is one of the primary sources of information on materials structure and functionality at the nanometer and atomic scales. The data generated through microscopy is often contained in well-structured datasets, enriched with extensive metadata and sample histories, although not always with the same level of detail or storage format. The broad incorporation of Data Management Plans (DMPs) by major funding agencies ensures the preservation and accessibility of this data. However, deriving insights from these rich datasets remains challenging due to the lack of established code ecosystems, standardized benchmarks, and integration strategies. Correspondingly, the efficiency of data usage is very low, and time expenditures at the analysis stage are enormous. In addition to post-acquisition data analysis, the emergence of application programming interfaces (APIs) by major microscope manufacturers now creates opportunities for real-time ML-based data analytics to enable automated decision making, and particularly ML-agent controlled real-time microscope operation. Despite these opportunities, there is a significant gap in integrating the ML community with the broader microscopy community, limiting the value that these methods bring to physics and materials discovery and materials optimization. Hackathons address these challenges by fostering collaboration between ML experts and microscopy professionals, encouraging the development of innovative solutions that leverage ML for microscopy and preparing the workforce of the future both for microscopy-intensive domains areas, instrument manufacturers, and ML scientists interested in real world applications for fundamental research, materials optimization, and manufacturing. The hackathon generated benchmark datasets and digital twins of microscopes that further contribute to the development of the field and establish data analysis ecosystems. All the codes can be found at GitHub(https://github.com/KalininGroup/Mic-hackathon-2024-codes-publication/tree/1.0.0.1) and Zenodo (https://zenodo.org/records/15579940).

## I. Opportunity

Microscopy and more generally physical and chemical imaging form one of the primary toolsets for exploring matter in fields ranging from materials science and condensed matter physics to chemistry, biology, electrochemistry, and their extensive subfields[1]. Many of the imaging tools such as electron microscopy (EM) e-beam diffraction[2, 3], certain modalities of scanning probe microscopy (SPM)[4, 5] and time-of-flight secondary ion mass spectrometry (ToF-SIMS)[6, 7] offer high-fidelity information on local properties and compositions. Structural imaging methods such as scanning transmission electron microscopy (STEM)[8, 9] allow visualization of the structure of matter atom by atom with few picometer precision[10], sufficient for direct mapping of the chemical expansion in electrochemically active materials[11-13] and ferroelectric[14, 15] and ferroelastic[16, 17] order parameter fields. Overall, microscopy datasets contain a wealth of information on materials structure and functionalities, processing histories, fundamental physical laws governing materials formation and property emergence, and hence are the keys for materials and physics discovery and materials optimization.

Yet whereas scientific domains such as astronomy, genomics, mass-spectrometry, X-Ray scattering[18, 19], thermodynamics, and many others have by now a well-established tradition of data analysis and integration for downstream applications, this is largely absent in microscopy. While simulation frameworks[20-25] are ample in scanning probe and particularly electron microscopy fields, data analysis ecosystems are scarce and disjointed. We identify the two primary factors behind this as (a) extreme heterogeneity of experimental microscopy workflows spanning fields from condensed matter and quantum physics to biology and (b) resultant dearth of downstream applications. In other words, no subfield of microscopy (outside of cryo-EM[26-28] and electron diffraction) has critical mass of researchers and applications within a single group required to build robust data analysis and software ecosystems. We pose that ML/AI methods are now positioned to change this paradigm and organized the first Hackathon on ML for Microscopy on December 16-17, 2024 (with the second planned for December 2025) to spark this transition by bringing together microscopists from multiple communities and organizations together with ML experts and enthusiasts to learn and build the ML-based solutions for electron and probe microscopy.

### I.A. Microscopy matters

Electron and Scanning Probe Microscopies are foundational tools in materials science, condensed matter physics, chemistry, catalysis, electrochemistry, and other fields. Scanning Transmission Electron Microscopy (STEM) allows imaging and characterization of materials at the nanoscale extending to atomic resolution, providing insights into the atomic and molecular structures of materials. By now, STEM has become one of the primary tools in multiple academic and industrial research labs worldwide[9, 29-33]. The versatility of STEM is further enhanced by its integration with techniques such as electron energy loss spectroscopy (EELS)[34-36], which allows for precise analysis of the chemical composition[37], electronic structure[38] of materials, and low energy quasiparticles[39]. The ability of STEM to image materials at nanometer and atomic levels makes it a crucial tool for advancing understanding of the structure property relationship in wide range of material systems. This combination is particularly beneficial in the development of new materials for technological applications, including semiconductors, solar cells[40], catalysts[41], and battery materials [42].

Similarly, the extensive application of scanning probe microscopy (SPM) has opened the doors to explore and modify the nanoworld. Compared to other materials characterization tools, SPM offers a desktop footprint, low cost, and versatility in operating in multiple environments[43, 44]. It provides a wide range of functional imaging capabilities, extending from basic topographic imaging [45-49] to probing of electronic[46, 50], magnetic[45, 51, 52], mechanical[45, 46], biological [49, 53-57], and chemical[58, 59] properties. Furthermore, SPM supports multiple spectroscopy techniques in a variety of imaging modes, enabling comprehensive understanding and manipulation of materials at the nanoscale[60-62] and exploring phenomena such as single-molecule chemical transformation in biomolecules, polarization switching phenomena in ferroelectrics, and local electrochemical activity in a broad range of energy materials from batteries to fuel cells.

### I.A.1 ML for microscopy: single point data analysis

Microscopy tools now offer almost unbounded sources of information on the structure and property of matter in the form of images, spectra, and hyperspectral imaging data. Even for simple imaging modes, modern cameras and detectors offer capabilities up to 32k x 32k pixel images, which are already above the capability of a human operator to usefully examine. Similarly, high-dimensional data sets that are common across scanning probe and electron microscopies are generally outside the human ability to comprehend and interpret, necessitating lengthy iterative analyses. Currently, the efficiency of data utilization in microscopy is extremely low, often with 2-3 best images from a day of work being analyzed for publications of downstream applications. The data analysis itself often relies on custom multistep workflows developed based on operator intuition and best practices and often take weeks to perfect. These factors lead to enormous hidden inefficiencies and strong operator biases in microscopy use. Compared to success stories such as molecular structure discovery by CryoEM, they also suggest the tremendous potential to increase efficiency and impact of imaging tool on broad range of scientific disciplines if data analysis and potentially data acquisition methods are brought to the intrinsic physical limits of instruments.

Like other domains, the growth of data science and machine learning have stimulated interest into big data and machine learning methods in electron microscopy. A number of early works have been reported in 80ies and 90ies, including that of (co-author) Duscher[63] and the especially visionary work of Noel Bonnett[64]. However, at that time, the computational capabilities and hence potential to work with large data volumes was limited, as were the libraries of available data analysis tools.

In the general computer science community, the ML ecosystem has been growing exponentially for the last two decades, with new network architectures, methodologies, etc. becoming available almost monthly. The development of deep learning in 2012 has become an inflection point that brought these developments to the attention of the broader scientific community. As a natural sequence, disparate microscopy communities started to adopt these methods for applications such as semantic segmentation of images, unsupervised learning over spectral and imaging data, and multiple other applications. The first sustained effort in ML for EM can be dated to the work of Watanabe et al[65] in ~2005. At that time, grid-based spectroscopic measurements had become common on many tools, and the computation power and codes sufficient for simple data analytics were starting to become available outside of the computer science community. From that moment, the combination of hyperspectral imaging and machine learning for physics extraction and dimensionality reduction has become the new paradigm in STEM-EELS[35, 66, 67], 4DSTEM[68, 69], EDS[70, 71], and spectroscopic imaging techniques in SPM. The field has been steadily developing ever since, and several comprehensive reviews on ML based analysis of STEM – EELS data have become available recently[72, 73]. It is important to note that the range of data analysis tasks in microscopy goes well beyond simple dimensionality reduction and image segmentation, and requires active learning methods, integration of microscopy data for downstream physics discovery as well as upstream experiment planning. As mentioned above, despite close similarity between analysis tasks performed on dissimilar microscopies, currently the scientific landscape is dominated by software developed in individual groups, typically lacking benchmarks or community-wide development. The few exceptions reporting benchmarking for microscopy tasks have become available only recently[74].

### I.A.2 ML for microscopy: active experiments

Any SPM and STEM operator is well familiar with the classical scan paradigm, and at some point, asked the question whether rectangular scanning and grid-based hyperspectral imaging orchestrated by a human operator are indeed the only or the best way to explore new materials. The progress of big data methods in areas such as robotic vision[75] and autonomous driving[76] brings forth the question as to whether similar methods[77-80] can be useful for building automated microscopes. So far, these were preponderantly realized in the form of workflows in which execution of the codes is driven by immediately available targets via fixed policies. For example, this can include the use of the deep convolutional networks or simpler image analysis tools for the identification of the *a priori* known object of interest such as atoms in scanning tunneling microscopy[81], identification of single DNA molecules[82], spectroscopy of grain boundaries [83], and ferroelectric domain walls[84-88]. These developments are paralleled by the development of sampling methods such as compressed sensing[89]. More complex examples entail inverse workflows, in which the goal is to discover the structural features that maximize the desired aspect of the spectral response [90]. The number of examples of ML integration into active microscopy workflows has been growing rapidly over the last 2-3 years[60, 91-95]. However, despite several early successes in active learning in SPM and STEM[68, 96], the amount of work on active learning is so far very limited, due to relatively early stages of the control APIs, challenges of real time (as compared to post-acquisition) data analysis, and particularly lack of workflow design tools and experience across microscopy communities.

### I.B. Data and software ecosystems

Complementary to post-acquisition data analysis and implementation of on-the-fly data analytics and active learning on individual instruments in support of single operator work, data integration across multiple data generation facilities and teams brings additional opportunities. Below, we summarize these in the light of the current open data mandates.

### I.B.1. Data management plans

For almost a decade, major funding agencies, including the National Science Foundation (NSF) and the Department of Energy (DOE), have mandated the inclusion of Data Management Plans (DMPs) in research proposals. A DMP is a detailed document outlining how data generated during the research project will be handled, preserved, and shared. It ensures that data is stored in accessible formats, with appropriate metadata to facilitate future use by other researchers. This requirement aims to establish a culture of data sharing and transparency in scientific research, implementing FAIR principles across multiple domains.

By now, there is broad acceptance within the scientific community that data should be shared online. This practice enhances reproducibility, enables further discoveries, and maximizes the return on investment in research. However, while data sharing is slowly becoming the norm, there remains a significant need for methods that enable **downstream use** of this data. For fields such as X-ray crystallography, genomics, and thermodynamics, well-established software ecosystems exist to facilitate the analysis and utilization of shared data. Projects like the Materials Project have further improved the ease of accessing and applying data and connecting it across domains, and by now are sufficiently advanced to be used as a part of graduate and undergraduate level education. For example, one of the authors regularly uses the Materials Project API as a part of his *Intro to ML in Materials Science* course.

In contrast, the field of microscopy has yet to develop a comparable software ecosystem for the downstream use of its data. Despite the extensive datasets generated through advanced microscopy techniques, researchers often face challenges in extracting meaningful insights due to the lack of standardized tools and benchmarks. The lack of shared data precludes development of the downstream use applications, and the lack of the latter in turn disincentivizes data sharing (note that simply sharing the data and sharing the data in the form that encourages downstream use are very different). This gap highlights the need for concerted efforts to develop and integrate machine learning (ML) and artificial intelligence (AI) tools specifically tailored for microscopy data to establish a **virtuous circle** of downstream applications and data sharing.

### I.B.2. Software ecosystem

Progress in ML in different scientific applications has been largely driven by the availability of software frameworks, many of which were initially developed by industry (the main examples being TensorFlow and PyTorch). Within the microscopy domain, given the strong heterogeneity of the space, packages exist for certain sub-domains but are virtually all utilizing classical/traditional analysis methods. These packages include efforts in electron microcopy, such as hyperspy[97], and in atomic force microscopy, such as with TopoStats[98]. However, packages to develop machine learning workflows for microscopy datasets are limited (notable examples include AtomAI and GPax[99]).

In parallel to development of the specialized ecosystems for microscopy data analysis, certain successes have been achieved with the broad use tools developed by ML community. For example, GUIs that can be easily used to train models for image segmentation and object detection abound, such as iLastik[100] and RoboFlow[101], are available. The recent push towards foundation models has also seen the deployment of universal segmentation models, such as the 'SAM' and 'SAM2' model from META[102]. The limitation of these is that they are optimized for real-space imaging from ordinary cameras, and do not always function as expected on microscopy datasets (e.g. atomically resolved images will be identified as chainmail or cloth).

Perhaps more than software codebases, a fundamental challenge is that **benchmarking** a new ML workflow or architecture remains difficult in the current ecosystem. We require better digital twins and 'ground truth' datasets, on which typical applications such as segmentation, reconstruction, and more advanced human-in-the-loop workflows can be reliably tested to gauge the performance of newly developed algorithms.

### I.B.3. Application Programming Interfaces (APIs)

APIs are the key bridges between microscopy and codes that enable the application of ML for automated microscopy. Only a few years ago these were generally lacking across microscopy manufacturers. While there were some exceptions, such as Nion[103] in EM and Nanonis[104] in SPM, what APIs existed were often not well known or well documented, and third party tools such as SerialEM(https://bio3d.colorado.edu/SerialEM/) or DigitalMicrograph(http://www.dmscripting.com/publications.html) were frequently used. Fortunately, this trend is now changing drastically, with both availability and functionality being improved.

For SPM, the programming interface API from SpecsGroup[105] is built for LabView[106] and requires another translation layer to work with Python and ML. The Python API from NanoSurf[107] offers direct communication to Python codes, but still doesn't provide full control and data access. On the user side, there have also been attempts to interface microscopy with ML with APIs. The AEcroscopy API[108] developed at ORNL enables the full control of SPM manufactured by Asylum Research[109] but requires custom hardware. The DeepSPM API[110] is based on the programming interface by SpecsGroup and thus only works Nanonis controllers. While the official APIs for SPMs are still scarce and come with limited control and data I/O for their microscopes, there is a rapid trend to their emergence and operationalization, and corresponding pressure from the customer community to build and share the in-house versions.

The APIs have similarly become more advanced in the electron microscopy community over the last several years. Companies such as Thermo Fisher Scientific[111], Nion[103], and JEOL[112] have recently introduced APIs that allow for hardware control through code. However, there is a need to increase community awareness of these tools and their potential benefits for machine learning. Similarly, the APIs built with different custom hardware and built for different brands prevent the share of ML workflows across the microscopy community. This hackathon will play a pivotal role in fostering this understanding and adoption and stimulate the development of pure-software generic API that works universally on different microscopy has been long awaited and can promote the development of automated microscopy.

### I.C. Summary and needs

Overall, the microscopy community has high anticipations for ML and has demonstrated multiple use cases for the applications of machine learning, both at the stage of post-acquisition data analysis, real time data analytics, and even ML/AI driven decision making during the experiment. However, these efforts are largely fragmented across the community, prototype workflows are developed within individual groups, and generally do not form large projects and software ecosystems required for benchmarking, cross-task and cross-domain use, and generally mature downstream applications.

The outstanding need in the field now is to integrate the emerging efforts in ML for scanning probe and electron microscopy. This is possible since the basic principles of operation, instrument control hyper languages, and many data analysis workflows are similar between electron and scanning robe techniques. Similarly, there is a clear opportunity to bridge the ML and microscopy communities, where the availability of APIs offers an advantage for real-world testing of the ML algorithms and microscope serving a prototype model for much more complex decision making and real-world operation environments such as robotics. The proposed hackathon aims to address this need by engaging students to develop a new ML-conversant workforce in microscopy, start building consensus on benchmark datasets, standard implementations, and approaches, developing multi-institution partnerships and a nexus of talent to seed future collaboration and development, and make a value proposition for manufacturers, community and funding agencies toward broader adoption.

## II. Recent events

The proposed hackathon series will to our knowledge be the first event of this type for the microscopy community. However, there have been several recent hackathon series examples organized by our colleagues at Argonne/UChicago and Acceleration Consortium on different topics that we use as a model, and several events organized by our team that give us confidence in our capability to execute on this funding and expect strong resonance with the microscopy and ML communities. These events are described below.

### II.A. ML for Microscopy Schools

Our team has extensive experience organizing the schools and tutorials in machine learning and scanning probe microscopy starting from 2006 (PFM workshop series started in 2006 and run in over 20 locations internationally; now PFM is a part of the major ferroelectric conference ISAF). Most recently, we have organized schools on machine learning in electron microscopy in 2023 and 2024.

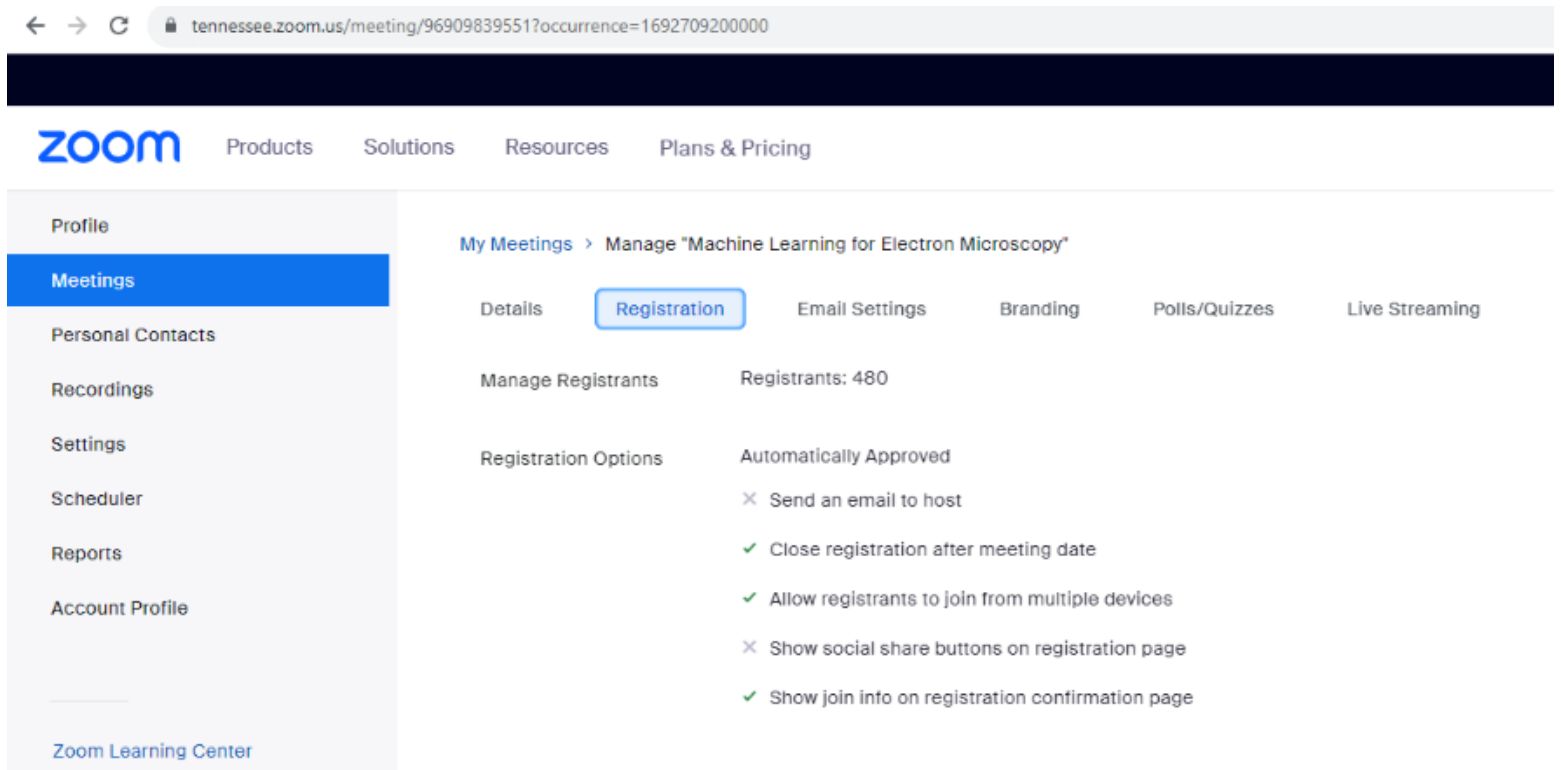

**Figure 1.** The ML for Microscopy school (2023) organized by Kalinin, Duscher et al has attracted 480 registrations.

**The 2023 school** was fully virtual, with ~480 registrants. The school was given as 2-hour lecture/hands-on tutorial 2 times a week for a month, to the total of ~50 hours of recorded materials. The event was organized by UTK faculty (Kalinin, Duscher), and tutorials have been given by the organizers and their colleagues at ORNL (Ziatdinov, Vasudevan, Roccapriore), PD scientists (Ghosh) and graduate students (Valleti, Wong). The information on the course and course materials is available on GitHub[113]. The recorded lectures are posted on the YouTube[114].

**The 2024 school** was given in a hybrid format, with ~50 attendees in person and 200 on-line registration. The event was organized continuously over full 5 days, combining lectures by UTK faculty (Duscher, Kalinin), guest lectures by ORNL staff (Vasudevan), and graduate students (Pratiush, Austin). The course information is available on GitHub[115]. It is also important to note that Duscher has been an early pioneer in adopting Colab notebooks and Python for teaching STEM to students starting from 2019, as evidenced by history of his GitHub repository[116].

In addition to these two events, we have run multiple smaller on-line schools and tutorials on ML for automated experiment (held in Spring 2023 for UTK faculty and students, Summer 2023 for general community, and Spring 2024 under aegis of ACerS). The typical topics covered were (1) Gaussian Processes (2) Bayesian Optimization (3) Bayesian Inference, (4) Structured Gaussian Processes, (5) Hypothesis learning (6) GP, BO, sGP, and HypoAI beyond 1D, (7) Linear dimensionality reduction methods, (8) Variational autoencoders, (9) Deep Kernel Learning, (10) DKL for process optimization, (11) Forensics and human in the loop automated experiment, all illustrated on practical microscopy and materials discovery problems with a broad range of hands-on exercises. The available lectures and the Colabs have been posted on the GitHub[117] and YouTube (for Summer 2023 workshop)[118] and Expertise gained from these events are now included in the new course MSE404/504 (Introduction to Machine Learning for Materials Science and Engineering) taught by PI Kalinin for undergraduate and graduate students at UTK.

**II.B. Previous experience by the team – Local hackathons**

Since 2020, Gerd Duscher (co-PI) and Rama Vasudevan (ORNL) are co-maintainers of a hackathon of the PyCroscopy eco-system[119]. PyCroscopy is a collection of software tools that include traditional and machine learning based data analysis. Common to all included packages is a data-structure that allows all tools to work on data originating from a variety of microscopes, including converters of data and metadata from manufacturer-specific formats to hdf5 form. Thus, trying to mediate the above stated problem and allow microscopists to expand their data-analysis portfolio. Lately, packages to allow for automatization of microscopes was added.

This hackathon is an international online meeting every Friday from 3-5 EST. The number of participants varies **from 5 to 30 per week**. After a brief discussion of the various issues raised on GitHub, programing tasks are distributed to the participants. Key ingredient of a hackathon is the ability to ask questions while trying to code specific microscopy related algorithms. The questions range from beginner Python-questions to expert instrument specific problems. The ability to use breakout rooms or discuss the raised questions in a larger group are part of the fun and further learning for all. We recruited regular participants of this hackathon (Vasudevan, Slautin, Valleti, etc) to help out with the hackathon to build on the expertise achieved in PyCroscopy. It is essential to transfer this kind of community spirit to the larger hackathon.

**II.C. Similar events – domain specific hackathons**

In building the hackathon program, we also drew the inspiration and experience from two different hackathon series events organized by the ML for materials community. These include:

- LLM for Chemistry & Materials Hackathon[120] – Organized in 2023 and 2024 led by Argonne/Uchicago – Several companies like Anthropic, Neo4J, Radical AI came forward for funding the awards and providing access to LLM AI's. The two day event ended with around 40 submissions. The event was hosted in a hybrid mode[121].
- Bayesian hackathon for chemistry and materials – Organized in 2024 by the Acceleration consortium (University of Toronto) and planned for continuation. Over 50 projects were submitted out of which top 10 projects were decided and awarded. MATTERHORN studio[122] was the primary sponsor for this event. The event was hosted in a virtual mode[123].

We have reached out to Ben Blaiszik, organizer of the LLM hackathon series, and had 1:1 consulting meeting to transfer the lessons learned. In addition, Utkarsh Pratiush, the member of the organizing team for the proposed hackathon, was a member of the winning team on the LLM hackathon (and regular participant of our local hackathon events).

**III. Hackathon Organization**

**III.A. Organizing committee**

The hackathon organization was led by Prof. Sergei V. Kalinin (PI, chair) and Prof. Gerd Duscher (co-PI, co-chair). The organizing team will include Dr. Rama Vasudevan (ORNL) and Dr. Maxim Ziatdinov (PNNL), Dr. Boris Slautin (U. Duisburg-Essen), Dr. Steven R. Spurgeon (NREL/Mines), Dr. Yongtao Liu (ORNL), Dr. Yu Richard Liu, Mr. Utkarsh Pratiush, and Mr. Austin Houston (all UTK). In addition to the core organization committee, we formed the jury for the submissions formed by the organizers, representatives of industry (silver sponsors and partners), and our colleagues worldwide.

The data sets and digital twins for the hackathon were based on the PyCroscopy software ecosystem. The code and data developed in this hackathon are published on GitHub and are available to the whole of the microscopy community under the MIT license. The participants were encouraged to develop the test cases to independent projects and join extant projects if these match their interests or to start new community wide projects for novel applications.

**III.B. Location and advertisement**

The hackathon series were planned to run in the hybrid format, with the in-person meeting held at the Institute for Advanced Materials and Manufacturing (IAMM) building on December 16-17, 2024. The on-line components were monitored and moderated by the organizing team (Section III.A)

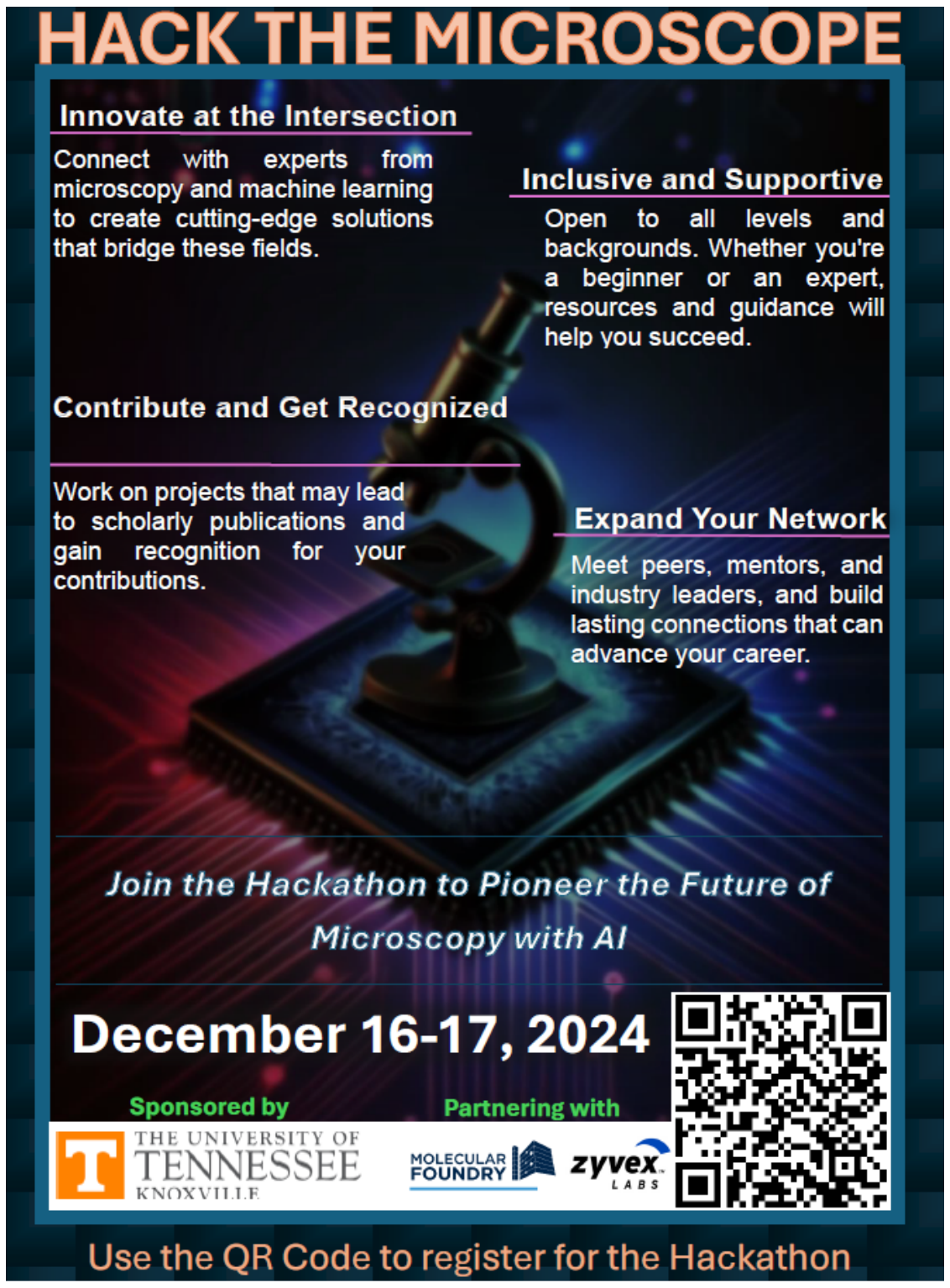

The information on the meeting was distributed through the

- Social media (LinkedIn, X/Twitter) of the UT Knoxville TCE, organizing team, and social network
- Targeted e-mails to the thought leaders in the microscopy community. Approximately ~300 individual emails were sent
- Microscopy list-server and e-mail lists formed based on the ML schools (500-1000 names).
- The hackathon poster was mailed individually to the leading SPM and STEM groups in US, Japan, Korea, and Europe
- For future, we will engage the e-mail lists of the DOE user facilities that we plan to partner with (CNMS at ORNL, Molecular Foundry at Berkeley, etc.)

Beyond this point, we anticipate the onset of the network effect for the community.

### III.C. Hackathon operation and results dissemination

Hackathon organization is a multistage process including pre-hackathon organization and training, hackathon per se, and post-hackathon judging and manuscript preparation. These stages are illustrated in Figure 1. Below, we describe these individual stages:

### III.C.1. Hackathon preparation

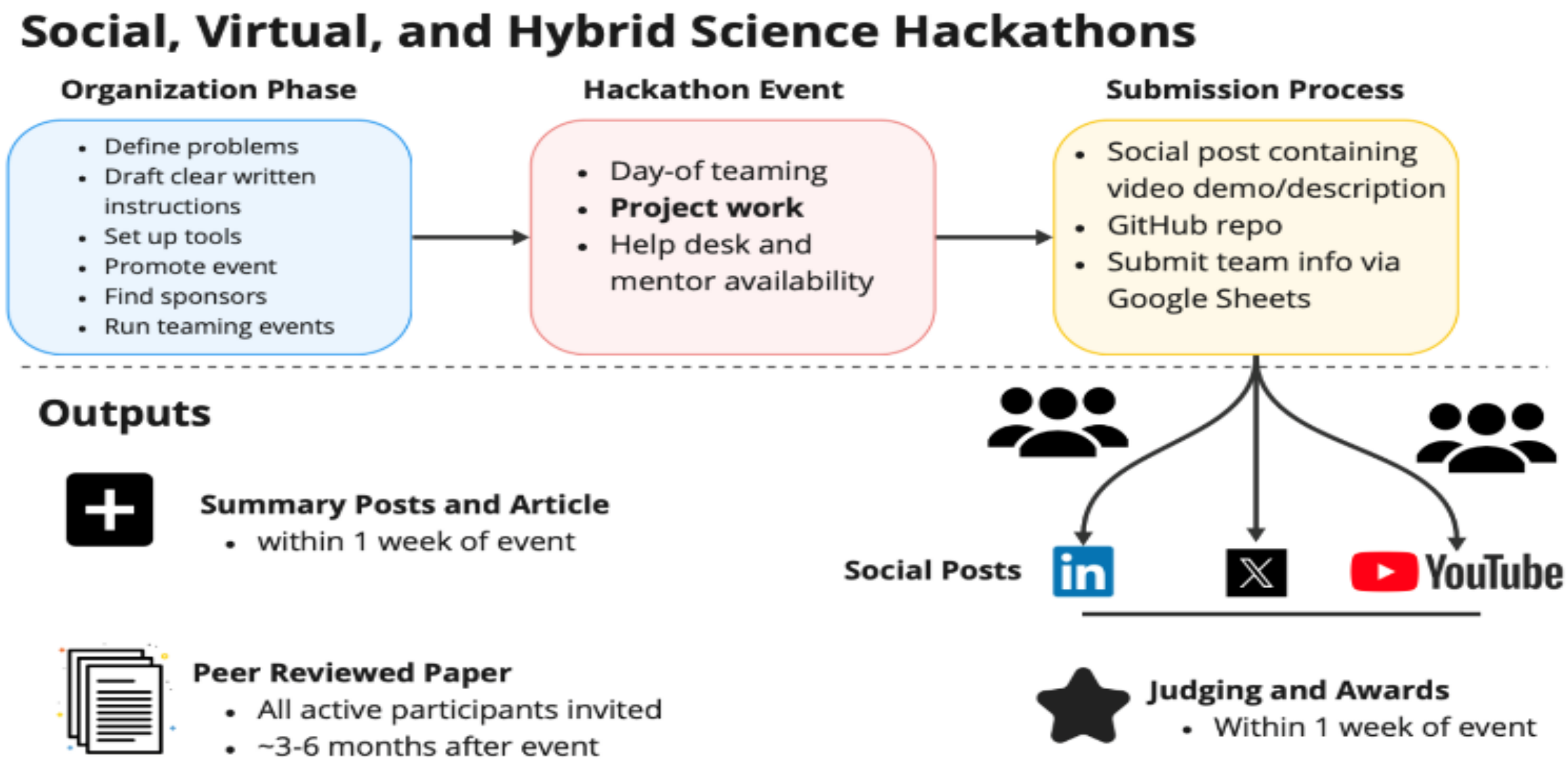

**Figure 2.** Hackathon organization progression. Figure courtesy of Ben Blaiszik, organizer of the LLM hackathon.

Hackathon preparation includes the social media campaign to disseminate the event information as described above, the development of the hackathon materials, community engagement, and participant training and team formation.

**III.C.1.a. Hackathon preparation – Logistics**

Six months before the event, we launched a dedicated **website** (https://kaliningroup.github.io/mic-hackathon/ ) that served as the central hub for all information regarding the hackathon. This website was designed not only to disseminate key details about the event but also to act as the primary communication channel with both **sponsors** and **participants.** The site featured everything from schedules and event guidelines to updates on new developments, ensuring that everyone involved was well-informed and aligned on the goals and expectations of the hackathon. Registration was made straightforward through a **Google Form** linked directly from the website. This simple yet efficient system allowed participants to sign up with ease while giving us a clear overview of the number of interested candidates and their backgrounds.

Over the course of the months leading up to the event, we also focused on community engagement. We held two **Zoom meetings** that provided an interactive space for participants to understand what to expect during the hackathon. These sessions served not only as a platform for training but also as a forum for participants to ask questions and connect with one another. We uploaded the videos on the website so anyone who was unable to attend could access the information. To further stimulate collaboration and creativity, we set up a **MIRO board** (https://miro.com/ )where participants could share their project ideas and work together even before the event started. Additionally, we produced and posted several instructional videos covering topics such as **GitHub** basics and STEM microscopy, which helped bridge the gap between traditional microscopy and modern computer vision techniques.

**III.C.1.b. Hackathon preparation – Digital Twin microscope and data sharing**

A pivotal component of the hackathon was the integration of the Digital Twin microscope, known as **DTMicroscope**(https://github.com/pycroscopy/DTMicroscope ), developed as part of the **Pycroscopy ecosystem** of packages(https://github.com/pycroscopy/pycroscopy ), this digital replica provided participants with a simulated microscopy environment, complete with starter challenges to kickstart their problem-solving processes. By offering a virtual, model of the microscope with the same scripting interface as the real microscope, participants could experiment freely without the constraints of physical equipment, thereby deepening their understanding of both microscopy and machine learning applications.

In parallel with the digital twin, we implemented a comprehensive data-sharing strategy. A variety of datasets—including STEM (scanning transmission electron microscope), AFM (Atomic force microscope) and STM (Scanning tunneling microscope) were made available. The samples included - CoAg, MoS2 nanowire and monolayer, SnSe, nanoparticles SI, abTEM simulations, and graphene data—were curated and made available through **Google Drive**. Each dataset was complemented by detailed **Jupyter Notebooks** (e.g., PZT_Data.ipynb, STEM_CoAg_Data.ipynb), which provided context, analysis instructions. Importantly, we secured the necessary permissions for each dataset, ensuring ethical use and respecting data ownership. This integrated digital ecosystem enabled participants to access real-world data and apply machine learning techniques in a hands-on, collaborative setting, making it a cornerstone of the hackathon's success. Dedicated Slack channels were created for each provided dataset and suggested problem, enabling newcomers to easily get started with a problem and find a team.

### III.C.2. Hackathon execution

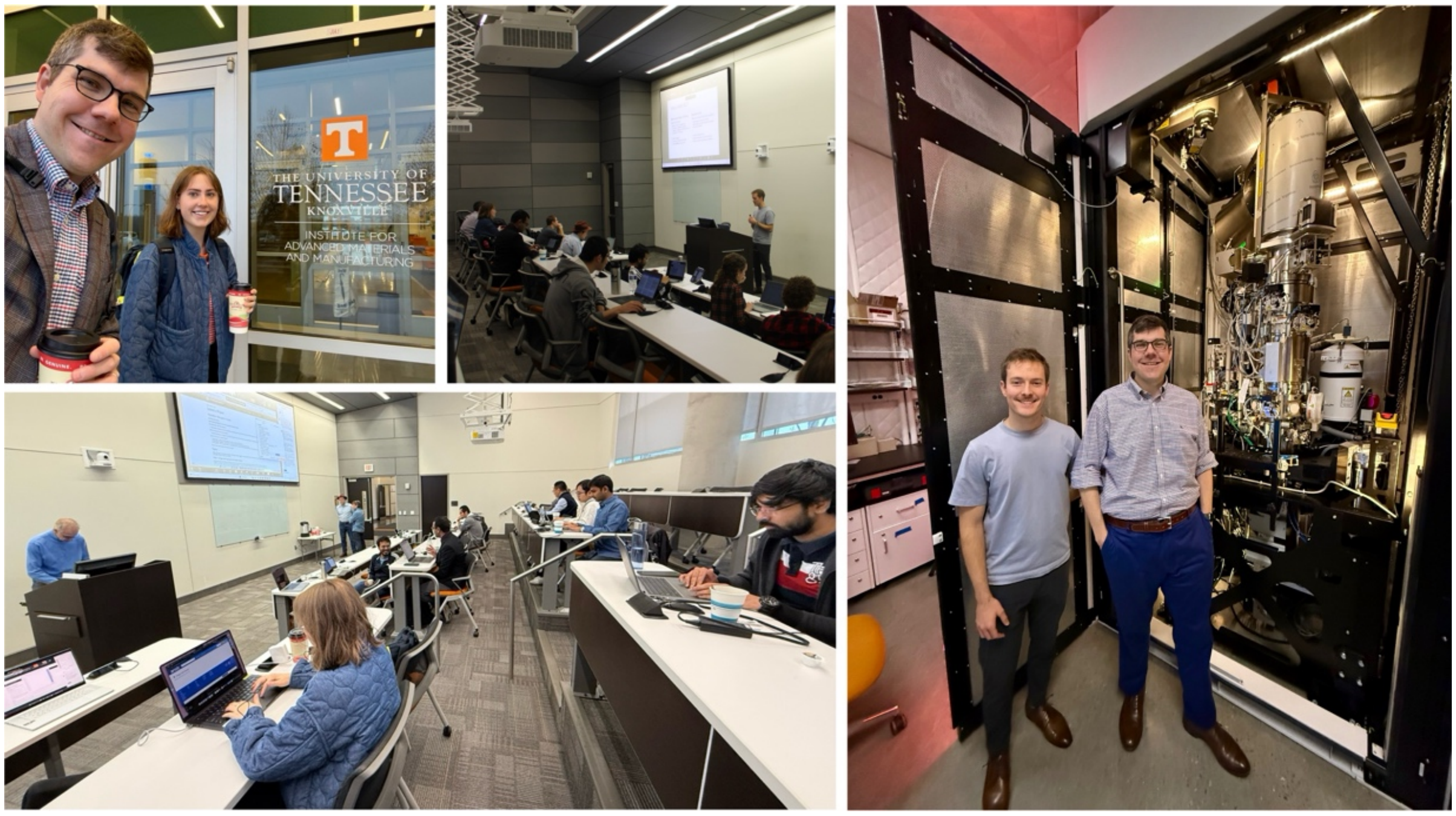

**Figure 3.** (a) Guests arriving to the in-person meeting. (b) Lecturer detailing MSA StC award during hybrid session. (c) Representative photo of the lecture hell during the hackathon working hours. (d) Two enthusiasts touring the University of Tennessee Electron Microscopy Core after the Hackathon. Photos courtesy of Steven Spurgeon.

On the day of the hackathon, Slack served as the primary communication channel, ensuring real-time connectivity among all participants. The onsite portion of the event took place over two days at IAMMM (UTK), where dedicated rooms were set up to facilitate brainstorming and collaboration. For the virtual participants, a continuous connection through Zoom was maintained for the event duration. To ensure a smooth and productive experience, the organizers led a series of structured activities, including introductions and overview talks. Then, a one-hour teaming session allowed participants to meet, exchange ideas, and form teams based on complementary skills and interests. Throughout the hackathon, two organizing team members remained on standby to provide support—whether resolving technical issues, answering challenge-related questions, or offering guidance on team dynamics. The organizers also oversaw the event's overall flow, ensuring that everything stayed on schedule and that participants had access to all necessary resources. They monitored progress, provided updates, and promptly addressed any logistical concerns. Additionally, they facilitated the submission of final projects and coordinated the judging process to ensure a fair and thorough evaluation. By providing structured support and guidance, the organizers fostered an environment that encouraged innovation, collaboration, and an enriching hackathon experience for all participants.

## IV. Project Summaries

The participants submitted ~20 projects, with a total of 80 active participants. The full summaries, codes, and YouTube presentations are available on the hackathon website. Here, we provide general overview of the submitted projects.

### IV.A. Image segmentation and shape analysis (Teams 3, 8, 10, 11)

Microscopy images allow us to visualize structures at micro, nano, and atomic scales, such as grain structures, nanoparticles, and defects in materials. These structures can affect the mechanical, electrical, and chemical properties of materials, as such, understanding these structural characteristics are crucial for optimizing and improving materials. However, traditionally it often relies on manual identification of interesting structures and manual analysis of structural correlations, which are time-consuming and prone to human bias. To overcome these limitations, four projects in this hackathon have integrated ML into microscopy analysis workflows to enable automated detection, classification, and analysis of structural features with enhanced efficiency and accuracy. Welborn et al. applied cell segmentation model to automate the detection of grain structures in AFM images of halide perovskite thin films. The cell segmentation model cellSAM[124] was originally developed for biological cells but proved effective for grain structure analysis in functional materials. Additional efforts were made to sorting grains based on size and nearest neighbors to discover regions of interest for further zoom-in imaging measurement. Kiefer et al. developed a python package, named Micrography, to analyze electron microscopy images in order to find regions of interest. By constructing a graph representation of atom positions and their nearest neighbors, the project aimed to learn more meaningful relationships about materials and predict unseen information to guide the microscope focus on regions of interest. Pinto et al. applied ML models to automatically identify the locations of nanoparticles and integrated a microscope navigator to actively search nanoparticles in the sample. They also developed an intuitive interface allowing users to view the nanoparticle search results in real time. By automating these tasks, the system significantly reduced the manual effort required in nanoparticle research. In another project, Wang et al. developed a workflow for real-time TEM image analysis of nanoparticles. A U-Net segmentation model was applied to detect nanoparticles. The workflow enabled real-time analysis of nanoparticle area, aspect ratio, convexity, etc., followed by unsupervised classification of contour shapes. A common theme among these projects is the emphasis on streamlining features of interest identification in microscopy images. The integration of ML not only accelerates the identification of these features but also enables more sophisticated analysis, such as shape classification, defect prediction, and automated search experiment.

### IV.A. Optimization of image analysis workflows (Team 1, 4, 14)

Sub-images of microscopy data are frequently used as inputs for ML analysis, where data augmentation (e.g., image cropping) is crucial for ML performance. Appropriate sub-image parameters, such as window size and step size, are essential for ensuring that relevant features are preserved while minimizing noise and redundancy. Three projects in this hackathon systematically investigated and optimized sub-image parameters to improve analysis accuracy. Barakati et al. utilized a reward-driven optimization method to balancing compactness and separation of user defined objectives for classification. By analyzing the Pareto front, the authors can identify a window size providing best results of phase separation in a 2D polycrystalline Pd-Se film. Similarly, Guinan et al. also optimized window size and step size to ensure effective classification of ferroelectric domain structures in STEM images of BiFeO3. Additionally, Wen et al. analyzed the effects of window size and step increment on the predictive accuracy of im2spec to enhance the structure-property prediction.

### IV.C. Artefacts in SPM (Team 8 and 16)

Unlike electron microscopy that uses an electron beam to scan samples, SPM uses a physical tip to scan samples. In this case, tip artifacts significantly impact the accuracy of SPM characterization, as these artifacts arise due to imperfections of probe tip (e.g., asymmetric or double tips) can obscure fine structural details, causing convolution effects that blur, duplicate, or alter the perceived surface structures and leading to misinterpretations of results. To address these problems, two projects in this hackathon leveraged ML approaches to reverse tip effect and reconstruct original surface structures. Nick et al. trained three ML models with a mean absolute error (MAE) loss and/or a structural similarity index (SSIM) loss to learn and reverse the distortions introduced by imperfect tips. They explored the feasibility of ML-based approach on tip shape estimation and image reconstruction. Narasimha et al. tested multiple ML architectures for tip artifacts removal, including a U-Net model, a hybrid ResNet-UNet autoencoder, as well as point spread functions (PSFs) to predict tip conditions. The U-Net model performs well for moderate distortions. However, it over-sharpens the image or recovers wrong features under conditions with extreme bluntness or strong double-tip effects. The hybrid autoencoder struggles for reconstructing specific structures (e.g., spiral patterns). Both hybrid autoencoder and PSF approaches struggle for double tip effects. While both projects demonstrate success in reducing distortions, challenges remain in ensuring generalizability to reverse tip artifacts under diverse experimental conditions.

**IV.D. Physics-based analytics (**Team 2, 5, 6)

Like the tip conditions that are unseen in real-time experiments, many scientific problems struggle with challenges of analyzing crucial properties that are unmeasurable due to practical limitations or resource-intensive to assess using traditional methods. In microscopy experiments, for example, inelastic scattering in diffraction requires computationally expensive models to simulate and ferroelastic-ferroelectric correlations are too subtle to measure directly. To address this challenge, two projects in this hackathon leverage ML in combination with measurable data to infer these otherwise inaccessible properties. Yoon et al. applied a U-Net model to predict inelastic scattering patterns in TEM using elastic diffraction patterns and atomic number distributions. Although the method requires further testing across diverse conditions, the results demonstrate the potential of ML to enable cost-effective prediction for inelastic scattering by learning the relationship between elastic and inelastic diffraction patterns. Bulanadi et al. applies a generative adversarial network (GAN) to infer ferroelastic-ferroelectric domain structures from standard AFM topography. Conventional piezoresponse force microscopy (PFM) measurements of ferroelectric domain structures require the application of a voltage to the sample, which can introduce unwanted modifications. GAN bypasses this issue by learning structural correlations between AFM topography and PFM imaging channels. Additionally, Addison-Smith et al. applied a dynamic mode decomposition (DMD) customized model with VampNets3 in high speed AFM to map molecules below the substrates. These projects demonstrate how ML can offer new insights into material properties that are too complex or impractical to measure directly in traditional ways.

**IV.E. Automated experiment and LLM integration** (Team 12, 15, 18)

Language models have shown the potential to enhance microscopy experimentation.[125] Three projects in this hackathon also showcase the application of LLMs in microscopy. Yin et al. developed an AutoScriptCopilot leveraging LLMs and vision language models for automation of TEM workflows. The AutoScriptCopilot allows users to dynamically adapt experiment parameters based on image quality assessment with the assistance of foundational AI models, balancing automation and expert intervention. Bazgir et al. developed MicroscopyLLM-Bench that leverages state-of-the-art vision-language models to automate key microscopic imaging tasks, such as object detection, classification, and feature analysis. This approach reduces manual annotation efforts and accelerates analysis workflows of microscopy-based research. In addition, Kalinin et al. used LLM to suggest possible descriptors of image patches for feature extraction.

**III.C.3. Post-hackathon**

**III.C.3.a. Awards**

The official closing of the hackathon is the submission of all materials in the form of the (a) GitHub repositories and (b) 3-minute video summaries to a specified email address by midnight – this method of gathering submissions individually was likely not optimal. In total, 20 teams (representing ~80 participants) submitted complete projects by the deadline. Within a day, organizers compiled these submissions into a single Google Drive, for distribution to the award judges. Two groups of award judges were formed, to vote on the general prize (funded by the AI Tennessee Initiative) and the student prize (funded by the Microscopy Society of America Student Council [MSA StC]). The student prize was $500 and free registration to PMCx60 at the M&M conference, with a presentation of the hackathon results at this conference. This prize was independently judged and awarded by the MSA StC, with the criteria that the winning team must be led by a student and must align with and promote the themes of the PMCx60. The general prize panel of judges was formed from the event organizers, and this panel graded each project on a 0-5 scale along 5 dimensions: innovation, technical execution, difficulty, teamwork, and presentation. Completing and compiling judge scores took two days, and prizes were announced on the third day.

**III.C.3.b. Winning Projects**

III.C.3.b.i. First Place: GANder: Ferroelastic–Ferroelectric Domains Observed by Image-to-Image Translation

This project implemented a generative adversarial network (GAN) based on the pix2pix architecture to predict PFM amplitude channels from PFM height channels in lead titanate (PbTiO3) thin films, aiming to non-invasively identify ferroelastic–ferroelectric domain structures. The model showed promise in capturing a-domains but struggled with overfitting, sometimes generating inaccurate predictions of reversed c-domains, suggesting the need for a larger and more diverse training dataset and improved hyperparameter optimization.

III.C.3.b. ii. Second Place: AutoScriptCopilot

This project is a framework for automated TEM experiment control, combining Thermo Fisher's AutoScript Python API with the Nodeology framework to enable state machine-based workflows and foundational AI integration. The system manages complex experimental workflows, balancing automated operations with human intervention, and incorporates feedback loops for image quality assessment, parameter recommendation, and error handling to enhance reproducibility and efficiency in TEM experiments.

III.C.3.b. iii. Third Place: Structure Discovery through Image-to-Graph Machine Learning Model

This project used a transformer-based model called Relationformer to predict molecular graphs, including atom types and bond information, directly from non-contact AFM (nc-AFM) images. The end-to-end model, trained on high-resolution simulated AFM datasets from PPAFM, aims to enhance efficiency and accuracy in identifying atomic-scale nanostructures by automating the structure discovery process

III.C.3.b. iv. Student Award: Reward based Segmentation: Phase Mapping of 2D Polycrystalline Pd-Se Phases

This project uses unsupervised algorithms, specifically Sparse Linear Unmixing and Non-negative Matrix Factorization (NMF), to identify and map distinct phases in atomic-resolution images of Pd-Se thin films. By employing a reward-driven optimization method, the approach selects optimal hyperparameters for accurate phase mapping, enabling real-time analysis and phase identification in active instruments like STEM.

III.C.3.b. v. Honorable Mentions

The “Unmasking Biomacromolecular Conformational Dynamics” project used dynamic mode decomposition (DMD) and VampNets to analyze the 2D domain motions of SARS-CoV-2 spike proteins in high-speed AFM data, achieving predictive capabilities for receptor binding domain (RBD) conformational changes critical to viral infectivity. The “Automated Grain Detection in Perovskite Materials” project adapted the CellSAM model for automated grain structure detection in AFM images, leveraging interactEM operators to efficiently sort grains by size and nearest neighbors, enhancing automated microscopy workflows. Meanwhile, the “MicroscopyLLM-Bench” project implemented a modular AI-driven pipeline using vision-language models to automate object detection, classification, and histopathological analysis of microscopy images, providing a scalable and efficient solution for biomedical research.

#### III.C.3.c. Towards Future Hackathons

Post-hackathon activities focus on capturing the insights and outcomes from the event to maximize its impact and set the stage for future initiatives. We will prepare and submit a publication summarizing the lessons learned and notable project results, ensuring that the knowledge gained is shared with the broader scientific and engineering communities. This publication will highlight the innovative approaches developed during the hackathon, the challenges encountered, and the solutions devised by the participants.

The lessons learned from the December 2024 hackathon will be carefully analyzed and used to design the next event in the series, which is planned for December 2025. By building on the successes and addressing the challenges of this hackathon, we aim to continuously improve the experience and outcomes for future participants, further advancing the integration of machine learning in microscopy and related fields.

**Summary:**

The hackathon served as a dynamic platform for the enhancement and improvement of scientific, engineering, and educational activities in several key ways. By bringing together experts in machine learning (ML), electron microscopy, and scanning probe microscopy, the event fostered the interdisciplinary collaboration that is essential for addressing complex scientific challenges. Participants built teams and worked on real-world problems, such as robust segmentation, denoising, and reconstruction of microscopy data, which are critical for advancing the field of nanoscience. The hackathon catalyzed the development of new ML tools and techniques specifically tailored for microscopy, enabling more accurate and efficient data analysis, which is fundamental to scientific discovery and innovation.

The hackathon pushed the boundaries of engineering by encouraging participants to develop and implement ML algorithms that can be integrated into microscopy hardware. This integration will lead to the creation of more advanced, autonomous instruments capable of real-time data processing and decision-making. Such innovations will not only enhance the capabilities of existing microscopy technologies but also pave the way for the development of next-generation instruments that can operate with greater precision and efficiency.

Finally, the hackathon contributed to educational activities by providing a hands-on, immersive learning experience for participants. It served as a unique training ground for students, early-career researchers, and professionals, offering them the opportunity to apply theoretical knowledge in a practical setting. The mentoring sessions, workshops, and tutorials helped participants build valuable skills in ML, data science, and microscopy. Additionally, the hackathon produced open-source resources, such as code libraries and educational materials, that can be widely disseminated and used for teaching and training in academic and research institutions. Most importantly, the hackathon laid the foundation for the development and use of digital twins, transitioning AI/ML from a post-acquisition data-analysis tool to an active participant in the research process.

**Acknowledgements:**

The organizers gratefully acknowledge the support of Office of Naval Research, Thermo Fisher Scientific, and AI Tennessee initiative that made the hackathon possible. The PyCroscopy coding research received support from the Center for Nanophase Materials Sciences (CNMS), which is a US Department of Energy, Office of Science User Facility at Oak Ridge National Laboratory. This activity is (in part) sponsored by the Artificial Intelligence Initiative as part of the Laboratory Directed Research and Development (LDRD) Program of Oak Ridge National Laboratory, managed by UT-Battelle, LLC, for the US Department of Energy under contract DE-AC05-00OR22725. This work was authored in part by the National Renewable Energy Laboratory (NREL) for the U.S. Department of Energy (DOE) under Contract No. DE-AC36-08GO28308. M.V., A.S., and S.R.S. were supported by the Operando to Operation (O2O) Laboratory Directed Research and Development (LDRD) program at NREL. G.G. was supported by the U.S. Department of Energy, Office of Science, Office of Workforce Development for Teachers and Scientists (WDTS) under the Science Undergraduate Laboratory Internship (SULI) program. The views expressed in the presentation do not necessarily represent the views of the DOE or the U.S. Government. The U.S. Government retains and the publisher, by accepting the article for publication, acknowledges that the U.S. Government retains a nonexclusive, paid-up, irrevocable, worldwide license to publish or reproduce the published form of this work, or allow others to do so, for U.S. Government purposes.

**Appendix**: Individual project writeup's

| No. | Project name |
| --- | --- |
| 1 | Phase Mapping of 2D Polycrystalline Pd-Se Phases |
| 2 | Recovery of thermal diffuse scattering in diffraction patterns via deep learning |
| 3 | Automating AFM through model-driven image segmentation and classification |
| 4 | Benchmarking Image Cropping Techniques and Data Augmentation for Structure-Property Prediction Using Machine Learning |
| 5 | Ferroelastic–Ferroelectric Domains Observed by Image-to-Image Translation |
| 6 | Unmasking biomacromolecular conformational dynamics from 2D analysis of subdomains dynamic modes and molecular kinetics |
| 7 | Micrography: GNN for Defect Detection in STEM Images |
| 8 | Mitigating Tip-Induced Artifacts in AFM Images Using Autoencoder-Based Solutions |
| 9 | GAeN: Automated Nanoparticle Detection in Microscope Images |
| 10 | On-the-fly TEM Image Analysis for Nanoparticle Synthesis Characterization |
| 11 | CT4Batt: A pipeline for Analyzing X-ray Chromatography Battery Images |
| 12 | Agentic Workflow for TEM Experiment Automation |
| 13 | AI-powered Thermal Mapping in Electron-ion Trapping Experiment (EiTEx) for Single Photon Detection |
| 14 | Unsupervised Classification of Ferroelectric Domains |
| 15 | MicroscopyLLM-Bench: Benchmarking LLM Capabilities for Open-Source Microscopy Datasets |
| 16 | Removal of tip-shape induced artifacts in AFM images using deep learning |
| 17 | Mixed Oxidation State Characterization from XAS/EELS using Machine Learning |
| 18 | Using LLM to enable bag-of-features segmentation of the metallographic images |
| 19 | Structure Discovery through Image-to-Graph Machine Learning Model |

# 1. Phase Mapping of 2D Polycrystalline Pd-Se Phases

Kamyar Barakati[1], Aditya Raghavan[1]

[1] Department of Material Science and Engineering, University of Tennessee, Knoxville

**Abstract**

Understanding phase regions and interface structures in atomic-resolution images of polycrystalline thin films is critical for uncovering their structural and functional properties. A key challenge is differentiating phase components, as variations in atomic arrangements—such as periodicity shifts or defect structures—affect the material's properties. Conventional workflows in image analysis rely on a stepwise process involving extensive hyperparameter tuning, including image segmentation, FFT processing, matrix factorization, and sparsification. However, these approaches are time-intensive and physical interpretation of extracted results is challenging. In this study, we develop an unsupervised, physics-driven workflow to identify and map distinct phases in 2D polycrystalline Pd-Se thin films, that dynamically adjusts hyperparameters based on user-defined objectives, enabling real-time interpretability and improved results accuracy. Not only useful for this task specific experiments, but this workflow can also facilitate automated decision-making in material characterization, making it well-suited for real-time applications in scanning transmission electron microscopy (STEM) and beyond.

**Methodology:**

We employ a Sparse Linear Unmixing approach to identify and map distinct structural phases in atomic-resolution images of polycrystalline 2D Pd–Se thin films. First, the image is divided into overlapping patches using a sliding window of a specified size, and each patch is transformed via FFT[1] to capture local atomic periodicities and defects.[2, 3] These FFT patches are then assembled into a data matrix and decomposed using Non-negative Matrix Factorization (NMF)[4], starting from a larger number of potential endmembers (e.g., 20). To isolate the dominant phases, we progressively impose sparsity on the NMF coefficients until only two significant components remain, which, when back-projected, reveal the primary phase boundaries. Crucial hyperparameters—such as window size, sparsity level, and the number of phases—are optimized via a reward-driven framework[5-7] that balances user-defined objectives, typically favoring phase compactness and minimal overlap among components. By examining the Pareto front[8], we identify trade-offs that yield robust phase discrimination.

The workflow and results for this study has been presented in **Figure 1, and 2**.

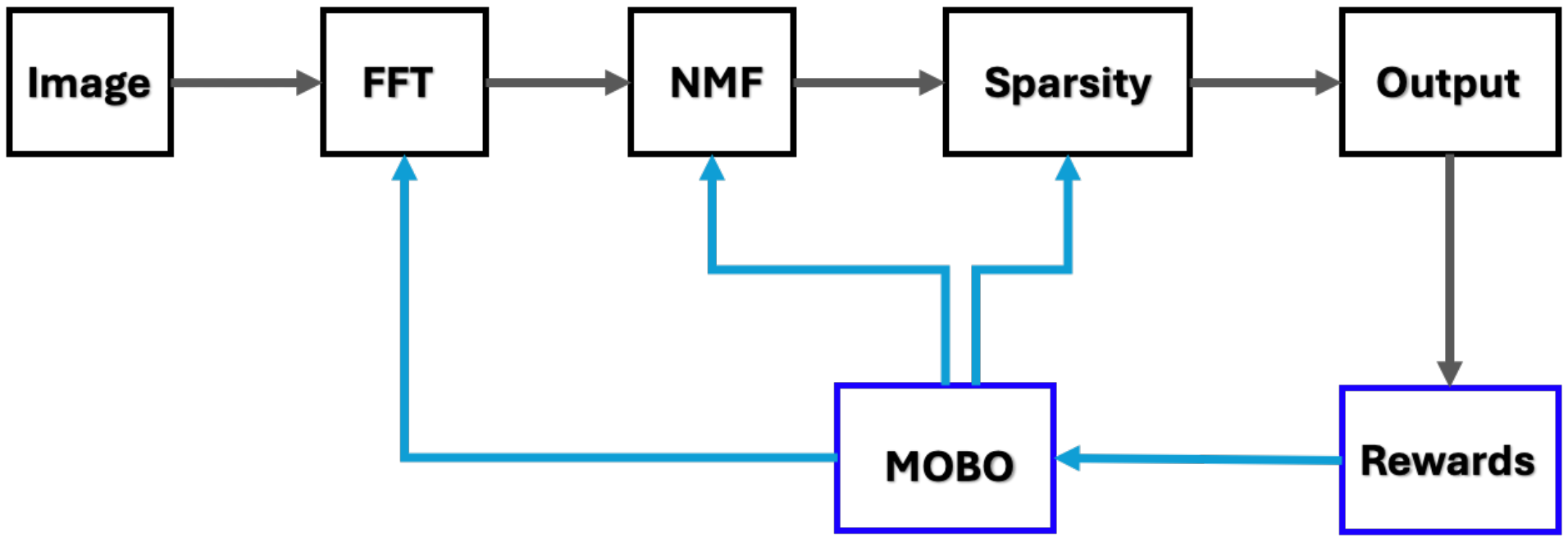

**Figure 1**:Reward driven image analysis workflow for phase mapping in Polycrystalline Pd-Se

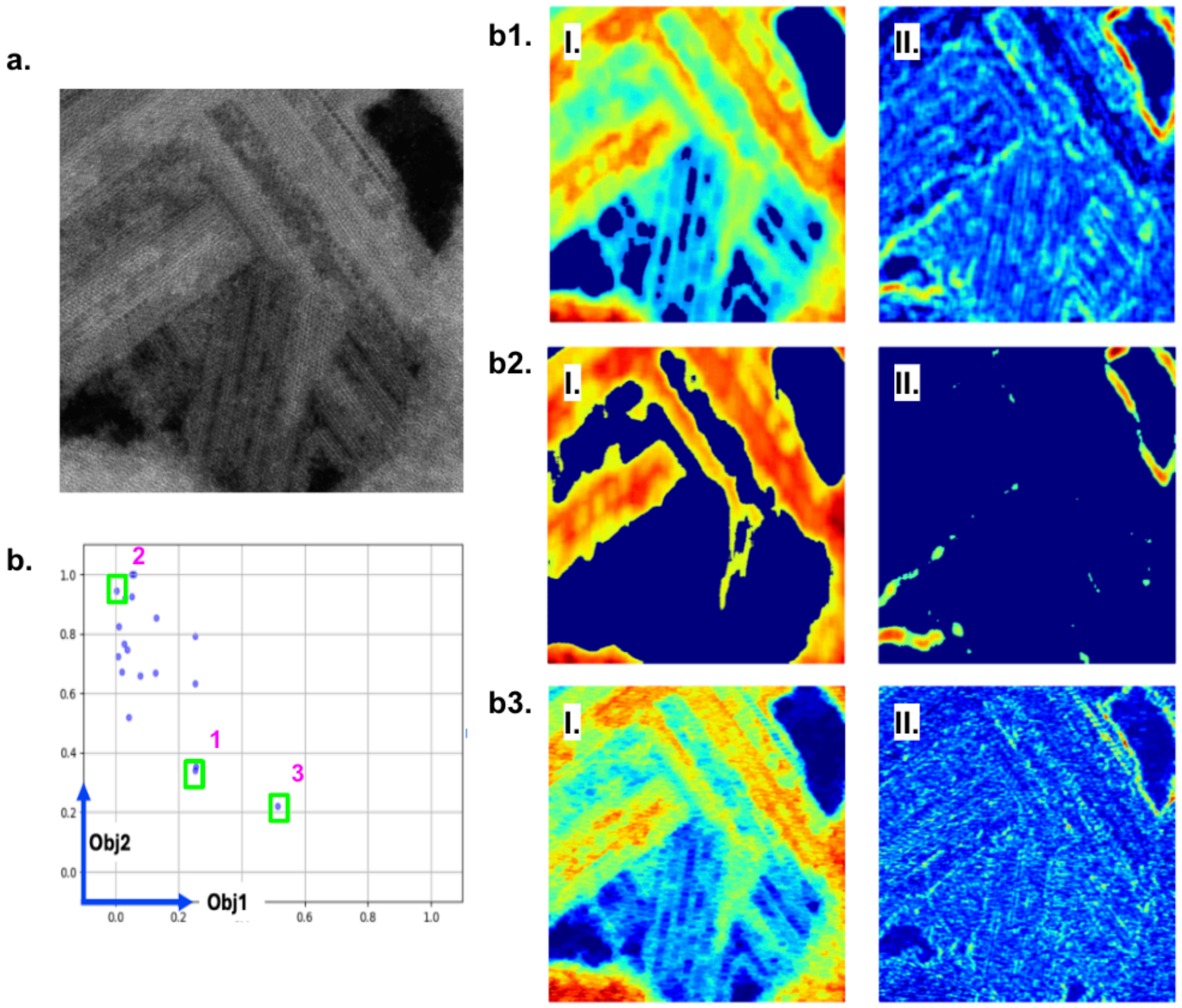

**Figure 2**: An overview of the reward-driven workflow outcomes, where (a) shows the input image, and (b) displays the Pareto front of solutions yielded by the workflow. Each point on the Pareto front encapsulates a set of hyperparameters (e.g., window size, sparsity level) alongside the corresponding reward values. In panel (b1), the solution reflects a balanced trade-off among multiple objectives, whereas in panel (b2, and b3), the workflow prioritizes one objective over others in alignment with the specific experimental goals.

## 2. Recovery of thermal diffuse scattering in diffraction patterns via deep learning

Dasol Yoon[1,2], Desheng Ma[1], Harikrishnan KP[1], Schuyler Shi[1,3], and Zhaslan Baraissov[1]

[1] School of Applied and Engineering Physics, Cornell University, Ithaca, NY

[2] Department of Material Science and Engineering, Cornell University, Ithaca, NY

[3] Department of Chemistry and Chemical Biology, Cornell University, Ithaca, NY

**Abstract**

Simulating realistic electron diffraction patterns requires a computational framework capable of accurately modeling thermal diffuse scattering (TDS). The widely used frozen-phonon method, which assumes uncorrelated atomic vibrations, effectively captures the dominant features observed in experimental patterns. However, its high computational cost often becomes a bottleneck in workflows involving large-scale or iterative electron scattering simulations. In this study, we propose a machine learning-based approach to reduce the computational burden of simulating thermal diffuse scattering. Using a supervised learning framework, we train a neural network to predict diffraction patterns that include thermal diffuse scattering, using purely elastic simulations as input and the averaged atomic number in the field of view as a label. In our test case, the model successfully learns the intensity redistribution from TDS contributions, offering a fast and efficient alternative to traditional frozen-phonon calculations. This approach can benefit a range of computational imaging applications, including ptychographic reconstructions.

**Methodology:**

We used the abTEM[1] simulation package to generate purely elastic and frozen-phonon diffraction pattern simulations for a fixed crystal symmetry, but with varying lattice constants and atomic species (crystal symmetry of strontium titanate but with different atomic species replacing the Sr, Ti and O atom sites). The generated data is divided into training and test sets by a ratio of 0.8. A conditional U-Net neural network[2] is then employed to predict the frozen-phonon diffraction patterns from elastic diffraction patterns, conditioned on the atomic number (Z) distributions. The U-Net architecture is designed to learn the mapping between elastic and thermal diffuse scattering, with a loss function that minimizes the difference between predicted and simulated frozen phonon patterns. A schematic of the network architecture and workflow is shown in Fig. 1. After training, the model predictions are compared with traditional simulations to evaluate accuracy.

To demonstrate the capability of the model to predict TDS, an example of the output from the trained model is shown in Fig. 2, comparing the (a) pure elastic simulation, (b) frozen phonon simulation and (c) ML prediction for the frozen phonon simulation. The azimuthally averaged intensities of the patterns in (a-c) are plotted in (d) showing a close match between the ML prediction and the simulated frozen phonon pattern. Here, we showed a proof-of-concept demonstration on sample structures with a specific symmetry and fixed thickness of one unit cell. Further testing across a broader range of sample symmetries, sample thickness and tilt is required to validate the model's generalizability. Future integration into phase reconstruction techniques, such as ptychography, could improve the accuracy of the forward model. This could potentially extend the capability of ptychography for the reconstruction of thicker samples and samples with heavy elements where phonon scattering effects are substantial and provide increased sensitivity to Z contrast.

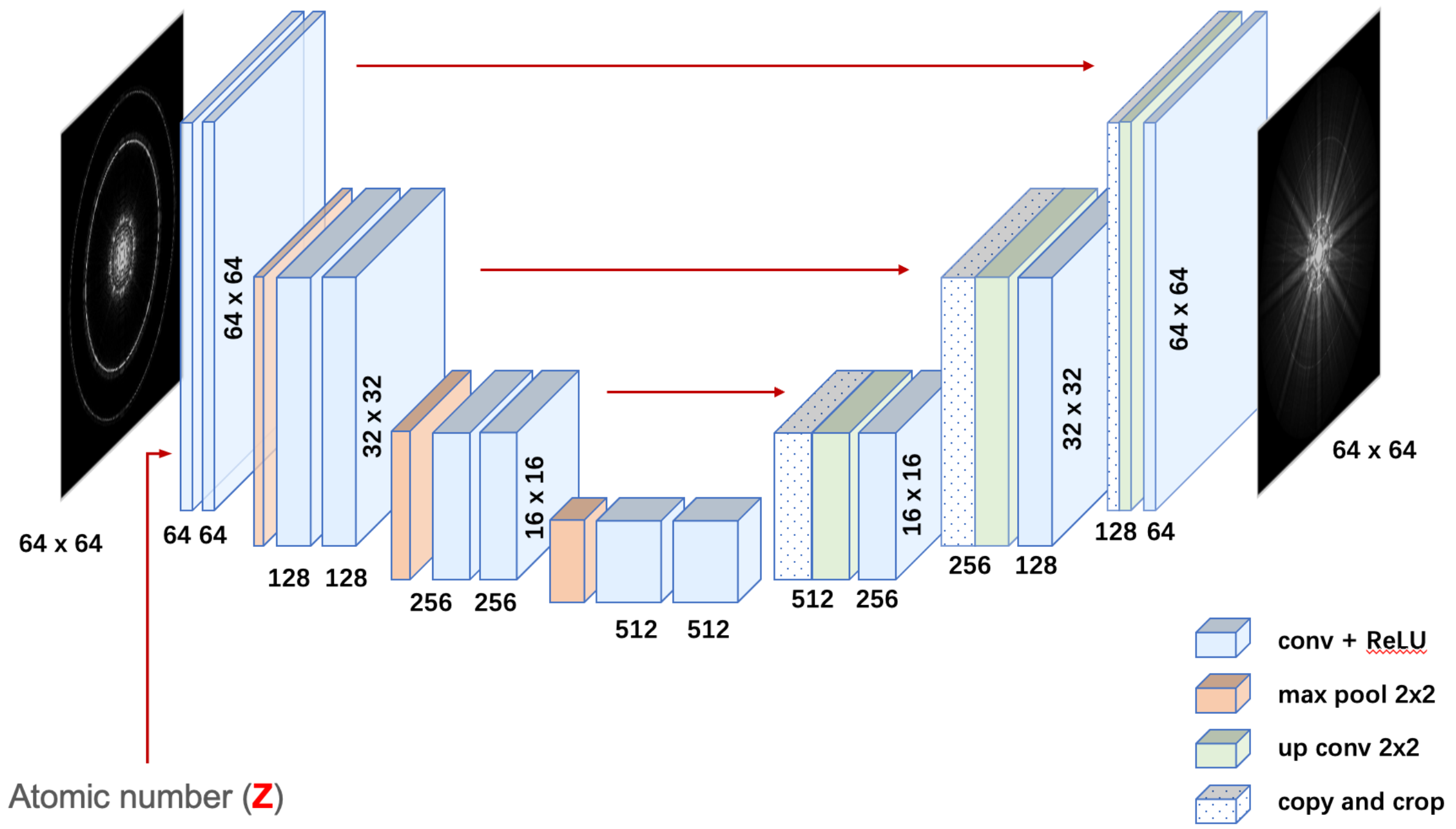

**Figure 1**: Conditional UNet architecture to predict diffraction patterns with thermal diffuse scattering from purely elastic diffraction patterns conditioned on the label of averaged Z number within the field of view.

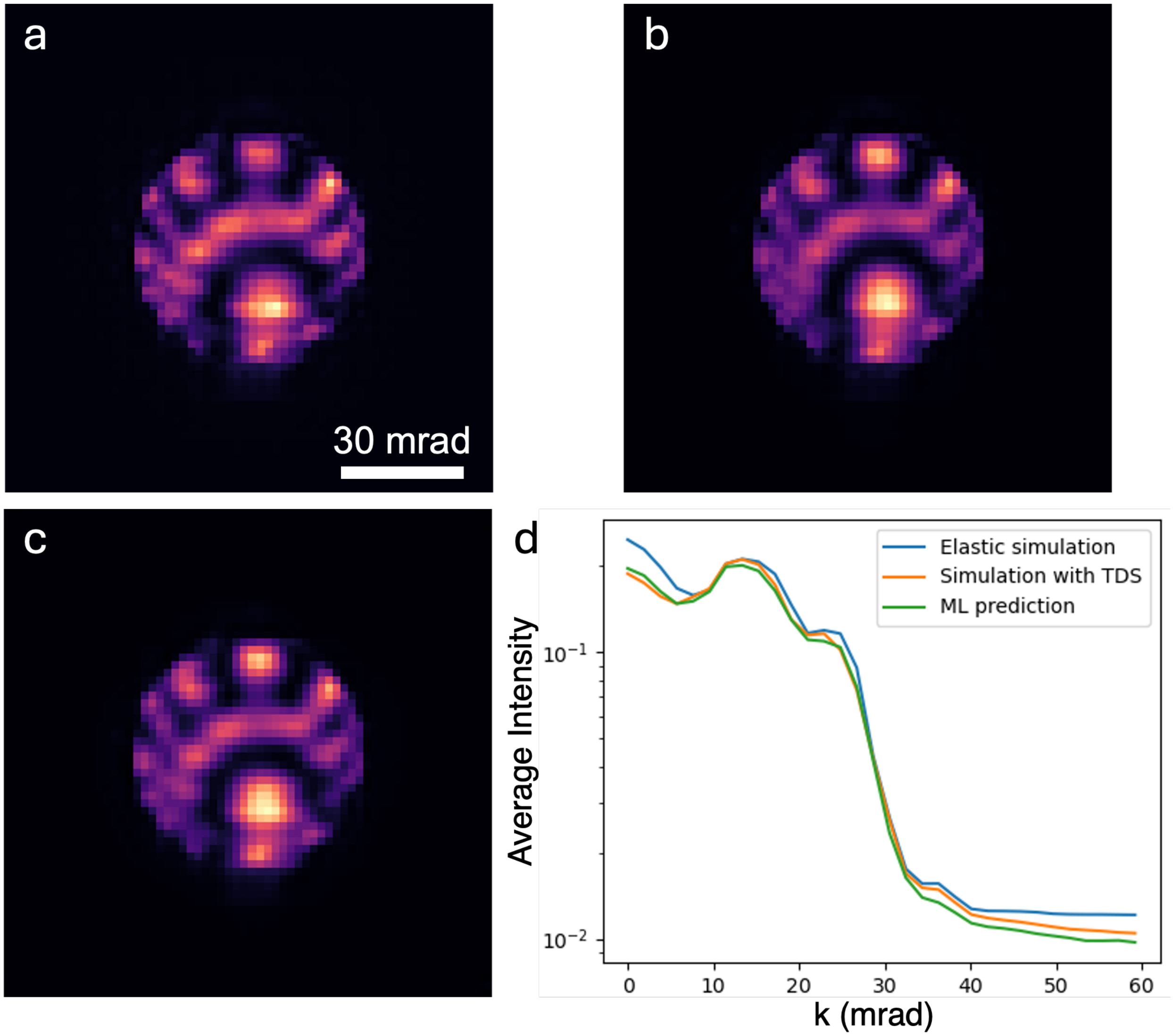

**Figure 2**: Example of simulated diffraction pattern with (a) purely elastic model and (b) frozen phonon model. (c) ML prediction of the diffraction pattern shown in (b) generated with (a) as input. (d) Azimuthally averaged intensity as a function of radial distance in diffraction space for (a-c). The ML prediction is a close match to the frozen phonon simulation with TDS.

## 3. Automating AFM through model-driven image segmentation and classification

Samuel S. Welborn[1], Mikolaj Jakowski[2], Shawn-Patrick Barhorst[2], Alexander J. Pattison[3], Panayotis Manganaris[4], Sita Sirisha Madugula[5]

[1] National Energy Research Scientific Computing Center, Lawrence Berkeley National Laboratory (LBNL), Berkeley, CA

[2] University of Tennessee, Knoxville, TN

[3] Molecular Foundry, Lawrence Berkeley National Laboratory (LBNL), Berkeley, CA

[4] Department of Nuclear Engineering, North Carolina State University, Raleigh, NC

[5]*Center for Nanophase Materials Sciences, Oak Ridge National Laboratory, Oak Ridge, TN 37831, USA*

### Abstract

Traditional microscopy relies on manual identification and targeting of features of interest, an obvious impediment to the goal of high-throughput automated microscopy. AI/ML tools are a promising alternative, but training such models for specific tasks typically requires prohibitively large amounts of training data. In this work, we repurpose cellSAM – an open-source foundation segmentation model trained to segment optical microscope images of cells [1] – to segment an atomic-force microscope (AFM) image of a perovskite, whose granular structure bears notable similarities to clusters of cells, and develop post-processing functions to select regions of interest based on various factors, such as grain size and density of interfaces. Finally, we incorporate these functions into interactEM as operators, enabling users to build custom analysis workflows tailored to the needs of their experiments.

### Methodology

We use cellSAM to segment images of perovskites, as shown in **Figure 1**. We then use different analysis functions to analyse the segmented image and/or select regions of interest for different kinds of experiments. One function bins the grains by size and selects random grains from each bin for further analysis. This is intended for studies examining the relationship between size and morphology [2]. Another function performs edge detection on the segmented image to locate grain interfaces, uses a 2D kernel to determine the local interface density, then selects the areas with highest interface density for further analysis. This is intended for studies of grain interfaces [3,4]. Another suite of functions performs statistical analysis on the population of grains, calculating the average size, average orientation and uniformity of the grains as well as providing alternative methods for grain location in the segmented image. These segmentation and analysis functions are incorporated as operators in *interactEM*,[5] an interactive flow-based programming platform that enables users to construct custom experimental workflows by chaining together various operators. Using additional operators to both receive information from and send instructions to the AFM, we create an experimental workflow capable of receiving data, analysing the image and selecting regions of interest for follow-up analysis. This workflow is shown in **Figure 2**.

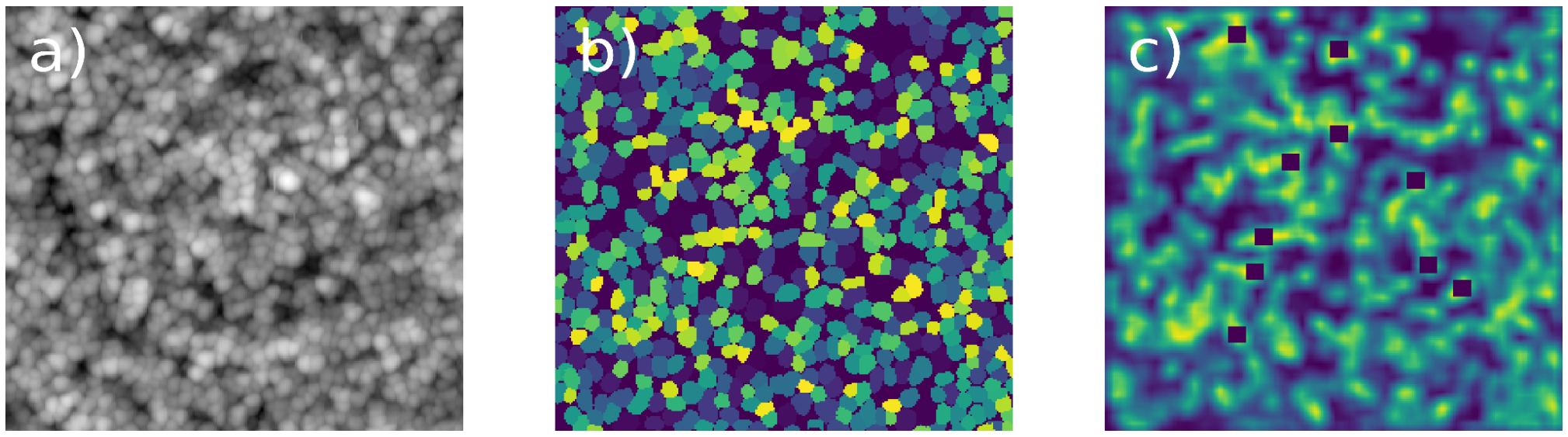

**Figure 1:** a) AFM image of perovskite. b) Image a) segmented by cellSAM. c) Map of interface density with regions of interest marked by black squares.

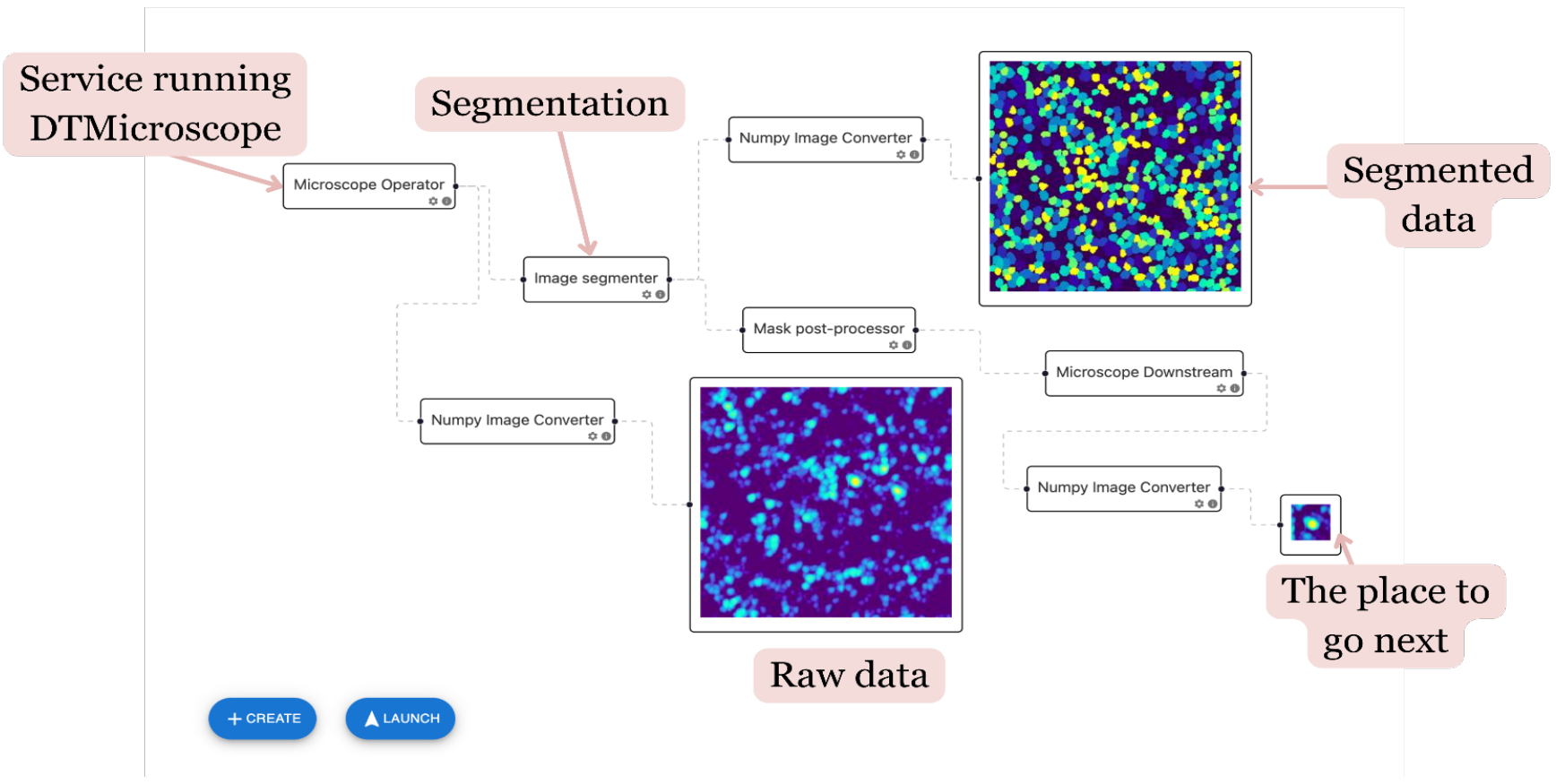

**Figure 2:** interactEM workflow for AFM perovskite image

## 4. Benchmarking Image Cropping Techniques and Data Augmentation for Structure-Property Prediction Using Machine Learning

Ningxuan Wen, Sai Venkata Gayathri Ayyagari, Andrew Balog, Vishal Kennedy, Pei Liu

### Abstract:

Recent advancements in computational power and microscopy instrumentation have enabled the generation of extensive datasets, often surpassing the analytical capacity of conventional manual methods. This is particularly true for disordered systems, where localized structural variations produce subtle yet significant differences in material properties, rendering datasets too intricate for manual analysis. As such, deep learning and machine learning methods have emerged as powerful tools to address the analytical challenges posed by complex microscopy datasets. Rigorous benchmarking is essential to elucidate impact the model's ability to generalize and accurately predict structure-property correlations. Here we report that random cropping is the most effective segmentation strategy for optimizing the performance of the encoder-decoder model to predict nanoparticle properties using EELS spectra.

### Introduction:

To enable automated analysis and the training of predictive models a method was developed by Roccapriore et al., who proposed a framework to correlate structural features of nanoparticles with their plasmonic responses [1]. Their approach employs an autoencoder network, designed to extract latent variables that capture the relationship between local particle geometries (spatial descriptors) and EELS spectra (spectral descriptors). Specifically, the model utilizes sub-images of microscopy data as inputs, paired with EELS spectra extracted from the center pixel of each sub-image, to train the network and establish these correlations. Despite these advancements, the influence of data augmentation strategies - particularly image cropping techniques - on the performance of neural networks and the relationships they infer between structural features and material properties remains insufficiently explored. Rigorous benchmarking helps in understanding how variations in cropping approaches, such as window sizes and step increments, impact the model's accuracy.

In this study, we conducted a comprehensive benchmarking of diverse image cropping methodologies to optimize the performance of the encoder-decoder model developed by Roccapriore et al. We trained using input data generated through various cropping techniques. These included systematic approaches such as grid cropping and sliding window cropping with incremental step sizes (e.g., 1, 2, 4, 8, and 12 pixels), as well as stochastic methods like random cropping with varying sample counts (e.g., 500, 1000, and 2714). To ensure consistency across datasets and maintain compatibility with the model’s input dimensions, all cropped images were resampled uniformly. We systematically evaluated the performance of various image cropping methodologies using an encoder-decoder model trained on electron microscopy datasets. Each cropping technique was benchmarked for its impact on the model's predictive accuracy, training loss, and test loss over 10 and 30 epochs. The results are summarized below:

**I) Sliding Window Cropping**

- **Step Size 1:**
    - **10 Epochs:** Mean error: **0.25897**; Training loss: **0.0007613**; Test loss: **0.0007841**.
    - **30 Epochs:** Mean error: **0.25897**; Training loss: **0.0007613**; Test loss: **0.0007841**.
- **Step Size 2:**
    - **10 Epochs:** Mean error: **0.23580**; Training loss: **0.0007429**; Test loss: **0.0007138**.
    - **30 Epochs:** Mean error: **0.03009**; Training loss: **7.6e-06**; Test loss: **7.2e-06**.
- **Step Size 4:**
    - **10 Epochs:** Mean error: **0.23507**; Training loss: **0.000751**; Test loss: **0.0013836**.
    - **30 Epochs:** Mean error: **0.03712**; Training loss: **6.7e-06**; Test loss: **1.94e-05**.
- **Step Size 8:**
    - **10 Epochs:** Mean error: **0.23885**; Training loss: **0.0007505**; Test loss: **0.0050422**.
- **Step Size 12:**
    - **10 Epochs:** Mean error: **0.23755**; Training loss: **0.0007305**; Test loss: **0.0069145**.

**II) Grid Cropping**

- **10 Epochs:** Mean error: **0.24629**; Training loss: **0.0007423**; Test loss: **0.0480156**.
- **30 Epochs:** Mean error: **0.06922**; Training loss: **6.6e-06**; Test loss: **0.0021871**.

**III) Random Cropping**

- **Full Dataset (2714 samples):**
    - **10 Epochs:** Mean error: **0.24948**; Training loss: **0.0007651**; Test loss: **0.0007217**.
    - **30 Epochs:** Mean error: **0.02988**; Training loss: **8.2e-06**; Test loss: **5.9e-06**.
- **500 Samples:**
    - **10 Epochs:** Mean error: **0.24218**; Training loss: **0.0007524**; Test loss: **0.0010061**.
    - **30 Epochs:** Mean error: **0.03398**; Training loss: **7.2e-06**; Test loss: **1.2e-05**.
- **1000 Samples:**
    - **10 Epochs:** Mean error: **0.24318**; Training loss: **0.0007631**; Test loss: **0.0007824**.
    - **30 Epochs:** Mean error: **0.03292**; Training loss: **7.6e-06**; Test loss: **8.6e-06**.

Among the cropping methodologies analyzed, **random cropping emerged as the most effective technique, demonstrating superior performance in terms of predictive accuracy and generalization.** Specifically, random cropping with the full dataset (2714 samples) achieved the lowest mean error of 0.02988 after 30 epochs, surpassing all other methods, including sliding window cropping with smaller step sizes. This success can be attributed to the inherent diversity introduced by random cropping, as it ensures that the neural network is exposed to a broad range of spatial features and structural variations, which are crucial for accurately capturing complex structure-property relationships. Unlike systematic techniques such as sliding window or grid cropping, which are constrained by fixed patterns and step sizes, random cropping introduces stochasticity that likely reduces overfitting and enhances the model's ability to generalize across unseen data.

Notably, grid cropping and sliding window cropping with larger step sizes, while computationally efficient, showed higher errors, likely due to insufficient coverage of fine-grained structural features in the data. In conclusion, the results highlight that random cropping not only achieves the best balance between accuracy and generalization but also offers flexibility in adapting to varying dataset sizes. Its ability to capture diverse spatial features makes it a reliable choice for training models in scenarios where structural complexity and variability are significant factors. These findings underscore the importance of stochastic augmentation techniques for optimizing model performance in challenging microscopy datasets.

## 5. Ferroelastic–Ferroelectric Domains Observed by Image-to-Image Translation

Ralph Bulanadi[1]*, Michelle Wang[2], and Kieran J. Pang[3]

[1]Department of Quantum Matter Physics, University of Geneva, 1211 Geneva, Switzerland

[2]Department of Electrical and Photonics Engineering, Technical University of Denmark, 2800 Kongens Lyngby, Denmark

[3]Department of Experimental Psychology, Justus Liebig University Giessen, 35394 Giessen, Germany

*ralph.bulanadi@unige.ch

### BACKGROUND AND INTRODUCTION

Ferroelectric materials maintain a spontaneous electric polarisation; ferroelastic materials maintain a spontaneous elastic strain. Many functional materials, such as lead titanate thin films ($PbTiO_3$, PTO), simultaneously exhibit phases that are both ferroelectric and ferroelastic, and this combined ferroicity has been used in transducer or sensor applications.

Ferroelectric polarisation can readily be observed at the nanoscale through piezoresponse force microscopy (PFM)—a variant of atomic-force microscopy (AFM) in which a scanning probe is placed in contact with a sample, an oscillating voltage is applied through the sample, and the consequent (piezoelectric) mechanical oscillations are read. However, the application of a voltage into the sample can cause ferroelectric switching, charge injection, or other chemical changes in the system. Further, the forces applied by tip–sample contact could change or alter the sample surface. Methods that could classify ferroelectric domains or polarisations with neither external bias, nor tip–sample contact, could therefore be useful for the study of ferroelectric materials.

Generative adversarial networks (GANs) are machine learning models which create synthetic outputs through alternating competition between a discriminator model trained to distinguish between output from a generator model and real data, and a generator network trained to fool the discriminator. Such frameworks could provide one such method to study ferroelastic–ferroelectric materials. In materials where the ferroelastic and ferroelectric domains are highly correlated, synthetic polarisation maps could be generated from initial mapping of sample topography to visualise ferroelastic–ferroelectric domains.

We have therefore trained a GAN based on pix2pix [1], on both the Height and Amplitude channels of typical PFM measurements of ferroelastic–ferroelectric PTO thin films. These films contain both *c*-domains (with an out-of-plane polarisation and out-of-plane tetragonal distortions) and *a*-domains (with an in-plane polarisation and similarly in-plane tetragonal distortions). As both *a*- and *c*-domains have different polarisation axes and strain axes from one another, we expect our model to differentiate between these domains.

Meanwhile, oppositely polarised *c*-domains, which express identical tetragonal distortions to one another, are not associated with as significant changes in topography. In these cases, we expect our trained model to be unable to determine changes in ferroelectric polarisation. This would allow differentiation between ferroelastic–ferroelectric domains, and purely ferroelectric domains.

## RESULTS

The trained model can be found at Ref. [2], while associated code is available on GitHub [3].

Training outputs of the GAN run on Height channels create synthetic Amplitude channels that visually resemble the initial height channels. An ideal output is shown in Fig. 1. Here, the topography channel (Fig. 1[a]) presents orthogonal striations that are attributed to *a*-domains. In the amplitude channel (Fig. 1[b]), which is treated as the 'Ground Truth', we likewise observe orthogonal lines we attribute to regions of ferroelastic–ferroelectric *a*-domains, distinct from surrounding *c*-domains. We also observe curved regions of reversed out-of-plane *c*-axis-oriented polarisation (which maintain the same ferroelastic strain as other *c*-domain regions). In the GAN output (Fig. 1[c]), we observe the orthogonal lines that we attribute to the ferroelastic–ferroelectric *a*-domains, but *not* the purely ferroelectric reversed *c*-domains. This suggests that our model is indeed fully capable of measuring specifically ferroelastic–ferroelectric correlations.

A less ideal example is shown in Fig. 2. Once more, the topography channel (Fig. 2[a]) shows some orthogonal striations that are attributed to *a*-domains, while the amplitude channel (Fig. 2[b]), shows both ferroelastic–ferroelectric *a*-domains and reversed out-of-plane *c*-axis-oriented polarisations. In the GAN output (Fig. 2[c]), however, we see both *a*-domains *and* *c*-domain reversal. This could suggest that the model is currently overfitted; many of the training datasets image similar regions as the test datasets, and so the model may attempt to match the test datasets too strongly to a dataset used for training. This could produce the reversed *c*-domains, even if there is no physical indication of such in the Height Channel. A larger training dataset with independent images could be used to mitigate these effects.

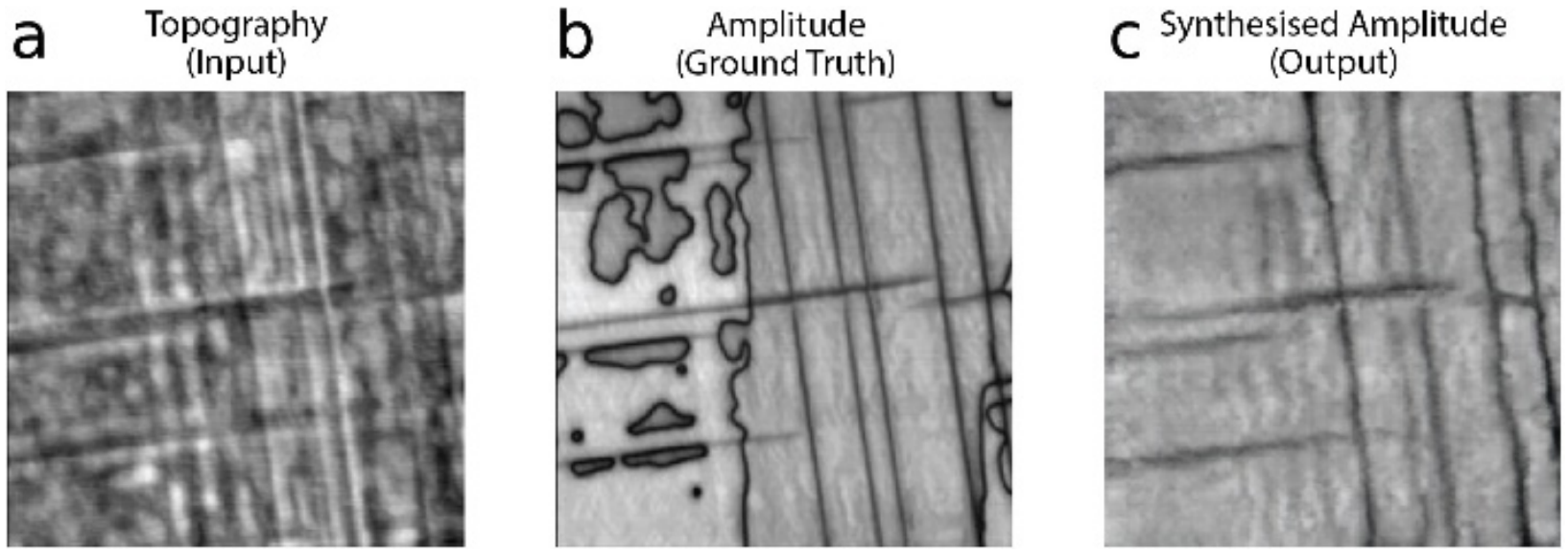

**Figure 1:** *Representative GAN output showing identification of ferroelastic–ferroelectric domains.* (a) PFM topography channel, used as GAN input. (b) PFM amplitude channel, treated by model as 'Ground Truth', showing both *a*-domains and *c*–*c* domain walls. (c) GAN output, showing the *a*-domain walls as in the PFM amplitude, but without the *c*–*c* domain walls.

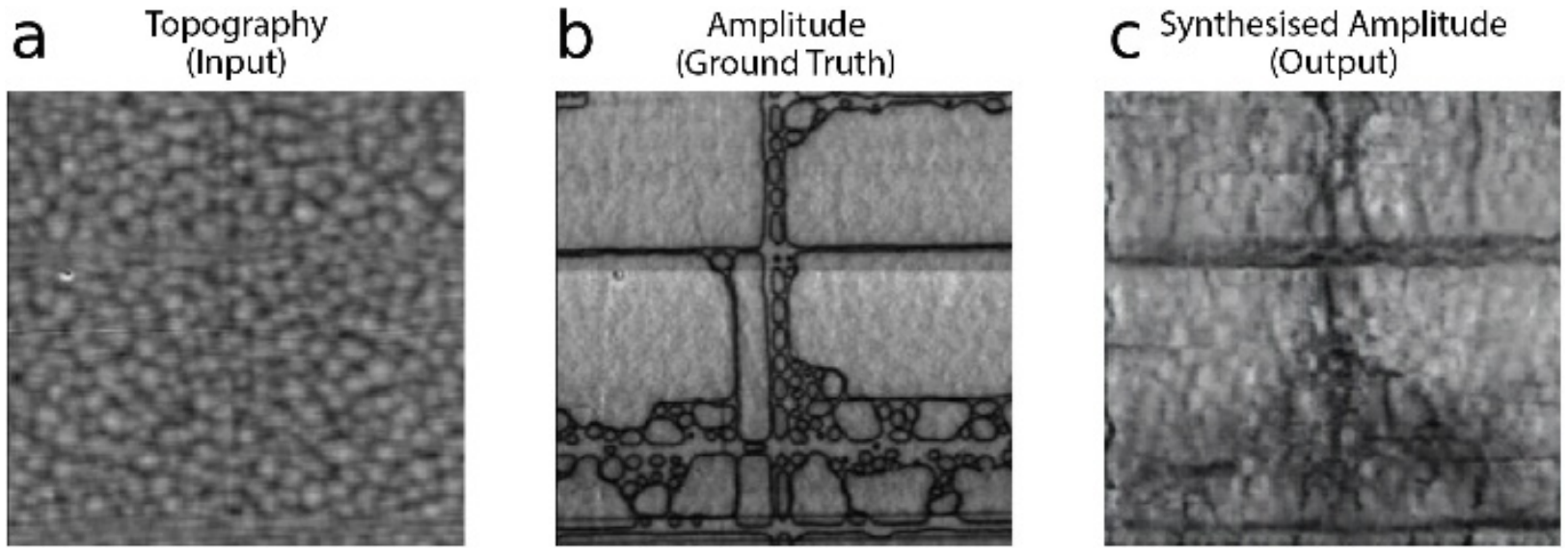

**Figure 2:** *Representative GAN output showing less ideal correlation.* (a) PFM topography channel, used as GAN input. (b) PFM amplitude channel, treated by model as 'Ground Truth', showing both *a*-domains and *c–c* domain walls. (c) GAN output, showing both the *a*-domain walls as in the PFM amplitude, along with *c–c* domain walls.

## CONCLUSIONS, OUTLOOK, AND FUTURE WORK

The present work shows promise in extracting ferroelectric–ferroelastic correlations, but further advancements could be performed on our model. Beyond improved training data, hyperparameter optimisation [4], alongside more complex network evaluation, could also be performed to improve the quality of results. Alternative physics could also be observed and tested on; ferroelectric–*ferromagnetic* correlations could be observed to study magnetoelectric applications, or scanning-probe microscopy data could be correlated with second-harmonic generation microscopy or nano-x-ray diffraction to unveil new physics. Synthetic images may also be used to supplement datasets for classification or regression models, or act as artificial training data where the authentic alternative is costly or sensitive, as is currently performed in medical imaging fields [5] [6].

## METHODS

*Sample and Image Preparation:* PTO (140 nm)/SRO (20 nm)/STO (001) thin films were grown as previously detailed in Ref. [7]. In brief, five samples were grown simultaneously by pulsed laser deposition and bombarded with varying levels of $He^{2+}$ ions. PFM was then performed in dual AC resonance tracking mode using *OPUS* OSCM-Pt tips in an *Asylum Research* Cypher with a 500 mV drive voltage. Images varied in scale from approximately 2 $\mu$m to 10 $\mu$m.

*Preliminary Data Processing:* HeightRetrace and Amplitude2Retrace channels directly taken from .ibw files from the *Asylum Research* software were directly converted to grayscale .jpg files using the gray sequential colormap in the matplotlib Python package. The minimum height (amplitude) and maximum height (amplitude) were set to black and white respectively, with no further processing. These .jpg files were then rescaled to 256 × 256 pixels, and paired height and amplitude images were combined into a single 512 × 256 pixel image. These composited images were used for model training.

*Data Augmentation:* Our initial 78 PFM images were quadrupled by rotation (±20°, randomly sampled from a uniform distribution). Random cropping and jitter were further applied, retaining at least half of the original image on each image dimension, and half of the set, randomly selected, was mirrored horizontally.

*Model Training and Evaluation:* Training was performed on our augmented image dataset for 40 000 steps, which took approximately 3.5 hrs. Model loss decayed over the majority of the training period, before reaching a plateau (Fig. 3). Although loss functions can be an inconsistent measure of model performance, conventional GAN evaluation metrics [8] are unsuitable for our data; image similarity and distance may ignore structural or high-level features vital to field-specific interpretation, and perceptual quality or other human-based measures require blind expert knowledge.

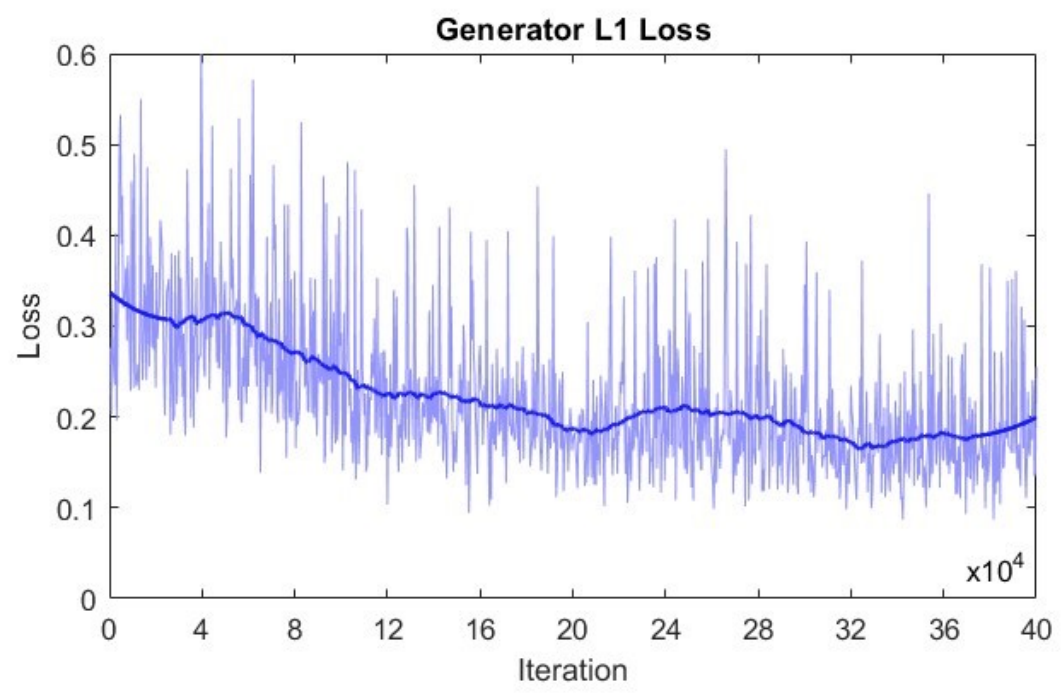

**Figure 3:** Generator L1 loss per iteration (light blue), and a Savitzky-Golay smoothed function generated over a 5000-step window (dark blue).

## ACKNOWLEDGEMENTS

$PbTiO_3$ samples were grown by Sahar Saremi and Lane W. Martin (University of California, Berkeley). R. B. acknowledges support from the Swiss National Science Foundation under Division II (Grant No. 200021-178782).

## 6. Unmasking biomacromolecular conformational dynamics from 2D analysis of subdomains dynamic modes and molecular kinetics

Ian Addison-Smith[1,3], Willy Menacho[2,3] and Horacio V. Guzman[2,3*]

[1]Department of Mechanical Engineering, Universidad de Chile, Beauchef 851, Santiago, Chile
[2]Institut de Ciència de Materials de Barcelona, CSIC, 08193 Barcelona, Spain
[3]Biophysics and Intelligent Matter Lab, E-08193 Barcelona, Spain

### Abstract

The adsorption of biomacromolecules onto substrates is critical for accurately interpreting high-resolution images acquired through tapping, multifrequency, or high-speed AFM modes. This challenge becomes particularly pronounced when proteins exhibit multiple domains with similar length scales but only subtle conformational differences, as observed in the receptor binding domain (RBD) "up" versus "down" states of the SARS-CoV-2 spike protein. In this work, we developed a dynamic mode decomposition (DMD) framework[1] using the PyDMD library[2,3] to analyze extensive high-speed AFM (HS-AFM) experimental measurements[4], alongside short molecular dynamics simulations of one spike protein laid over polarizable model substrates[5,6]. By reducing the dynamics to 12 dominant modes, the DMD model demonstrated strong agreement with experimental measurements, as evidenced by matching trends in root mean square deviations (RMSD) and radius of gyration ($R_g$) metrics. We anticipate data driven methods based on Koopman operator[7] like DMD or VAMPnets[8] will further enhance the predictability of molecular states and microstates, enabling also the mapping of interfacial features in adsorbed biomacromolecules.

### Methodology:

In this work, we address the dynamic characterization of biomacromolecules through two complementary approaches: molecular dynamics (MD) simulations and the analysis of HS-AFM images of adsorbed spike proteins. For the HS-AFM measurements, multipage TIFF files were processed using the Python imaging library to extract individual image frames. Subsequently, these frames were reshaped into a data matrix wherein each column corresponds to a flattened image. The DMD framework was then employed on this matrix, again using a 12-mode reduction, to capture the dominant spatial and temporal features in the experimental data. The reconstructed snapshots from the DMD model were visually compared to the original AFM images at selected frames for two spike variants: WT (Figure 1 a-b) and Omicron (Figure 1 c-d). Furthermore, the nonzero pixel intensities—interpreted as a macromolecular distribution—were used to compute the center of mass and the radius of gyration for each frame. The temporal evolution of these structural metrics and RMSD calculations relative to a reference frame provided a quantitative measure of the model's performance. As a proof-of-concept for interpreting and mapping the experimental results into molecular resolution with MD simulations, we performed a short MD simulation of the spike protein in laid-over conformation, following the protocols used in ref.[5]. The structural and morphological metrics were computed to establish a baseline understanding of the system (Figure 1 e-h).

First, we identify the protein and glycans residues that influence adsorption onto the AFM simulated substrate[1]. Glycans are well distinguishable in simulations and their influence in protein dynamics have been well validated[9-10]. However, less is known about the role in the adsorption mechanisms. Here, we study a trajectory subjected to singular value decomposition to extract the principal components (PCs), effectively reducing the dimensionality of the system while retaining its essential dynamical features. Figure 2, shows the reconstructing the original trajectory with BOPDMD[11] and a snapshot of the used fragments, showing good RMSD agreement with both the semi-flexible protein and flexible glycan.

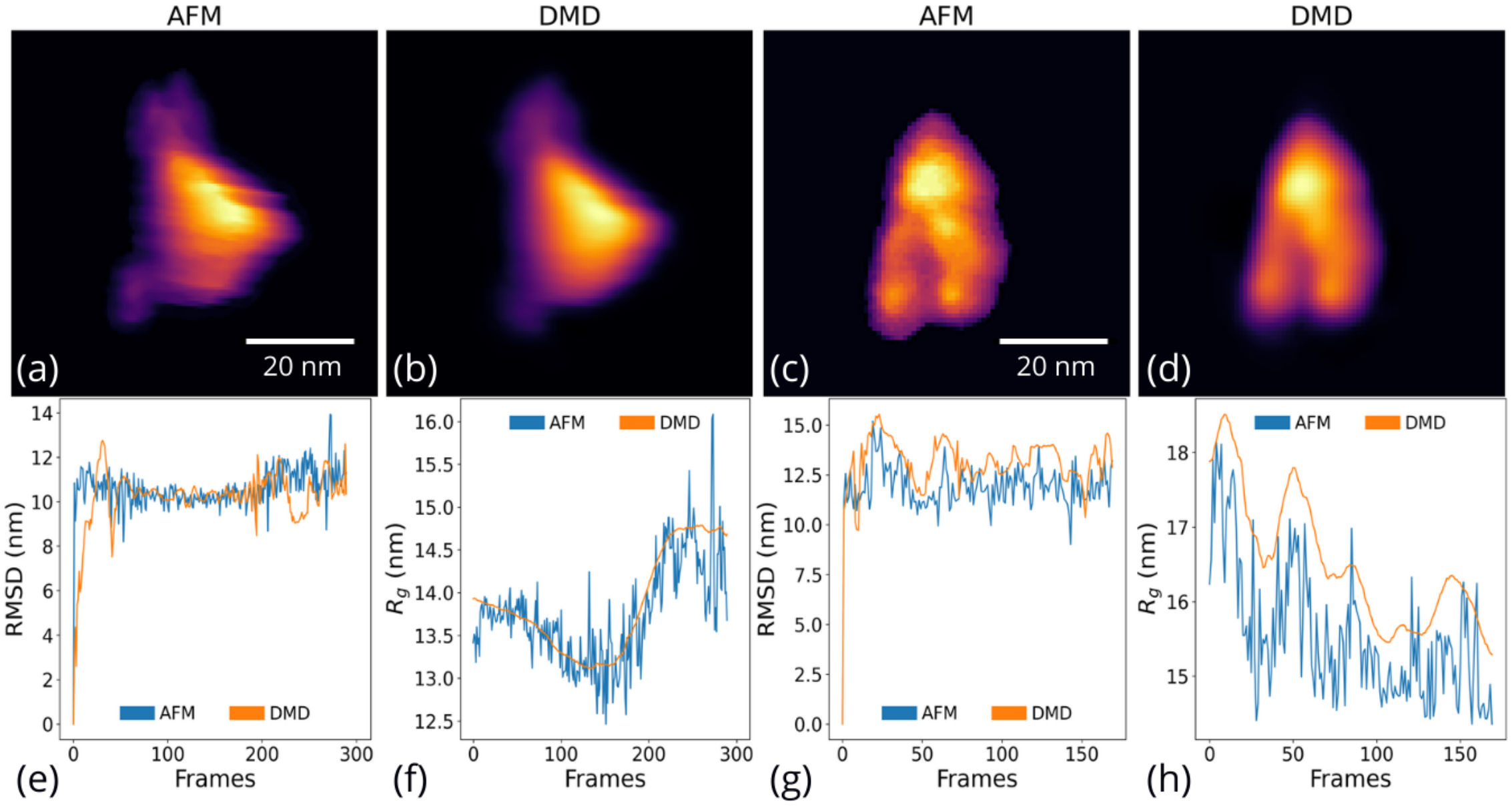

Figure 1. DMD applied to HS-AFM images of the SARS-CoV-2 spike protein. WT variant (a) AFM experiments and (b) DMD reconstruction. The Omicron variant is shown in (c) Experiments and (d) DMD reconstruction. (e, f) RMSD and $R_g$ vs. number of frames for AFM and DMD WT variant, respectively. (g, h) similar to (e-f) applied to the data for Omicron.

Collectively, these analyses demonstrate that DMD captures the essential slow dynamics of the experimental systems and offers a pathway toward integrating experiments with advanced techniques (e.g., VAMPnets and Koopman operator approaches) for improved mapping of states/microstates with molecular tractability.

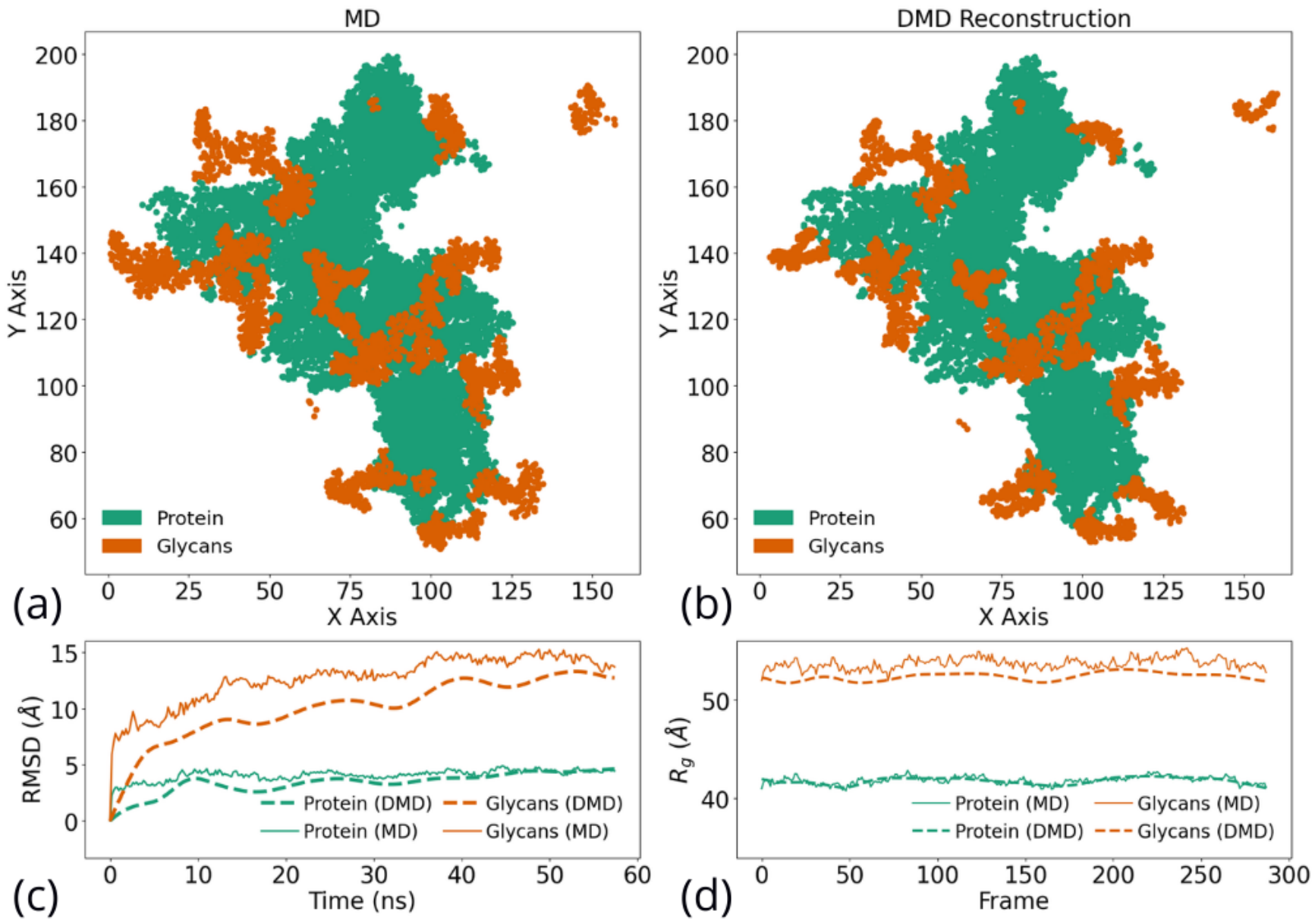

Figure 2. Structural comparison of the spike protein using MD simulations and DMD analysis. (a) Two-dimensional projection of the spike protein from MD simulations. (b) BOPDMD-based projection of the spike protein. (c) RMSD comparison between MD and DMD representations. (d) $R_g$ evolution for MD and DMD models, illustrating structural consistency between both approaches in terms of $R_g$, also for flexible Glycans.

## 7. Micrography: GNN for Defect Detection in STEM Images

Alexander Kiefer[1], Nicholas Furth[1]

[1]University of Tennessee - Knoxville

**Abstract**

Micrography introduces a novel graph-based framework for analyzing electron microscope images of materials, addressing fundamental limitations of traditional pixel-based approaches. While conventional methods like convolutional neural networks and autoencoders treat images as pixel arrays, our approach reconstructs the physical arrangement of molecules as graphs, where nodes represent individual molecules with their spatial and morphological features, and edges capture their nearest-neighbor relationships. This transformation preserves critical structural information that is often lost in pixel-space transformations. By applying graph neural networks to these molecular arrangements, we can detect subtle structural anomalies, characterize material properties, and identify defects with significantly improved interpretability. Our comparative analysis demonstrates that graph-based representations capture material properties that remain hidden to traditional methods, particularly in regions with structural irregularities. The framework lays groundwork for intelligent, guided microscopy that can autonomously direct imaging resources toward regions of scientific interest, fundamentally changing how electron microscopy is utilized in materials science research. We demonstrate this on an open access dataset of Sm-doped BFO STEM images [1].

### Methodology

Our methodology combines image analysis with graph theory to create a more powerful and interpretable framework:

1. **Image Preprocessing and Molecular Identification:**
    - Raw STEM images undergo binarization, erosion, and dilation to isolate distinct molecular regions
    - Connected component analysis labels individual molecules
    - Size-based filtering removes artifacts and noise
    - Statistical features (centroid coordinates, spatial distribution, size) are extracted for each molecule
2. **Comparative Segmentation Approaches:**
    - Gaussian Mixture Models provide initial classification of molecular types
    - A Variational Autoencoder enhances segmentation, particularly in boundary regions
    - Statistical analysis of both approaches validates molecular classifications
3. **Graph Construction and Topological Analysis:**
    - Molecules become graph nodes with their extracted features as node attributes
    - K-nearest neighbor connections (K=4) establish edges between molecules, capturing physical proximity
    - Edge weights reflect the Euclidean distances between molecular centroids
    - Nearest neighbor distance distributions reveal material lattice properties and structural patterns

4. **Graph Neural Network Enhancement:**
    - A Graph Autoencoder learns embeddings that preserve both molecular features and their spatial relationships
    - The embedding space enables identification of structural motifs and anomalies not visible in pixel space
    - Unsupervised clustering in the embedding space reveals regions with similar characteristics
    - Graph visualization techniques provide intuitive interpretation of material properties
5. **Structural Insights Through Graph Analysis:**
    - Unusual nearest-neighbor distances highlight potential defect regions
    - Deviations in graph structure identify areas of material strain or lattice distortion
    - Neighborhood feature distributions reveal local material composition variations
    - Spatial correlation of graph features provides insights into long-range order and disorder

This graph-centric approach significantly augments traditional methods by explicitly preserving the spatial relationships between molecules, enabling multi-scale analysis from individual molecules to global structure, and providing a more physically interpretable representation that directly reflects the material's actual structure rather than pixel-based abstractions. The resulting framework not only improves defect detection but creates the foundation for guided microscopy where regions with unusual graph properties can be automatically targeted for more detailed examination.

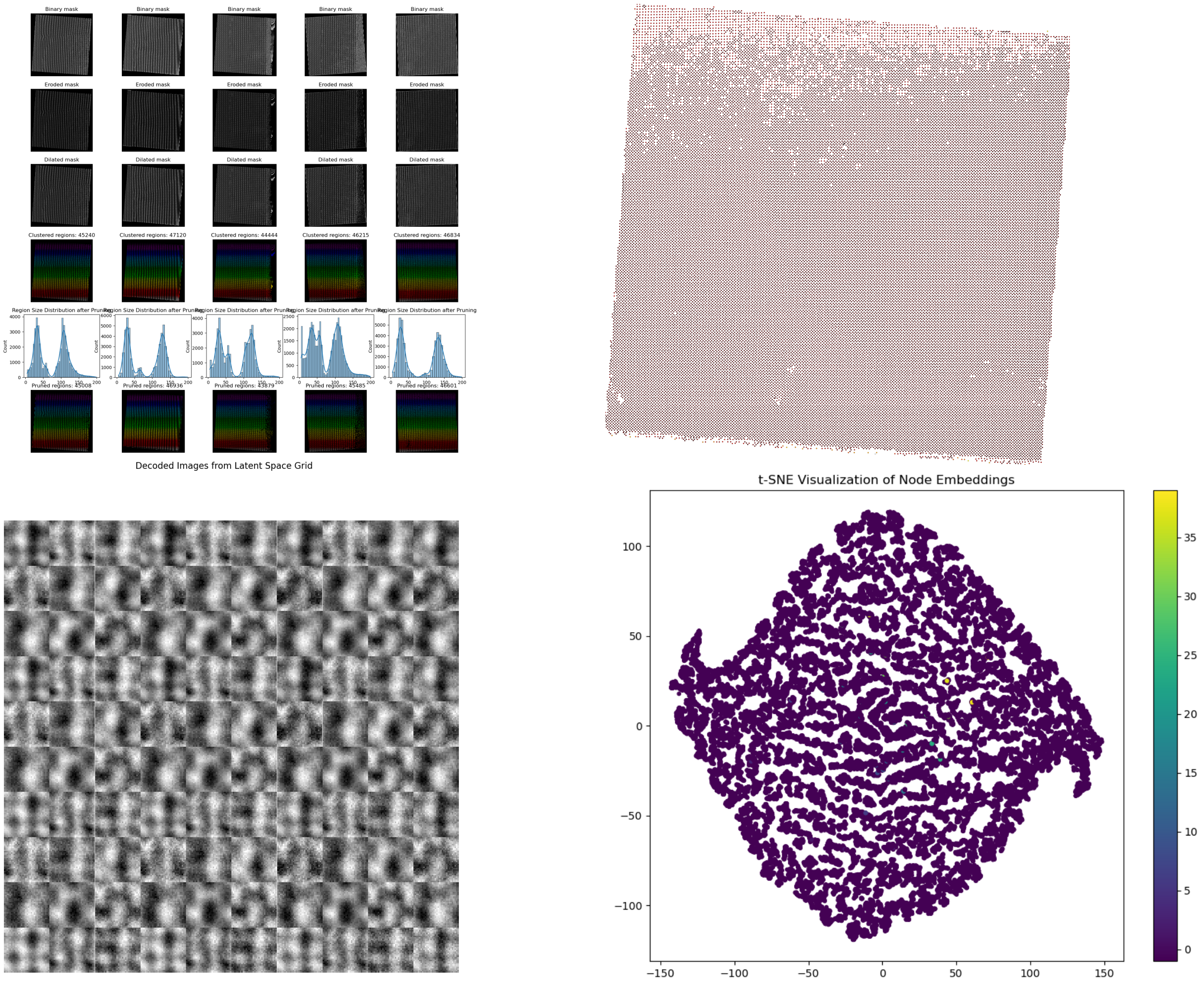

*Figure 1: Processing Pipeline*

## 8. Mitigating Tip-Induced Artifacts in AFM Images Using Autoencoder-Based Solutions

Nikola L. Kolev[1,2], Mikhail Petrov, Viktoriia Liu[7], Sergey Ilyev[5], Srikar Rairao[6], Tommaso Rodani[3,4]

[1]University College London, [2]London Centre for Nanotechnology, [3]AREA Science Park, [4]Università degli Studi di Trieste, [5]Moscow Institute of Physics and Technology, [6]University of Tennessee, Knoxville, [7]Aspiring Scholars Directed Research Program, Fremont, California.

Abstract

Nanoscale imaging via Atomic Force Microscopy (AFM) and Scanning Tunneling Microscopy (STM) is essential for analyzing atomic or molecular level details, but image quality[1-2] is often degraded by probe tip-induced artifacts. These distortions lead to unreliable data interpretation, impacting materials science, nanotechnology, and biological imaging. This project addresses the need to automatically mitigate such artifacts. Traditional correction methods are often manual, inefficient, or incomplete. Deep learning, particularly autoencoder-based approaches, offers a promising data-driven alternative for effective image restoration.

Methodology

This work focused on mitigating tip-induced artifacts in AFM images using autoencoder-based deep learning models, with preliminary exploration in STM.

**Dataset:** For AFM studies, synthetic training data were generated from a single AFM scan by convolving the scan with Gaussian kernels of randomly sampled standard deviations (tip radius, range [0.2,0.8)) and center coordinates to simulate blunt tip effects as shown in Figure 1. The use of a single source image for training is a limitation acknowledged to potentially lead to overfitting. For STM, an experimental dataset was used, where blunt tip defects were artificially introduced to test transferability.

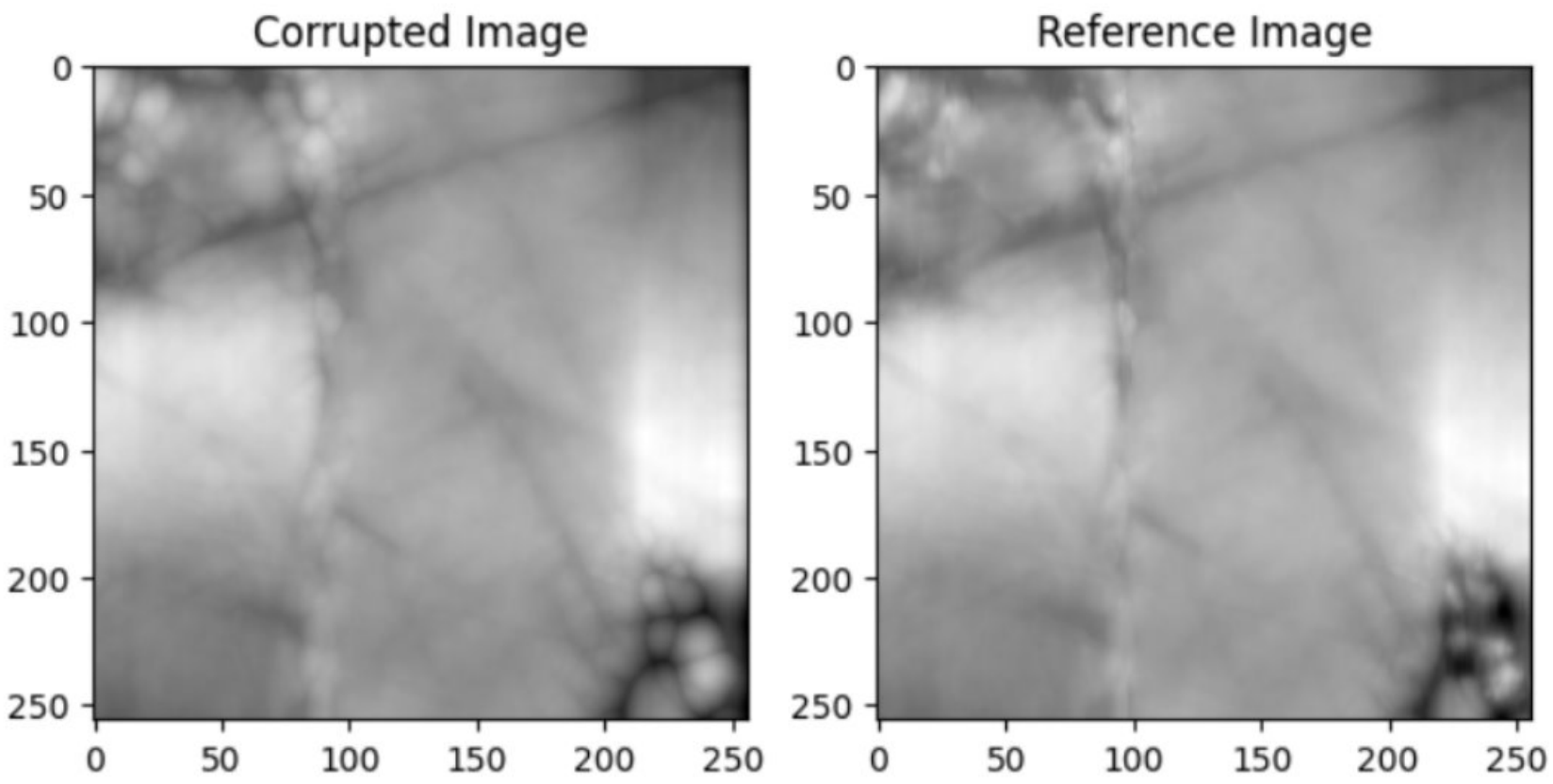

Figure 1: AFM image analysis: corrupted (left), ground truth or reference (right).

**Models & Approach:** Three autoencoder architectures were explored:

- **Model 1:** Encoder-decoder architecture without skip connections, trained for 70 epochs on the difference (noise) between corrupted and uncorrupted AFM images[3]. Loss:

combined Mean Absolute Error (MAE) and Structural Similarity Index Measure (SSIM) loss.

- **Model 2:** A compact autoencoder (33k parameters) trained for 1000 epochs on 1200 STM images (128×128px). Loss: MAE. Based on the architecture from the recent article.[4]
- **Model 3:** A larger autoencoder (6.7M parameters) with skip connections, trained for 75 epochs on 100 STM images (256×256px). Loss: combined MAE, SSIM, and a custom gradient loss.

Performance was benchmarked against Gwyddion, a standard SPM image analysis software[5]. The primary objective was to investigate the feasibility of these autoencoder solutions for artifact correction.

## Results and Discussion

This section evaluates our autoencoder-based models for mitigating tip-induced artifacts in AFM and STM images, benchmarked against Gwyddion software.

Two synthetically corrupted AFM images were analyzed. *Corrupted Image 1* was generated with a simulated tip radius of 0.0580 and center at [0.4837,0.2528]. *Corrupted Image 2*, was generated with a tip radius of 0.0750 and center at [0.3292,0.4286].
Quantitative metrics for Corrupted Image 1 and 2 are in Table 1. Gwyddion achieved a PSNR of 16.5. Model 1, yielded a lower PSNR (15.2) but a higher VIF (0.107) than Gwyddion (0.086), suggesting potentially better perceptual information preservation. Model 2 performed poorly in image-wise metrics (PSNR 4.04, VIF 0.008).

| **Method** | **PSNR Score** | | **VIF Score** | |
|---|---|---|---|---|
| | **Image 1** | **Image 2** | **Image 1** | **Image 2** |
| **Corrupted Image** | 16.9 | 24.8 | 0.086 | 0.069 |
| **Gwyddion** | 16.5 | 22.1 | 0.086 | 0.063 |
| **Model 1** | 15.2 | 15.2 | 0.107 | 0.057 |
| **Model 2** | 4.04 | 4.04 | 0.008 | 0.008 |

**Table 1: Comparison of results against Gwyddion baseline**

Gwyddion achieved a PSNR of 22.1 and VIF of 0.063. Model 1 scored lower (PSNR 15.2, VIF 0.057). Model 2's metrics remained low. Model 1's variable performance relative to Gwyddion suggests sensitivity to artifact characteristics.

### STM Image Artifact Mitigation

Model 3 was tested on an STM dataset with artificially introduced blunt tip artifacts. Figure 3 shows a corrupted STM image (tip radius: 0.0800, center: [0.3998,0.4218]), its reference, and Model 3's reconstruction.

| Metric | Value |
|---|---|
| Validation Loss (75 epochs) | 0.026 |
| SSIM | 0.9534 |

**Table 2: Model 3 Validation Results**

In summary, autoencoder models show promise. Model 1 had varied AFM performance against Gwyddion, notably achieving a higher VIF for one image. Model 3 demonstrated effective structural correction for STM images. Further research is needed to enhance performance and generalization.

**Code:**

**Model 1:** https://github.com/nickkolev97/Tip_deconvolution_hackathon

**Model 2:** https://github.com/ilev-sergey/Tip_deconvolution_hack/blob/main/STM_deep_learning_deconvolution.ipynb

**Model 3:** https://colab.research.google.com/drive/1rH3imv9pJ356FbI1zod1dp3PmIhCJkQI

## 9. GAeN: Automated Nanoparticle Detection in Microscope Images

### Abstract

High-resolution instruments such as Transmission Electron Microscopes (TEM) are indispensable for nanoparticle analysis, offering detailed imaging of these structures. However, manual operation of these systems can be slow and inefficient, especially when processing large datasets. One of the main challenges is the detection and precise localization of nanoparticles within a sample, a task that becomes tedious and time-consuming when performed manually. To address this issue, we propose an automated nanoparticle detection system based on Machine Learning. The system employs two convolutional neural network (CNN) models to autonomously detect and locate nanoparticles within the sample. The first model, based on the ResNet-34 architecture, identifies the presence of nanoparticles, while the second model, implemented using U-Net, determines their center of mass (COM) coordinates. A motion algorithm, integrated with these models, allows the system to navigate the sample, center nanoparticles, and acquire images automatically. The entire system is demonstrated through a Graphical User Interface (GUI) that showcases its performance on a simulated sample, thereby enhancing research efficiency by automating data acquisition.

### Methodology

This study is divided into two main phases. The first phase focuses on developing a Machine Learning (ML) model capable of detecting nanoparticles (NPs) and determining their spatial coordinates. The second phase involves implementing a microscope motion algorithm that leverages the trained models to enable automated navigation and image acquisition within a simulated sample, with potential future integration into a real microscope system.

For nanoparticle detection and localization, we employ two ML models based on Convolutional Neural Network (CNN) architectures. The first model, responsible for detecting the presence of nanoparticles, is based on the ResNet-34 architecture.[1] It was trained using images containing NPs and background images without NPs. After training, the model outputs the probability that a given image contains a nanoparticle. The second model, designed to determine the center of mass (COM) of a detected NP, is implemented using U-Net, a widely used CNN for image segmentation in electron microscopy.[2] This model is trained on NP images where the COM has been manually labelled. After training, it provides the coordinates of the NP COM within the image.

Regarding the motion algorithm, it begins at an initial point of interest defined by the researcher. From this starting point, the sample is divided into a grid, generating a sequence of coordinates that guide the navigation. These coordinates follow a spiral pattern, optimizing exploration and minimizing unnecessary movements. As long as no nanoparticles are detected, the system performs systematic horizontal scanning within each grid cell. Once a cell is fully scanned, the algorithm moves to the next coordinate in the predefined spiral sequence (Figure 1a). If the ResNet-34 model detects a nanoparticle during this process (with a probability above 0.9), the algorithm enters a centering phase (Figure 1b). It analyzes the probability of an image contains a nanoparticle and iteratively adjusts the microscope's position to align with the highest probability region, ensuring full NP capture (Figure 1c). Once the highest probability region is reached, the second model (U-Net) is activated to provide the COM coordinates of the NP. Finally, the microscope centers the image on the NP's COM and acquires the image (Figure 1d). After acquisition, the microscope resumes global motion until a new NP is detected.

The graphical user interface (GUI), developed in Python (Figure 2), is structured as follows: the right panel contains search parameters, including the number of nanoparticles to detect, a start button and the probability of NP in the microscope view image. The central-right section provides a live feed from the microscope, while the central-left section displays the sample. Captured images are shown at the top. To create a realistic testing environment, the simulated sample consists of cropped nanoparticle images combined with noise patches, ensuring robust evaluation of the detection system.

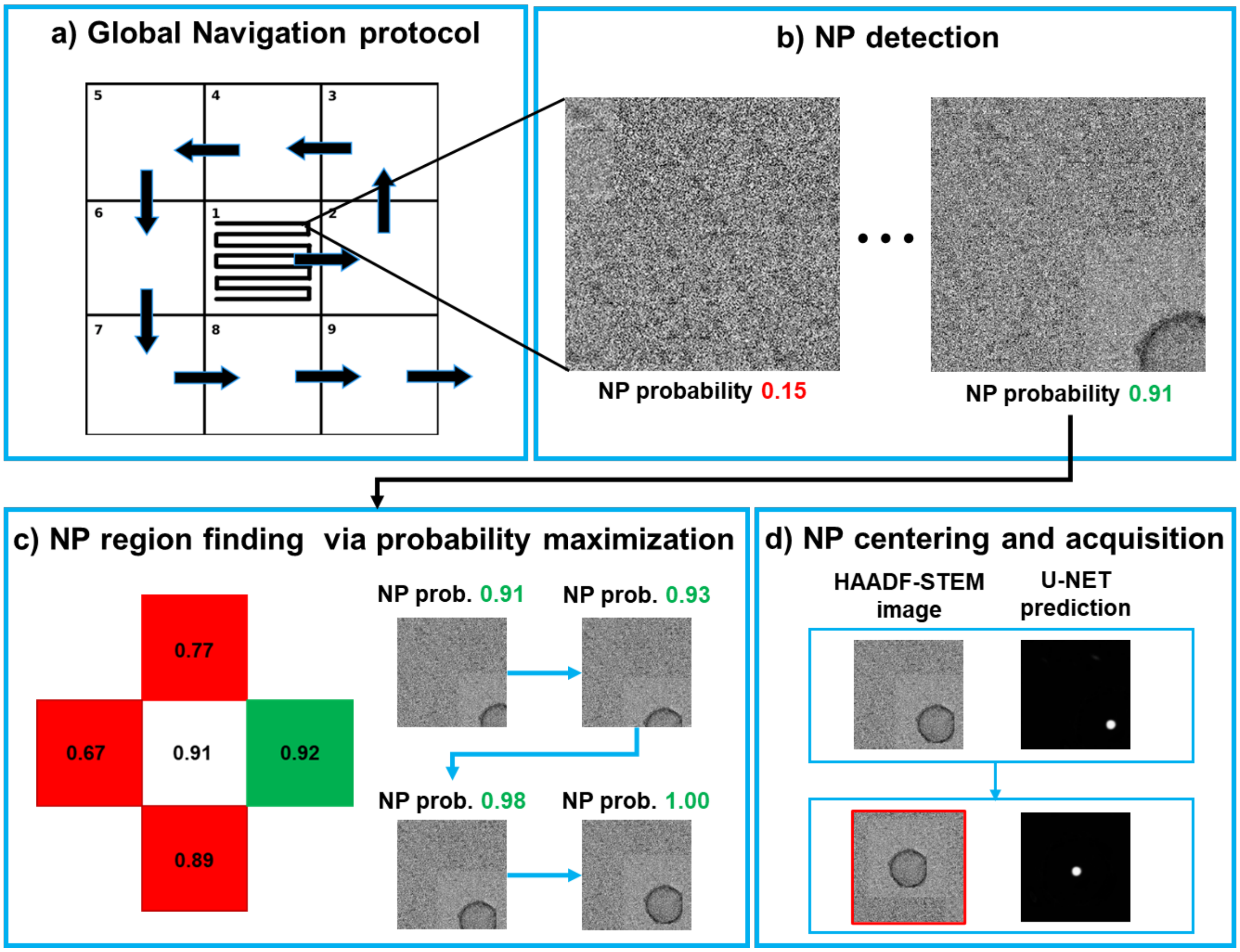

**Figure 1**: (a) Diagram illustrating the global movement algorithm, showing how the system navigates the sample in a structured manner. Initially, the system scans the entire grid horizontally; upon completion, it proceeds to the next grid following a spiral pattern. (b) The trained ResNet-34 model provides the probability of nanoparticle presence in each movement frame. When the probability exceeds 0.9, it triggers a refined movement strategy. (c) The refined movement strategy directs the system toward the region with the highest probability of containing the nanoparticle, continuing until the maximum probability is reached.(d) Finally, the trained U-Net model determines the center of mass (COM) of the nanoparticle and acquires the image at this position.

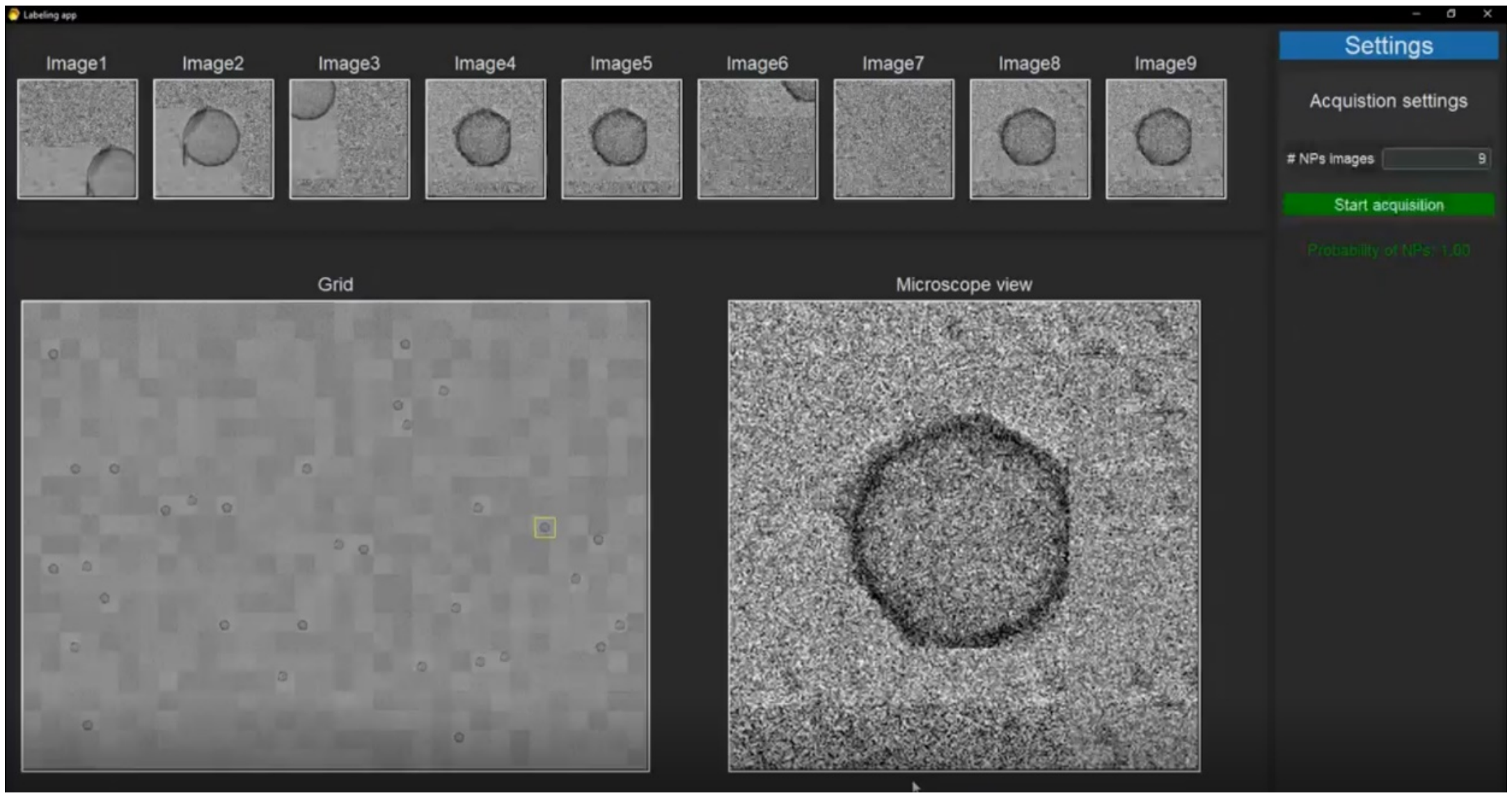

**Figure 2**: Graphical User Interface (GUI) layout. The right panel includes search options, the central-right section displays the live microscope feed, while the central-left section shows the sample. Captured images appear at the top.

## 10. On-the-fly TEM Image Analysis for Nanoparticle Synthesis Characterization

*Xingzhi Wang[1], Lehan Yao[2], Fanzhi Su[3], Pawan Vedanti[4], Zhiheng Lyu[1]*

1. Department of Materials Science and Engineering, University of Illinois at Urbana−Champaign, Urbana, Illinois 61801, United States
2. Pacific Northwest National Laboratory, Richland, WA 99354, United States
3. Department of Materials Science and Metallurgy, University of Cambridge, 27 Charles Babbage Road, Cambridge, CB3 0FS, UK
4. Department of Materials Science and Engineering, University of Pennsylvania, Philadelphia, PA 19104, USA

### Abstract

The properties of nanoparticles are highly related to their size and shape, exemplified by the varying colors of plasmonic metal nanoparticles with different sizes and catalytic performance variations on different facets. With electron microscopy (EM) widely used for characterizing nanoparticle shape and size, recent advancements in high-throughput image collection provides us with a convenient way to build large datasets for evaluation of sample distribution, product purity, and synthetic protocol efficiency. Targeting at fully automating EM in imaging capturing and on-the-fly decision making, there is a pressing need for real-time nanoparticle shape analysis with high speed and accuracy.[1–3] In this project, we developed an algorithm implementing real-time extraction and classification of nanoparticle shape (*e.g.*, area, aspect ratio, eccentricity, etc.) in transmission electron microscopy (TEM) imaging.

### Methodology

(*i*) *U-Net segmentation*. The input image (4096×4096 pixels) is resized to a standardized size of 512×512 pixels to ensure uniform processing. A U-Net model trained with 30 labelled images is used to segment the image into the three output channels.[4] Specifically, the pixels in the TEM image are classified to the non-overlaying regions of a nanoparticle (Channel 1), overlaying regions of nanoparticles (Channel 2), and the background (Channel 3).

(*ii*) *Contour delineation*. A modified watershed segmentation algorithm is applied to delineate the contours of individual nanoparticles from the outputs of U-Net. The Euclidean distance transform of the binary masks is calculated, and its negative is used to identify local minima as markers for the watershed algorithm. To refine the markers, a thresholding operation is applied to select regions of interest. Morphological erosion is performed using a 5×5 kernel to enhance separation between adjacent regions. The segmented regions are then cleared of boundary artifacts and labeled using connected component labeling. For each segmented particle, a combined mask is created by merging the non-overlapping particle region with overlapping regions. The combined mask is segmented again using the watershed algorithm to separate adjacent particles. Contours of the resulting segments are extracted.

(*iii*) *Feature extraction*. The delineated contours provide a representation of convex structures within the image, facilitating downstream analysis. Various morphological properties of detected features are then calculated, including area, aspect ratio, eccentricity, circularity, convexity, and solidity. Convexity and solidity are used to filter out contours that do not represent the actual shapes of the nanoparticles, and the other features are used for shape classification.

(*iv*) *Shape classification.* An unsupervised classification of the contour shapes based on the AutoDetect-mNP algorithm is applied.[5] The algorithm determines the optimum number of shape classes to classify the contours into, and applied a modified naïve Bayes classifier to assign each contour to a shape class.

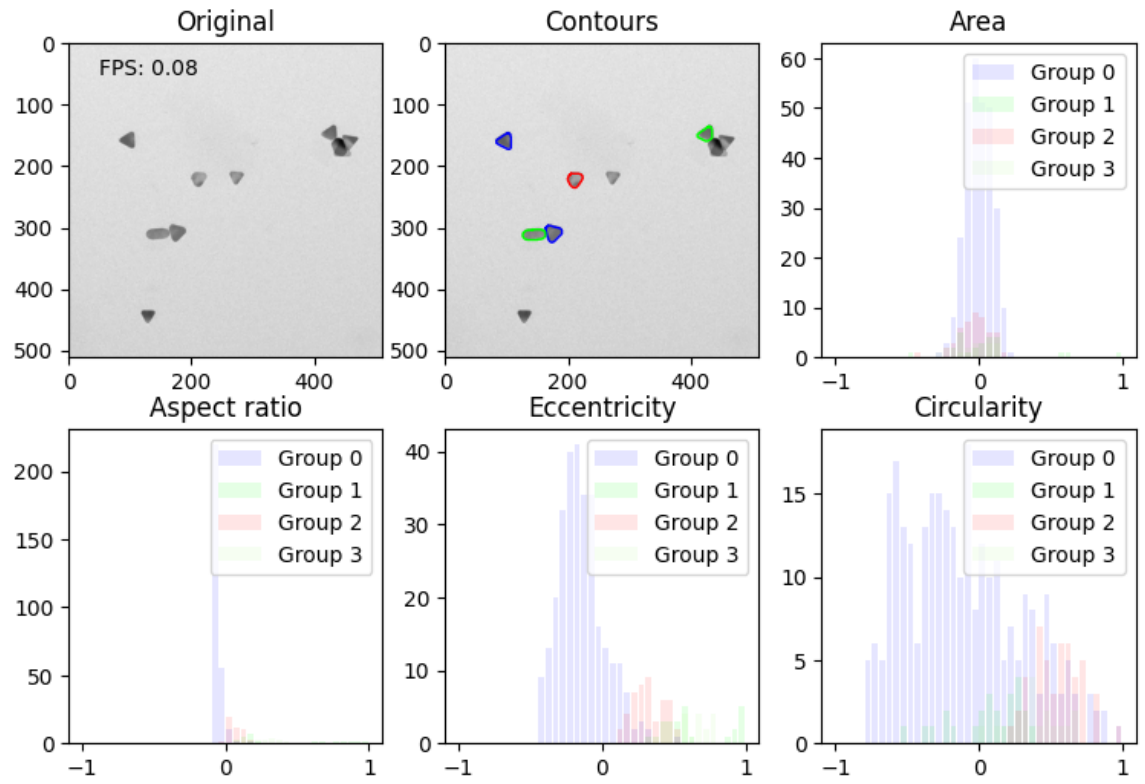

**Figure 1.** Real-time shape classification of nanoparticles.

The final output includes a real-time visualization with six panels showing: the original TEM image, the detected contours (color-coded by shape classes), and statistical distributions of the morphological properties by shape classes. This workflow runs on-the-fly as TEM images are being read into the memory, effectively creating a live analysis system that can characterize and classify features in TEM data as they are acquired. Future application of the algorithm with autonomous EM data acquisition and advanced in-situ imaging can be expected.

## 11. CT4Batt: A pipeline for Analyzing X-ray Chromatography Battery Images

Amir Taqieddin[1] and Forrest Laskowski[1]

[1] Technology Integration – Materials Informatics and Modeling Department, Solid Power Operating Inc, Louisville, CO 80027

**Abstract.** Manual detection of battery flaws using X-ray computed Tomography (CT) images is a labor-intensive and error-prone process, which significantly limits the efficiency of quality inspection. To accelerate both the evaluation of battery health and quality inspection, the analysis can be automated by leveraging machine learning techniques. We introduce CT4Batt, a two-step machine learning pipeline designed to efficiently analyze CT images of electrode stacks. The pipeline employs a convolutional neural network (CNN) integrated with attention mechanisms to automatically locate and isolate electrode tabs. Following isolation, electrode tabs are segmented using a dynamic watershed-based approach. Various user-controlled sliders ensure handling of images with variable sizes, numbers of tabs, and contrasts. Overall, CT4Batt pipeline enables accurate and precise straightforward post-processing, visualization, and quantitative analysis of automated electrode tab detection and segmentation in X-ray CT imagery for batteries.

**Methodology.** The CT4Batt pipeline is a two-step approach for analyzing X-ray images of electrode stacks, built using publicly available data from the X-ray-PBD dataset[9]. First, the framework employs a CNN with attention mechanisms to automatically assign a bounding box around the electrode tabs[10, 11]. By predicting bounding boxes directly from the image, the network ensures that the downstream segmentation process can focus on a well-defined region of interest, thereby simplifying subsequent steps. Two variations of the model were developed where both differ only in the final prediction step: one employs a Multilayer Perceptron (MLP) layer and the other uses a Kolmogorov-Arnold Network (KAN)[12] to predict the bounding vertices. Both models begin with a convolutional feature extractor, followed by coordinate attention and sliding window attention mechanisms to emphasize the electrode regions. Residual blocks were used to refine hierarchical feature representations, while a global pooling layer condenses these features into a compact representation. The performance of both models was evaluated using the Intersection over Union (IoU) loss function to determining the bounding box regression. The models were able to accurately detect the zoom-in region with a loss percentage as low as 2% after 200 epochs, with each epoch requiring an average training time of ~0.4 sec (see **Table 1**). **Figure 1** shows the obtained zoom-in contextual results using three different tested models.

**Table 1** Summary of the trained and tested models for performing contextual zoom-in identification around the electrode region.

| Model | Model A | Model B | Model C |
|---|---|---|---|
| Approach | CNN with KAN | CNN with MLP | Reduced CNN with KAN |
| Number of parameters | 23,326,976 | 4,442,884 | 3,646,976 |
| Relative loss at 200 epoch with respect to the loss at the first epoch (%) | 1.84% | 1.33% | 8.25% |

The second part of the pipeline performs a segmentation task to isolate individual electrode tabs. A watershed image segmentor is used to define the bulk battery region and then to segment the tabs. Various user-controlled sliders allow the approach to handle images with variable sizes, varying numbers of tabs, and variable contrasts. Each output channel of the segmentation model corresponds to a single tab, allowing the user to extract a 3D array (Number of Tabs × Height × Width) of binary masks, where each mask highlights one tab at a time as shown in **Figure 2**. By carefully crafting each stage—first bounding box prediction, then targeted segmentation—CT4Batt achieves a flexible and robust solution. The result is an automatic pipeline that can handle highly variable imagery: from subtle differences in electrode tab shapes/arrangements to diverse image resolutions/scales encountered in X-ray CT data. Looking ahead, this framework can be further extended with more advanced instance segmentation techniques, improved attention modules, and additional shape priors. Overall, this two-step solution provides a strong starting point for automated electrode tab detection and segmentation in complex X-ray imagery.

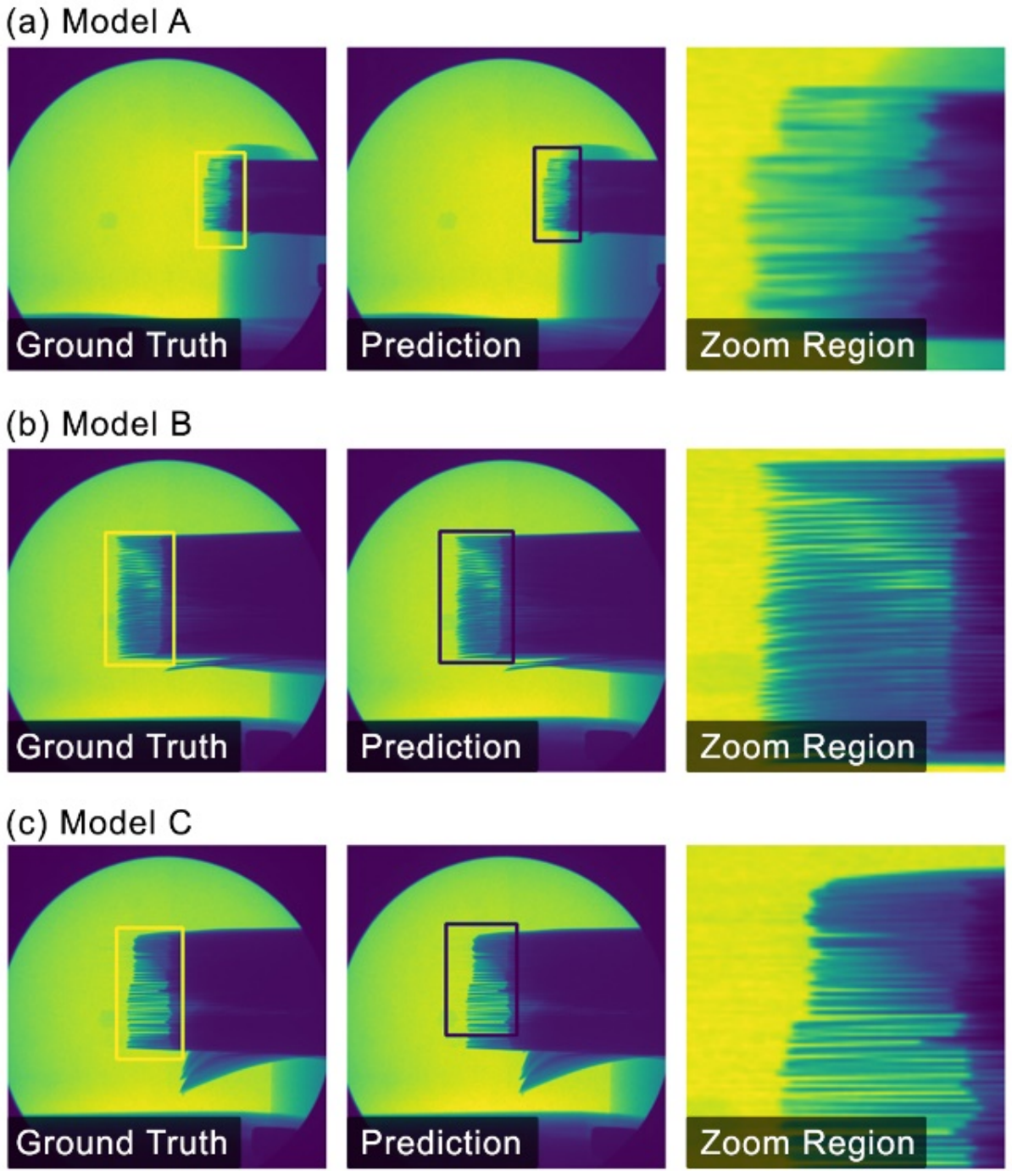

**Figure 3** Zoom-in results from step #1 on an image from the testing set using a) CNN with KAN, b) CNN with MLPs, and c) Reduced CNN with KAN layers.

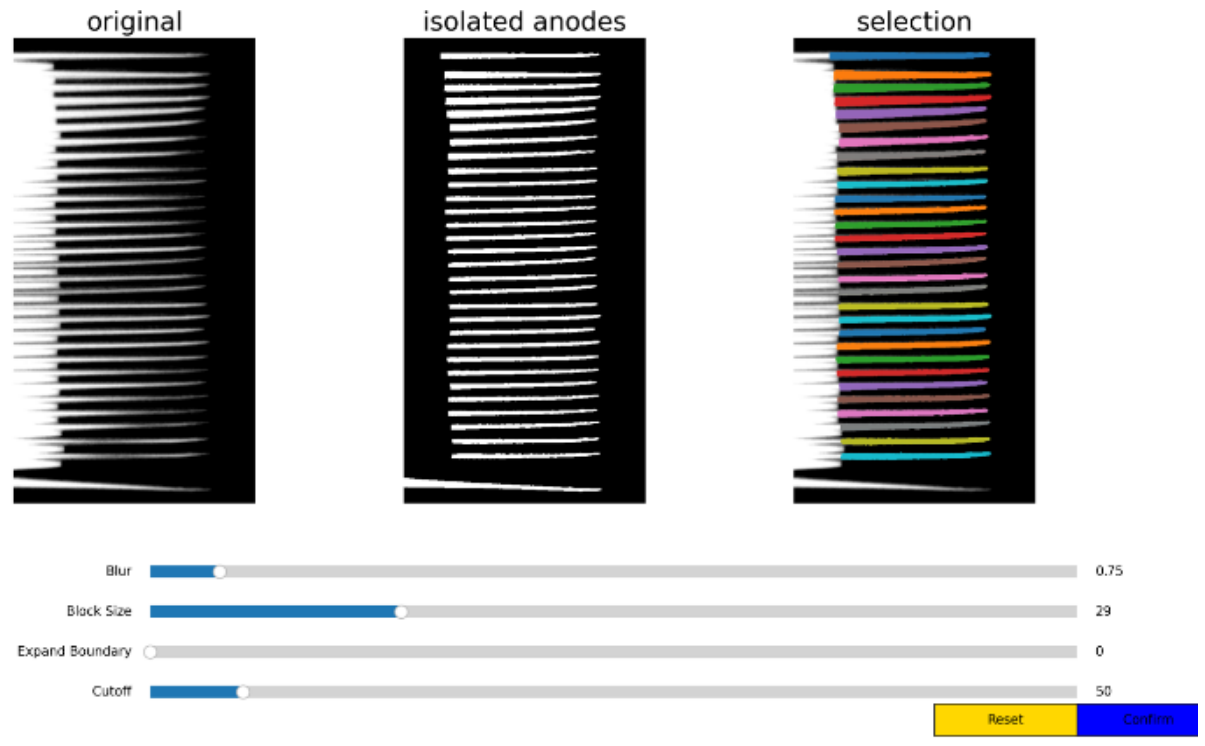

**Figure 2** Example of the individual electrode segmentation using CT4Batt.

## 12. Agentic Workflow for TEM Experiment Automation

Xiangyu Yin[1], Benjamin Fein-Ashley[2], Yu-Tsun Shao[3], Yi Jiang[1]

[1]Advanced Photon Source, Argonne National Laboratory, Lemont, 60439 IL, USA

[2]Ming Hsieh Department of Electrical and Computer Engineering, University of Southern California, Los Angeles, 90089 CA, USA

[3]Mork Family Department of Chemical Engineering and Materials Science, University of Southern California, Los Angeles, 90089 CA, USA

### Abstract

Foundation models, including Large Language Models and Vision-Language Models, hold significant promise for automating and accelerating scientific research [1–3]. Previously, we introduced Nodeology [4], a foundation AI agents orchestration framework, and demonstrated reduced trial-and-error efforts in a ptychography reconstruction workflow [5]. In this study, we extend this approach by developing an agentic workflow for automated Transmission Electron Microscope (TEM) experiment control. Traditional TEM workflows frequently involve complex manual steps—initialization, parameter tuning, data acquisition, and quality assessment—which can introduce inefficiencies and biases. Here, we present AutoScriptCopilot [6], an automated workflow, integrating Thermo Fisher Scientific's AutoScript [7] with the state machine-based orchestration layer powered by Nodeology. This workflow leverages foundation models for real-time parameter recommendations and image quality assessments, at the same time incorporating decision-making and error-handling points for human intervention. Our approach enhances conventional workflows by promoting reproducibility, adaptive parameter tuning, and structured data management. Furthermore, its modular architecture supports integration with external analytical functions, paving the way for highly automated end-to-end STEM/TEM workflows.

### Methodology:

The workflow employs a graph-node abstraction to effectively coordinate instrument control, AI-driven decision-making, and user oversight, with each functional element represented as a distinct node (Figure 1). Initially, the system sets up a microscope client via AutoScript Python API, retrieving key status parameters such as vacuum conditions, beam status, and detector readiness. Next, an automated validation node assesses system readiness, verifying that essential experimental prerequisite (e.g., vacuum levels, beam alignment) are satisfied. Upon successful validation, the workflow activates an AI-based parameter recommendation node, combining domain expertise (e.g., instrument limitations, sample constraints) with user-defined objectives (e.g., desired image quality, minimal beam damage). Recommendations can be reviewed and adjusted by domain experts through a dedicated confirmation node, allowing manual intervention when required. Once confirmed, image acquisition is executed by directly interfacing with AutoScript's instrument control capabilities to capture STEM/TEM images under specified conditions. During the initial acquisition phase, a sample image is captured, followed by an auto-alignment node. Subsequently, an image quality assessment node automatically evaluates critical image metrics, including focus, signal-to-noise ratio, and astigmatism, identifying potential issues such as drift or brightness/contrast imbalance. If the assessment indicates suboptimal conditions or significant astigmatism, the workflow returns to the alignment node to optimize experiment parameters further. Finally, the workflow advances to a data analysis node, where captured images and associated metadata undergo additional processing or integration with existing advanced analytical pipelines. Throughout, Nodeology ensures strict data validation, meticulous state management, and documentation of user interactions, facilitating reproducibility and retrospective analyses. With the modular and flexible design of Nodeology, the workflow can be expanded further to incorporate data analysis feedback into experimental condition optimization for a fully integrated experiment-analysis loops.

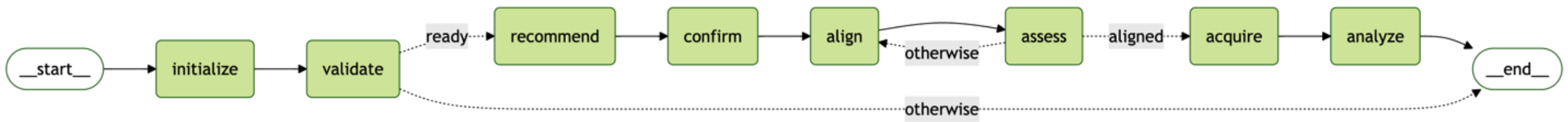

**Figure 1**: State machine-based TEM automation workflow in AutoScriptCopilot, illustrating nodes for initialization, validation, AI-driven parameter recommendations, image acquisition, alignment, quality assessment, and data analysis. Dashed arrows indicate conditional branches and optional human intervention points.

# 13. AI-powered Thermal Mapping in Electron-ion Trapping Experiment (EiTEx) for Single Photon Detection

Vineet Kumar[1*], Yogesh Paul[2], Himanshu Mishra[1]

[1]Dept. of Surface and Plasma Science, Faculty of Mathematics and Physics, Charles University in Prague, Czech Republic

[2]Institute for Neuromodulation and Neurotechnology, University Hospital and University of Tuebingen, Tuebingen, Germany

**Abstract**

The EiTEx at the Faculty of Mathematics and Physics, Prague, is designed for single-photon detection via electron-ion interactions.[1] Operating at ultra-high vacuum ($< 2.5 \times 10^{-10}$ mbar), EiTEx setup integrates 3D-printed traps, an atomic oven assembly, and an optical cavity. Radiative heat transfer from the oven poses a potential challenge, impacting both the trap and optical cavity, thereby influencing the localized electron-ion system. Additionally, Johnson-Nyquist noise and blackbody radiation from surfaces impact ion motional heating rates, further necessitating precise thermal control. To overcome these challenges, EiTEx employs a dual-approach thermal modeling strategy that combines general intelligence-based analytical modeling with artificial intelligence-driven data analysis. This hybrid framework facilitates thermal mapping, enhances predictive modeling, and improves system stability, ensuring optimal performance of EiTEx.

**Methodology:**

EiTEx is designed for single-photon detection via trapped electron-ion interactions, with detailed findings to be reported in upcoming publications.[1] Model drawings of the EiTEx subsystems are provided, with Figure 1 offering a comprehensive visual representation.

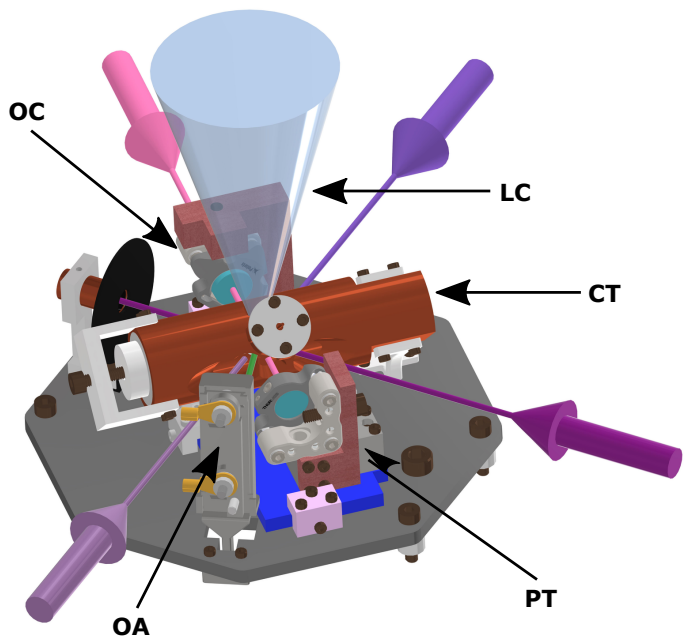

**Figure 1**: Schematic of the EiTEx chamber's octagonal base plate, highlighting the oven assembly (OA), coaxial trap (CT), optical cavity, and piezoelectric transducer (PT).

In EiTEx, heat from the oven aperture ($x = 0$) propagates toward the trap center ($x_c$), driven by input power $P_{dc}$ over duration $t_{oh}$, starting from $T_{lab} \approx 293$ K.
The GI-based modeling as a polynomial function, described as $T_{pol}(P_{dc}; x, t_{oh}) = \sum_{j=0}^{n_1} \sum_{k=0}^{n_2} \sum_{l=0}^{n_3} C_{jkl} x^j P_{dc}^k t_{oh}^l$, may include interaction terms, capturing combined variable effects (e.g., $xP_{dc}$, $xt_{oh}$). Coefficients $C_{jkl}$ are determined by fitting the model to observe FEM-simulated temperature data. Figure 2 illustrates the GI-based performance.
Due to the GI limitations ($R^2 \approx 0.708$ without interactions, $R^2 \approx 0.992$ with 75 terms for a 6th-order polynomial), we next adopted an AI-based approach. Figure 3 illustrates neural network (NN) and random forest (RF) performance. The code is available on GitHub.[2]

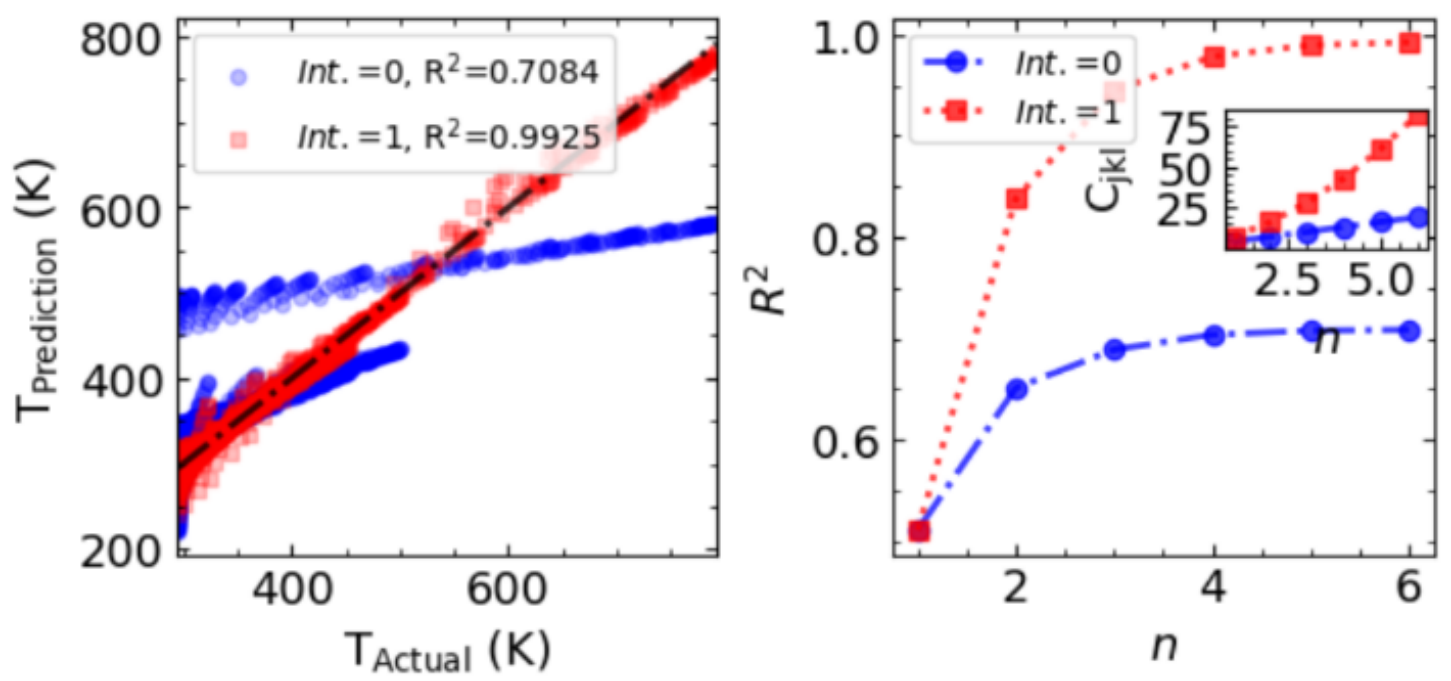

**Figure 2**: (Left) Prediction accuracy for models with (red) and without (blue) interaction terms. (Right) $R^2$ vs. complexity ($n$).

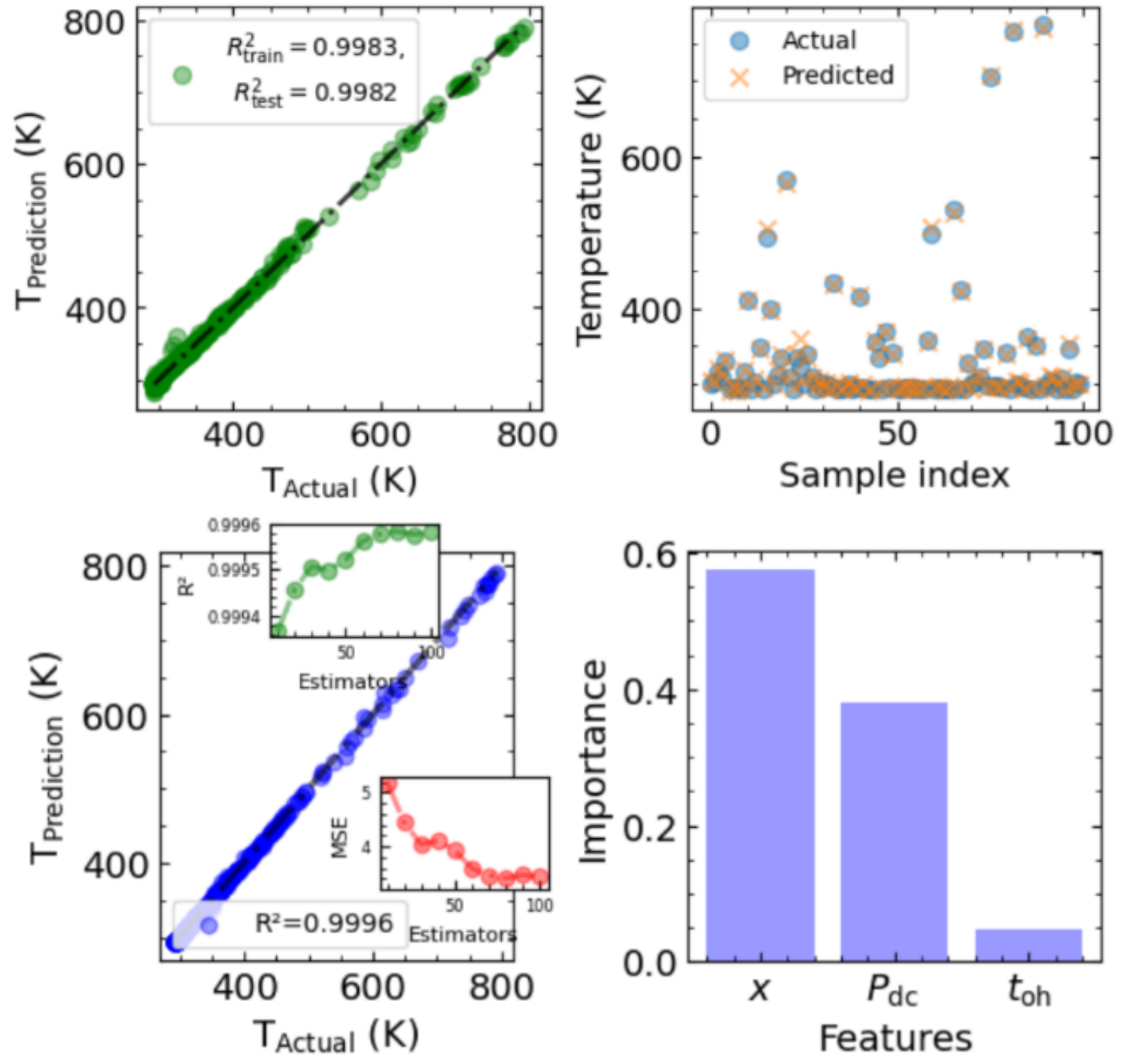

**Figure 3:** The upper panel presents NN results, showing a strong correlation between predicted and actual temperatures. The lower panel displays RF results, demonstrating high prediction accuracy as well. Insets highlight the increasing $R^2$ and decreasing mean squared error (MSE) with more estimators. The analysis of the feature importance indicates that $x$ has the highest influence, followed by $P_{\mathrm{dc}}$ and $t_{\mathrm{oh}}$.

## 14. Unsupervised Classification of Ferroelectric Domains

Grace Guinan[1], Addison Salvador[1], Madeline Van Winkle[1], Arman Ter-Petrosyan[2]

[1] National Renewable Energy Laboratory, Golden, CO

[2] Department of Materials Science and Engineering, UC Irvine

### Abstract

A major hurdle when applying machine learning (ML) to materials science is the limited amount of training data and minimal ground truth data. This makes supervised ML techniques difficult to apply out of the box, since significant effort must go into developing enough data to train the model and labeling ground truths to validate the model. Unsupervised learning techniques, such as clustering, aim to circumvent this issue by using patterns to group similar data without the need for labeled data. Here, we employ three different clustering methods for unsupervised classification of ferroelectric domains. For this project, the data analyzed was a high-resolution STEM image of pure ferroelectric BiFeO3. Additional datasets included images of BFO doped with varying percentages of Sm, but since the pure image had the most distinct domain boundaries; it was the starting point for the development of this algorithm. The images were obtained by Dr. Chris Nelson from Oak Ridge from samples prepared by Professor Ichiro Takeuchi at the University of Maryland. While this method was generally inconclusive for this data, this workflow can be applied to data that is better tailored to clustering techniques. Clustering presents many exciting possibilities for out of the box ML for segmentation and image classification in materials science.

### Methodology:

The first step in the process was to chip up the image into much smaller patches, forming a grid. This is known as the sliding window approach. While the image-chipping code was given, there were three key parameters to optimize: the window size in the x and y dimensions, and the step size, which determines the extent of overlap between adjacent patches. These are critical parameters for the success of the clustering methods. If too small, no patterns will be detected, but if made too large, the chips will not be able to precisely identify each region. Through iterative testing, it was seen how drastically the window size affects the results.

Then the images were run through three different clustering methods, k-means, spectral, and agglomerative. A standardized chip size was processed through each clustering method, and the resulting clusters were plotted. The final output was the pure BFO image classified using three different clustering methods along with PCA plots. From these outputs it was concluded that this combination of window shape and clustering algorithms is not an effective way to classify the domains for this material. Some suggestions for future development would be to try a circular clustering method or try embedding the data before clustering. Alternatively, a non-clustering method such as locating the atomic coordinates might be more successful.

The workflow and results for this study has been presented in **Figure 1.**

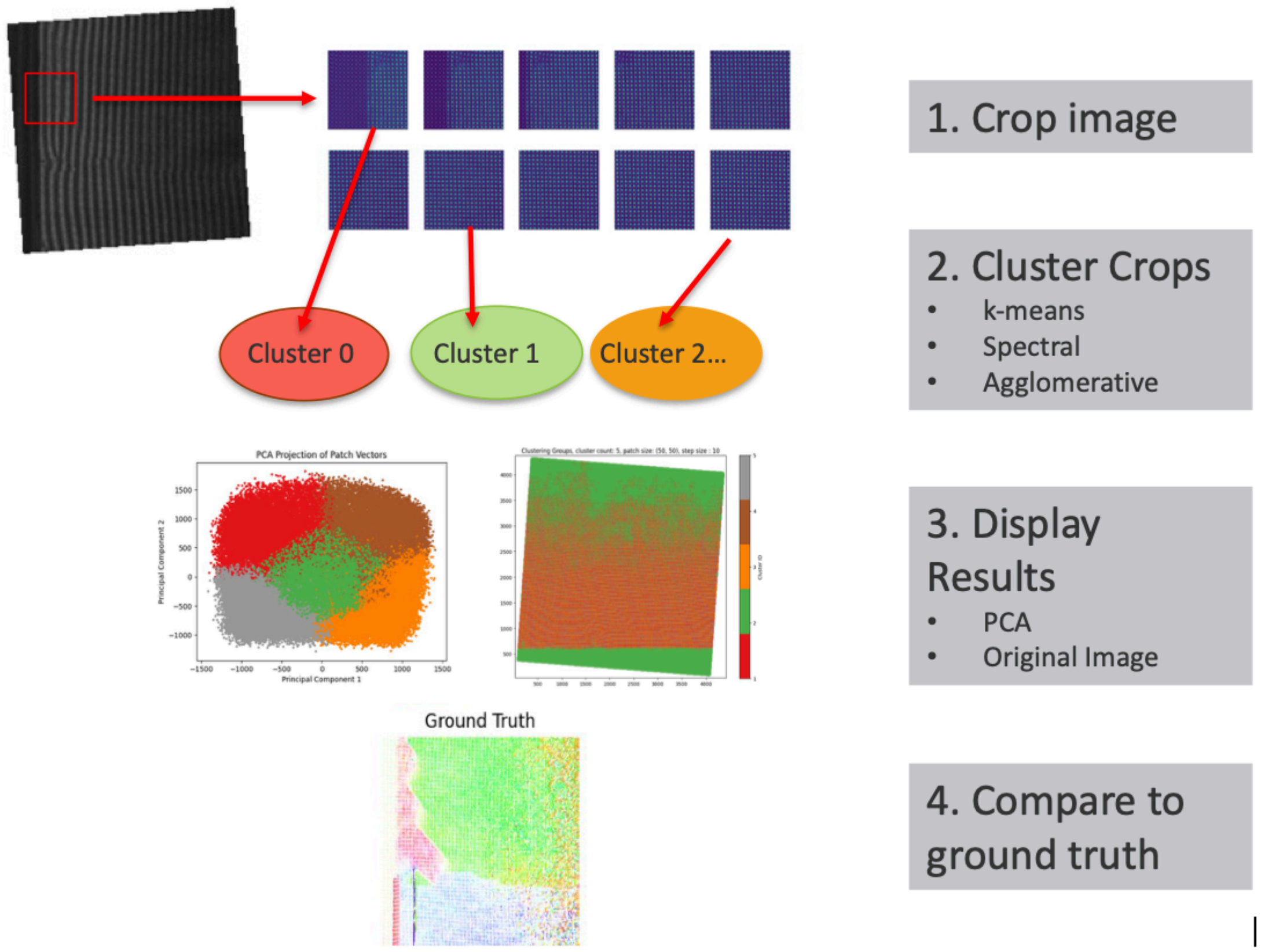

**Figure 4**: Cropping and Clustering Workflow

## 15. MicroscopyLLM-Bench: Benchmarking LLM Capabilities for Open-Source Microscopy Datasets

*Adib Bazgir[1], Rama Chandra Praneeth Madugula[2], Yuwen Zhang[1]*
[1] Department of Mechanical and Aerospace Engineering, University of Missouri-Columbia
[2] Department of Mechanical Engineering, New York University

### Abstract

Microscopic image analysis in biomedical research, histopathology, and diagnostic medicine can be slow and error-prone when done with traditional approaches. Here, we describe MicroscopyLLM-Bench, an AI-powered, automated pipeline that utilizes cutting-edge vision-language models (including Gemini-1.5 Flash, Gemini-1.5 Pro, Intern-VL2, Llava, and variants of GPT-4) for object detection, cell classification, and analysis of histopathological features. The approach also combines heterogeneous datasets with high-resolution images (Brightfield Microscopy SCC, Dreambooth Cell Images, and PubMedVision) in order to standardize input data and to increase model robustness. Benchmarks show that our approach detects and localizes important features, classifies cell morphologies, and generates textual descriptions of tissue organization. By decreasing the need for manual annotations and increasing reproducibility, MicroscopyLLM-Bench establishes a new standard for automated microscopy workflows and capabilities for real-time diagnostics.

### Introduction

Microscopy underpins pathology, cell biology, and translational research, but conventional computer-vision pipelines still struggle with the volume and diversity of modern image datasets, often demanding thousands of pixel-level labels and task-specific networks before yielding robust results [1]. While these deep learning systems accelerated segmentation and detection, their single-task focus and annotation appetite limit adoption in resource-constrained laboratories. Large vision language models have emerged as a more adaptable alternative. With nothing more than a few in-context examples, general-domain VLMs can now match bespoke networks on histopathology QA benchmarks [2], and rapid domain tuning, as demonstrated by LLaVA-Med, can be completed in under 15 hours on commodity GPUs [3]. Open-source initiatives such as InternVL 2.5 continue to close performance gaps with proprietary giants through systematic model- and data-scaling strategies [4].

Parallel progress in agentic AI is pushing beyond single-model reasoning. Multi-agent frameworks have been proposed for causal inference in biomedicine [5], cross-modal data fusion in materials science [6], and even hypothesis generation and experimental planning [8-11]. A recent hackathon survey captured 34 diverse LLM applications that automate everything from literature mining to real-time lab control, underscoring the breadth of tasks now within reach [7]. Despite this momentum, no benchmark systematically evaluates how these powerful, but largely uncalibrated, models perform on the day-to-day workflows of microscopy laboratories.

MicroscopyLLM-Bench meets that need by (i) harmonizing Brightfield Microscopy SCC, Dreambooth Cell Images, and PubMedVision under a common preprocessing pipeline; (ii) probing Gemini-1.5 Flash/Pro, Intern-VL2, LLaVA, and GPT-4 variants across detection, classification, and descriptive reasoning tasks; and (iii) emitting compact JSON predictions aligned with ground truth to streamline quantitative comparison and downstream automation. In doing so, it delivers a reproducible yard-stick for next-generation multimodal agents aimed at reducing annotation overhead and enabling real-time, patient-centric diagnostics.

## Methodology

### Data Curation and Preprocessing

The process starts with the aggregation of several microscopy datasets. Normalization and augmentation processes are to the original images to standardize the datasets, making fluctuations in imaging characteristics irrelevant and useful for the downstream AI models despite existing differences in the datasets. Figure 1 depicts a general schematic of the procedure that leads to the standardization of the original images from Brightfield Microscopy SCC, Dreambooth Cell Images, and PubMedVision to the final images that are appropriate for each task.

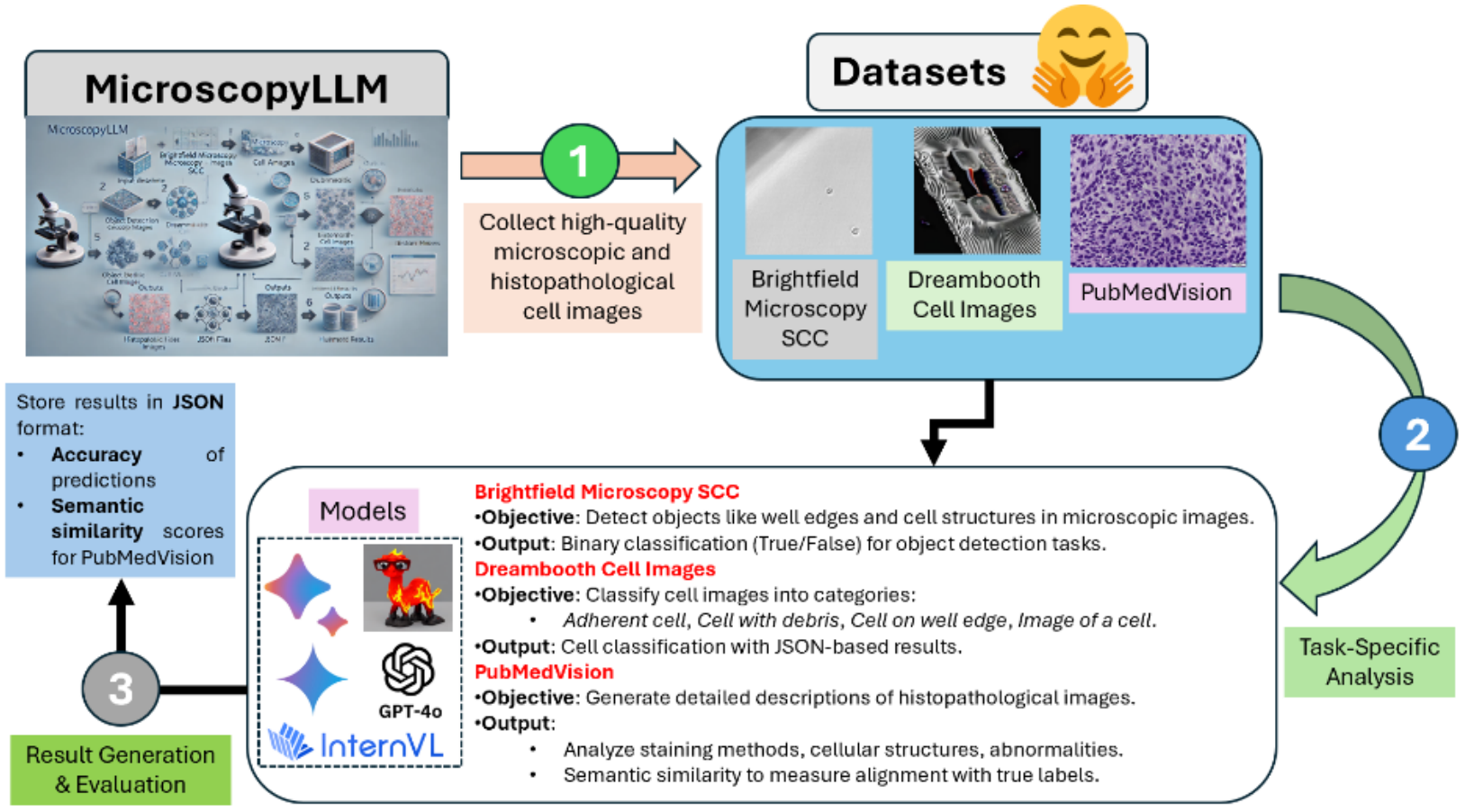

**Figure 1**: Overview of the MicroscopyLLM-Bench workflow, including data collection from Brightfield Microscopy SCC, Dreambooth Cell Images, and PubMedVision, followed by model inference and JSON-based output generation.

### Object Detection

The pipeline leverages the Brightfield Microscopy SCC dataset as a reference point for identifying important structural features, specifically edges of microtiter plate wells, which is the focus of the methods being compared. For this task, both detection methods were able to locate edges with good regularity, with Gemini-1.5 Flash achieving an 80% detection accuracy. When analyzing Intern-VL2's results, the model showed a high true-positive detection rate for edges, however it did occasionally mislabel areas of the background as edges or otherwise errors when detecting edges. It was noted when looking at the outputs of Gemini-1.5 Pro, while the overall detections were more balanced this output once again emphasized the importance of adequately calibrating the models to distinguish relevant outputs from undesirable background representations. Figure 2 shows the JSON based output when detecting objects labeled toward outputs, employing true/false (where "True" indicates a predicted well edge) in conjunction with a ground truth label.

This response format simplifies any subsequent evaluation of performance and additionally facilitates process integration in larger automated workflows.

```
{
    "prompt": "Given the image of a cell , answer the following multiple choice question.\n\n   \n      Question: Does the below Image contains elements of a microtiter plate well edge??\n\n      choices: \n      1)True \n      1)False\n\n      select the option and no need of explanation.\n \n      ",
    "response": {
      "vision": "1) True"
    },
    "true_label": True
  }
```

```
{
    "prompt": "Given the image of a cell , answer the following mutliple choice question.\n\n\n\n      Question: Does the below Image contains elements of a microtiter plate well edge??\n\n      choices:\n 1)True\n      1)False\n\n      select the option and no need of explanation.\n\n      ",
    "response": {
      "vision": "True"
    },
    "true_label": false
  }
```

**Figure 2**: Sample JSON output for an object detection task. The pipeline is queried on whether the image contains a microtiter plate well edge, returning both the model's predicted label and the corresponding ground truth.

### Cell Classification

For the task of cell classification, the Dreambooth Cell Images dataset was used to classify cells into several categories based on morphology (adherent, debris containing, edge-bound, or generic). Intern-VL2 achieved 85% precision in classifying the cells with debris, while Gemini-1.5 Flash generally performed well at classifying adherent cells although some cells were misclassified between edge-bound and generic. Figure 3 provides more examples of how the pipeline processes classification results in JSON format, including how it processed prompts such as "Is this an adherent cell?" followed by outputting the predicted label (e.g., "image of a cell") along with the "true_label" (e.g., "a grayscale microscopy image of an adherent cell"). These examples further demonstrate the pipeline can provide fine-grained differences between cell morphology.

```
{
    "prompt": "\" The given image is a type of cell image captured by microscopy technique. can you identify what kind of the following cell does the given image more representing for?\n      1)An adherent cell\n 2)A cell with debris \n      3)A cell on\u00a0the\u00a0well\u00a0edge\n 4)Image of a cell\n      please respond with the option with value, no need of explanation.\n\n      ",
    "response": {
      "vision":"4) Image of a cell"
    },
    "true_label": "a grayscale microscopy image of a cell"
  }
```

```
{
    "prompt": "\" The given image is a type of cell image captured by microscopy technique. can you identify what kind of the following cell does the given image more representing for?\n      1)An adherent cell\n 2)A cell with debris \n      3)A cell on\u00a0the\u00a0well\u00a0edge\n 4)Image of a cell\n      please respond with the option with value, no need of explanation.\n\n      ",
    "response": {
      "vision": "4) Image of a cell"
    },
    "true_label": "a grayscale microscopy image of an adherent cell"
  }
```

**Figure 3**: Example of a morphological cell classification prompt and response, where the pipeline identifies cell types (adherent, debris-containing, edge-bound, or generic). The "true_label" field indicates the ground-truth classification for validation.

### Histopathological Feature Analysis

In the final stage of analysis, stained tissue images from the PubMedVision dataset were examined in order to generate specific textual descriptions. Using Gemini-1.5 Flash, a semantic similarity score of 75% was achieved compared to the experts' annotations, successfully locating glomeruli and proximal tubules. The metric was increased to 78% when extending the analysis to Intern-VL2 and offered more detail by describing nuclear irregularities and slight variation in staining.A good example of the pipeline's descriptive power can be viewed in Figure 4, which presents the JSON output for a kidney cortex section stained with PAS (Periodic Acid-Schiff). The output text produced by the pipeline clearly identified mustard-colored brush borders and glomeruli, agreeing with the true_label output during the verification analysis, and resulting in a measure of semantic similarity of approximately 0.73. These examples illustrate the pipeline's impressive potential in contexts relevant to histopathological diagnosis where the interpretative verbal reasoning process is important.

```
{
    "id": "Alignment_VQA_14",
    "prompt": "Write a description of the given image.",
    "response": {
      "vision": "The image is a microscopic view of a tissue sample, specifically highlighting two distinct structures within the tissue: \"Proximal Tubules\" on the left and \"Glomerulus\" on the right. The text is written in black with a bright yellow arrow pointing to each of these structures. The overall background is filled with dense and spiky cellular structures, typical of tissue samples stained for detailed microscopic examination. The texture suggests the use of a histological staining technique, likely hematoxylin and eosin (H&E), commonly used in histology to distinguish various tissue components. The darker, oval-shaped structures highlighted in yellow are characteristic of proximal tubules, found in kidney cells. In contrast, the larger, more rounded structures are indicative of the glomerulus, a crucial part of the nephron where filtration occurs."
    },
    "semantic_similarity": {
      "vision": 0.7312068939
    },
    "true_label": "The image displays a histological section of kidney cortex stained with PAS (Periodic Acid-Schiff) staining method. This particular staining highlights the basement membranes, brush border, and other intrinsic structures in shades of magenta and purple colors. We observe multiple cross-sectional and longitudinal views of glomeruli, which appear as well-defined, spherical to ovoid structures. Each glomerulus is surrounded by a Bowman's capsule, which is less intensely stained and slightly lighter compared to the glomeruli. The surrounding renal tubules, specifically the proximal tubules, show a prominent thick red\/pink border encircling the luminal walls, indicative of intense PAS positivity due to the abundance of brush borders. The connective tissue in between the tubular structures appears lightly stained, providing a clear distinction from the highly stained tubules and glomeruli."
  }
```

```
{
    "id": "Alignment_VQA_13",
    "prompt": "Describe the composition and the subjects in this picture.",
    "response": {
      "vision": "This is a microscopic image, a micrograph, of a tissue sample.  The composition is primarily cells, stained a range of purples and light lavenders.  The cells are densely packed, with little visible extracellular matrix between them.  The staining technique used is hematoxylin and eosin (H&E), a common method in histology.\n\nThe subject is a histological section showing a neoplasm, likely a type of cancer. The specific type of cancer cannot be determined from the image alone; further analysis and clinical information would be needed for a diagnosis.  The image shows the cellular morphology (shape and structure of the cells), which is a key feature used by pathologists to classify tumors.  The cells appear relatively uniform in size and shape, suggesting a relatively monomorphic population.  However, there is some variation in the intensity of the staining, which could be due to differences in cellular activity or other factors.\n"
    },
    "semantic_similarity": {
      "vision": 0.6658643484
    },
    "true_label": "The image depicts a histopathological examination of an excised tumor under a microscope with H&E (hematoxylin and eosin) staining at 40x magnification. The tumor tissue shows a dense population of cells with significant pleomorphism. The cells are predominantly epithelioid with basophilic cytoplasm indicating high RNA content, likely reflecting active protein synthesis. Intermixed are occasional spindled cells. The nuclei of the cells vary in size and shape, showing irregular contours, coarse chromatin, and prominent nucleoli, suggesting malignant transformation. There are visible signs of mitotic activity among several cells, an indicator of rapid cellular proliferation. The overall features suggest a highly cellular and aggressive tumor morphology."
  }
```

**Figure 4**: Histopathological feature analysis demonstrating how the pipeline describes a PAS-stained kidney cortex image. The generated text is compared against an expert label, yielding a semantic similarity score that quantifies descriptive accuracy.

## Conclusion

MicroscopyLLM-Bench represents a significant advance in automated microscopy workflow, merging a variety of vision-language models into a customizable and scalable pipeline. The system is able to automate feature detection, classify cells, and describe histopathology, thus reducing reliance on manual annotation and promoting reproducibility in biomedical research. The exhibiting of the JSON-formatted outputs in Figures 2-4 are a transparent and organized method of reviewing and leveraging AI-enabled discovery. Collectively, the benchmark framework described will enable analysis workflows and create future prospects for real-time diagnostic applications in pathology.

## Supplementary Material

- **GitHub repository**: https://github.com/adibgpt/MicroscopyLLM-Bench
- **Video link**: https://drive.google.com/file/d/1u2DfXzWRL7fwN8ZG3Ycc4WTXg6lXjt11/view?usp=sharing

## 16. Removal of tip-shape induced artifacts in AFM images using deep learning

Yu Liu[1], Ganesh Narasimha[2], Zijie Wu[2]

[1] Department of Material Science and Engineering, University of Tennessee, Knoxville

[2] Center for Nanophase Materials Sciences, Oak Ridge National Laboratory, Oak Ridge, Tennessee, USA - 37831

### Abstract

Atomic Force Microscopy (AFM) is a powerful technique for investigating the structure and properties of materials at the meso- and nanoscale. High-resolution topographic imaging in AFM relies on scanning using an atomically sharp probe in either tapping or contact mode. However, probe-tip imperfections often compromise image fidelity, inducing imperfections, including bluntness, asymmetry, and double-tip configurations. These distortions obscure critical structural features, particularly in atomically resolved imaging, necessitating the need for image deconvolution. Moreover, conventional deconvolution methods are sensitive to hyperparameters and induce noise during image restoration. In this work, we explore the application of deep learning methods for the removal of tip-shape-induced artifacts in AFM images. We systematically simulate tip-induced distortions on various synthetic and experimental topographies and evaluate the performance of convolutional neural network architectures, including UNet and a ResNet–UNet hybrid. Additionally, we introduce a workflow that predicts the tip's point spread function (PSF) as an intermediate step, which significantly improves the robustness of image recovery under severe tip degradation.

### Methodology:

Our goal is to reconstruct the original image from a distorted image based on the method of deep learning-based feature identification. Our approach builds on recent advances in machine learning-based surface reconstruction[13], aiming to enhance deconvolution accuracy and extend applicability to real experimental data.

To obtain the training data, we start by simulating AFM images, followed by incorporation of real topographic scans. Firstly, to capture the effects of tip distortions, we simulated a diverse set of tip kernels - including single tips with varying diameters and double tips with different separations and height offsets. Then, to create standard sample features, we synthesized ground-truth patterns such as checkerboards, spirals, and atomic lattices of varying scales and densities. These were augmented with image patches extracted from real AFM topography maps of two combinatorial library samples. All simulations were conducted using the *SpmSimu* package[14].

We studied the effects of image deconvolution based on three methods. Our first model employed a standard UNet architecture trained on 1,000 image pairs for 100 epochs. The network learns to remove distortions by mapping simulated tip-convoluted images to their ground-truth counterparts. While the UNet performed well on synthetic and moderately distorted experimental data, it struggled with images affected by severe bluntness or strong double-tip effects, often producing over-sharpened or incorrect reconstructions (Figure 2a).

To enhance feature extraction and generalization, we implemented the second model which combines the ResNet–UNet architecture, consisting of ResNet blocks in the encoder with a U-Net decoder. This design facilitates efficient hierarchical representation and feature propagation, with skip connections preserving fine-grained details across layers. This model showed improved reconstruction fidelity across a wider range of distortions (Figure 2b).

Finally, we developed a predictive approach to infer the tip condition - represented as a point spread function (PSF) that is predicted from the distorted images of a standard sample. We simulated ~9,000 height maps generated from randomized tip degradations (e.g., blunted tips, varying double-tip asymmetries). A convolutional autoencoder was trained to map these distorted images to their corresponding PSFs. Incorporating the predicted PSF into the deconvolution pipeline enabled improved image recovery compared to using distorted images alone, particularly in recovering the dominant tip features. These results are summarized in Figure 2c.

The workflow and results for this study are presented in **Figure 1** and **Figure 2**.

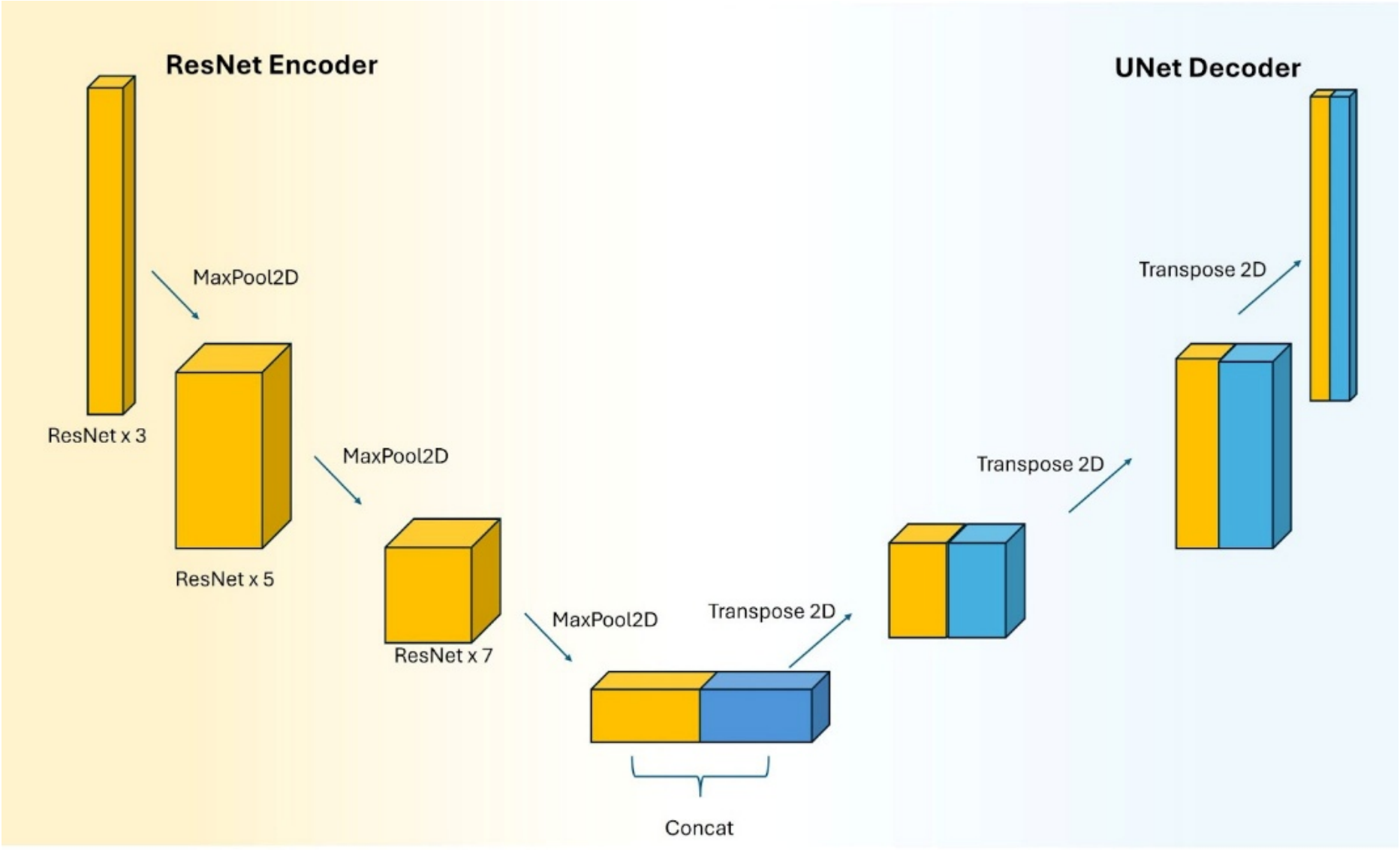

**Figure 5**: Schematic diagram of the ResNet encoder – UNet decoder setup.

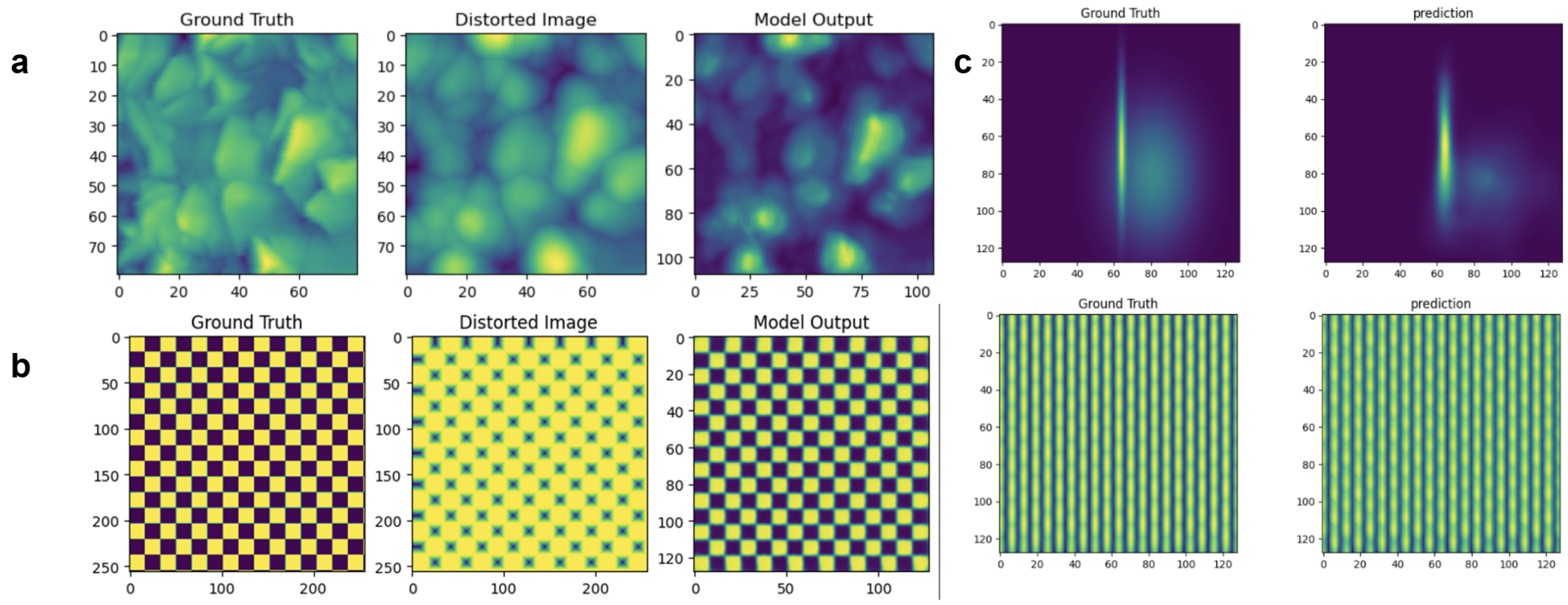

**Figure 6**: Reconstruction results of the DL based image deconvolution and PSF prediction **a)** The ground truth image, corresponding distorted image, and the UNet recovered output. This is an example of UNet based deconvolution model applied on a real AFM topographic image. **b)** Scan image reconstruction results based on the ResNet-UNet model. Here the ground truth is the synthesized checkerboard patterns. **c**) Ground truth and model-predicted PSF and of the tip.

## 17. Mixed Oxidation State Characterization from XAS/EELS using Machine Learning

Eric Montufar-Morales*[1], Jian Huang*[1] and Pravan Omprakash*[1]

[1]Institute of Material Science and Engineering, Washington University in St. Louis, St. Louis

*Equal contribution

### Abstract

Core loss Spectroscopy including, Electron Energy Loss Spectroscopy (EELS) and X-Ray Absorption Spectroscopy (XAS) are effective techniques for probing the elemental contents, bonding, oxidation states of samples at the nanoscale. However, both methods capture the electronic behaviour of the materials, matching with standard samples is required to extract physical descriptors such as oxidation state or coordination number. Moreover, for complex compounds with mixed oxidation states, widely present in 3d transition metal compounds, the match with standard samples can be time-consuming. Instrumental errors, varying composition and other experimental artefacts, introduce noise and change the spectra, making a systematic solution hard to achieve[15]. Thus, Machine Learning (ML) models trained on XAS or EELS spectra can be useful in spectral analysis to obtain relevant physical descriptors on the chemical structure. Previous models have focused on one specific element and predicted the mixed oxidation state and the fraction of integer oxidation state[16, 17]. In this work, we build a neural network to analyse the L2-3 edge of seven 3d transition metals (V, Cr, Mn, Fe, Co, Ni, Cu), to predict the oxidation state as well as the fraction of states found in these compounds. A generalized dataset based on multiple elements containing mixed valence states allows for a general-purpose model to be built, capable of distinguishing between elements and oxidation states.

### Methodology

Firstly, an EELS dataset was created by extracting existing data from the Materials Project[18] (L2-3 edge), consisting of 5840 samples. However, 90% of the samples were integer states and the rest 10% were mixed valence, leading to an imbalance. Thus, the next step was to create mixed oxidation state spectra by mixing integer spectra in various fractions[3]. Then the dataset was augmented with gaussian noise, instrumental broadening and shifted peaks to account for common experimental errors and artefacts. A fraction of integer states was retained along with the simulated mixed states. The dataset amounted to ~200,000 was randomly split into training and testing in 3:1 ratio. We designed a Fully Connected Neural Network to predict the mixed oxidation states of 3d transition metal compounds. The network consisted of 8 layers and 2,703,915 learnable parameters. The hyperparameters were tuned by trial and error to decrease the validation loss. A custom loss function was designed, which weights the errors between predicted and actual samples with the oxidation states, thus penalizing not only the fraction of states, but also the degree. The model has a Root Mean Standard Deviation (RMSD) of 0.18 for predicting average oxidation states, as shown in **Figure 2**, and an RMSD below 0.12 for predicting integer oxidation states (between 0-7), as shown in **Figure 3**.

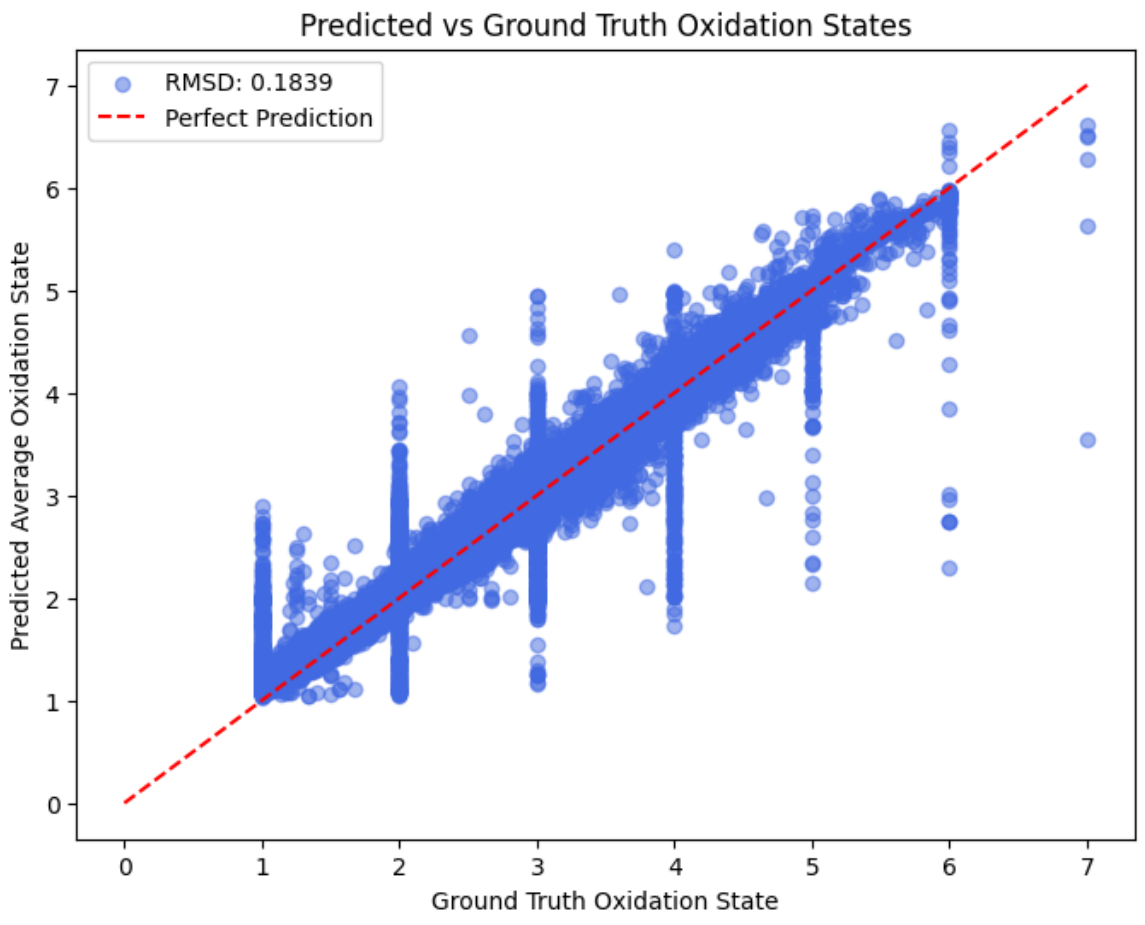

**Figure 1:** Comparison of true and predicted average oxidation states for 48934 testing samples.

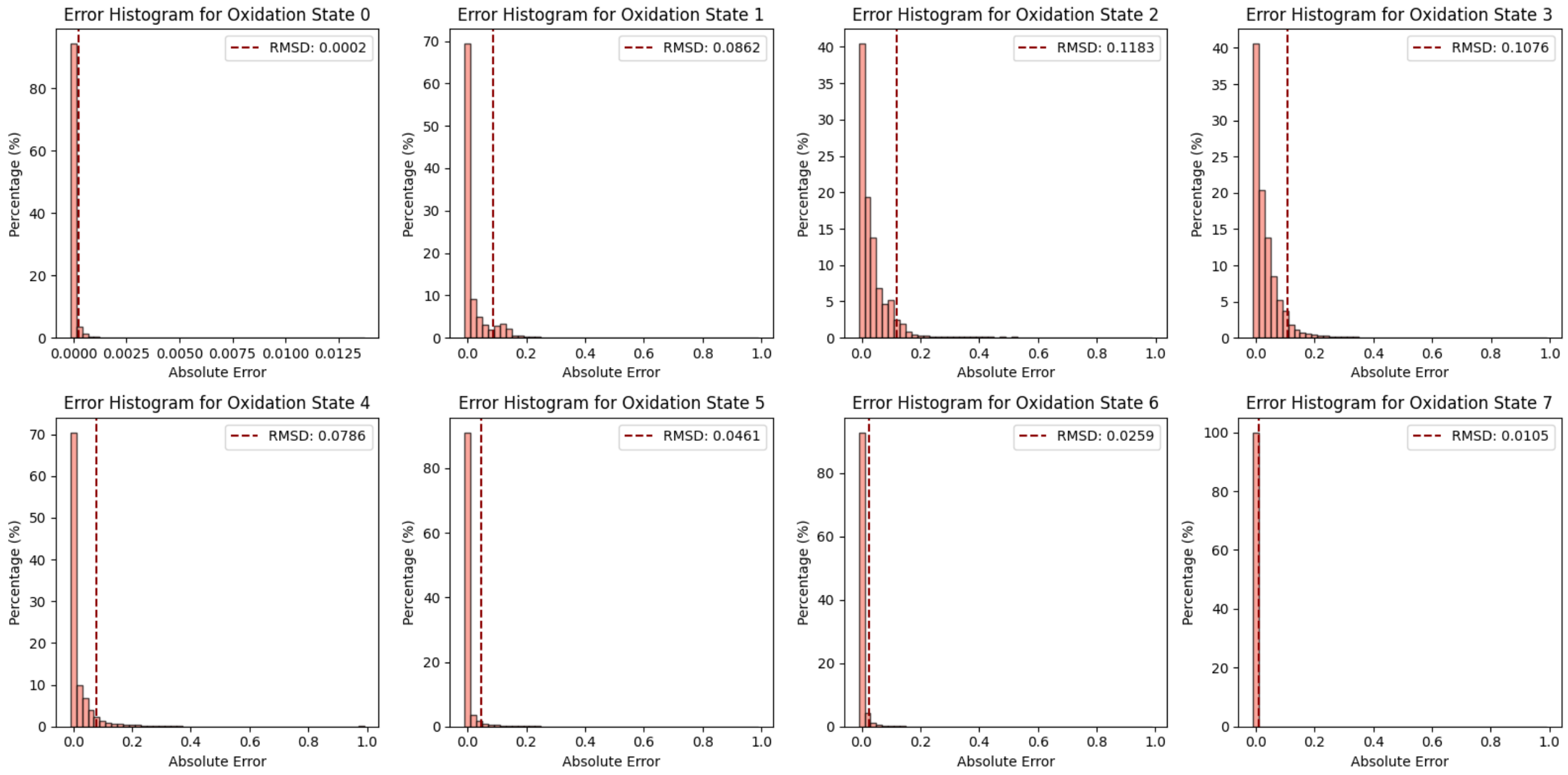

**Figure 2:** Error histograms for each oxidation state in the test dataset, with red dashed lines indicating the RMSD.

## 18. Using LLM to enable bag-of-features segmentation of the metallographic images

Sergei V. Kalinin and Vivek Chawla

Department of Materials Science and Engineering, University of Tennessee, Knoxville

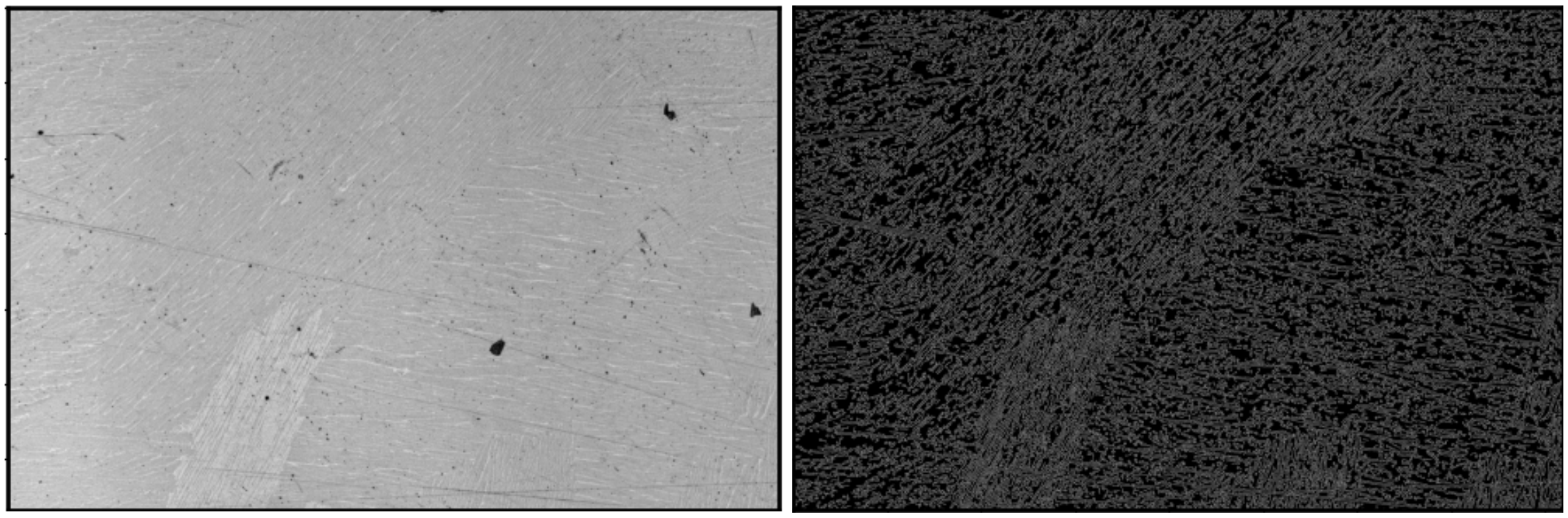

**Figure 1.** Original metallographic image and the Canny edge filter pattern.

Areas such as materials discovery, metallography, geology, etc. generate large volumes of imge data sets containing depictions of the materials microstructure in multiple possible modalities. The latter can include optical microscopy, scanning electron microscopy, and scanning probe microscopy. The resulting images contain patterns that are easy to identify by the human eye, but nonetheless present considerable challenges to the unsupervised segmentation analyses.

Here we develop an unsupervised workflow for the analysis of these images based on the LLM-assisted bag of features approach combining the human heuristic and ML. As a first step, we note that the human identification of the features is based on the clearly seen edge-like features, associated with the grain structure in the materials. Hence, the use of the adaptive edge filters (Canny filter here) allows to highlight these features. In order to segment the image, we use the uniform sampling grid to convert the image into the patch descriptors samples on the square grid. The variability of these features is sufficiently large so that the simple approaches based on invariant variational autoencoder (VAE) or simple clustering (GMM or k-means) do not allow to generate discernible segmented images. The limited success can be obtained with the GMM on the Radon-transformed image patches; however, the analysis requires extensive human optimization for the chosen patch size, etc.

We use the LLM (here ChatGPT) to suggest possible descriptors of the image patch including uniformity, edge density, perimeter length, area coverage, aspect ratio, circularity, compactness, complexity, average orientation, and edge symmetry. Note that the additional descriptors can be developed as necessary. Application of these descriptors to the patch collection converts each patch to the descriptor vector defined at each lattice site. Note that the realizations of the descriptor generators are such that the outputs can be unphysical; hence it is necessary to introduce the additional functionality that converts numerical and domain projects to zero. The resulting descriptor vector can be transformed using the principal component analysis (PCA) to give the loading maps that provide the statistically-significant representations of the data.

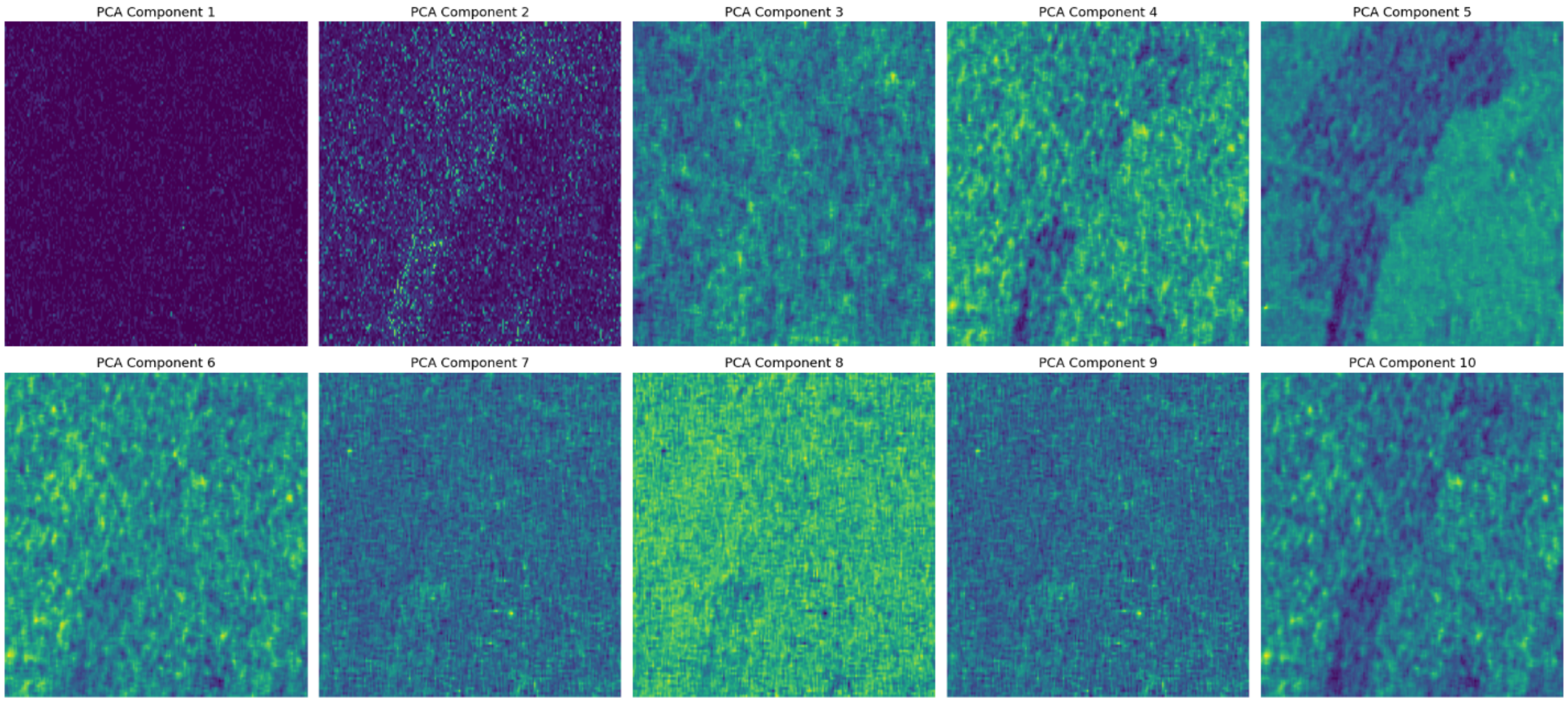

**Figure 2.** PCA components of the feature vector developed based on the LLM-generated descriptors.

We note that the proposed workflow is limited in that it requires extensive human design, starting from chosen filtering, edge detection method and its parameters, patch size, and feature vector construction. We pose that the use of the reward-driven approach will allow unsupervised creation and tuning these workflows, allowing for the fully-unsupervised implementation. This requires construction of suitable reward functions that can represent the properties of ideal segmentation. Once designed, this will allow fully unsupervised and human in the loop unsupervised segmentation, applicable to real-time imaging.

## 19. Structure Discovery through Image-to-Graph Machine Learning Model

Harshit Sethi, Jie Huang, Lauri Kurki

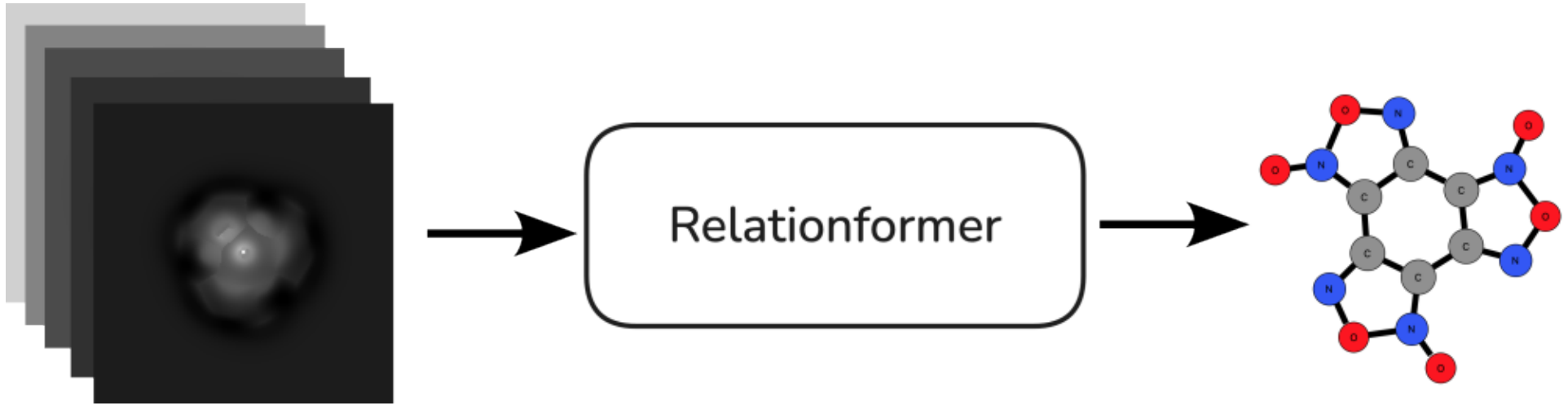

### Motivation

Atomic Force Microscopy (AFM) with a CO functionalized tip plays a crucial role in characterizing atomic-scale nanostructures. However, identifying structures from AFM images is a challenging task that relies heavily on human expertise. Simulations, particularly with Particle Probe Model (PPM) [1], offer a cost-effective solution for generating large volumes of AFM images through atomic structures, thereby providing a map from atomic structure to AFM image. Conversely, in practice, we always hope to know the atomic structure of a given AFM image, aiming to obtain the inverse map from AFM image to its corresponding atomic structures. By training state-of-the-art machine learning models with extensive PPAFM-generated datasets, the underlying molecular geometry can be predicted accurately.

In the previous works [2,3,4], we developed machine learning workflows based on a two-stage process involving two neural network models. The first model, a U-net-based PosNet, is used to predict the possible positions of all the atoms, while the second graph neural network (GNN), utilizes the output of PosNet to determine the species of atoms by examining the sub-regions of the AFM image. This two-model cascade structure presents practical difficulties including challenges in training and evaluation. To enhance this workflow, we aim to use an end-to-end model that simultaneously outputs the positions and types of atoms. Such an end-to-end model could simplify the training process and improves overall efficiency. This approach has great potential for application to experimental AFM images, enabling faster and more reliable structure discovery.

Relationformer [5] is an image-to-graph framework that has shown its efficiency in multiple tasks, including 2D road graph extraction from satellite images and 3D vessel graph predictions. Given its capabilities, we plan to try this image-to-graph model in the task of structure discovery through AFM images, where the atomic structure can be represented as a 3D graph.

### What we've done

During the hackathon, we successfully started training Relationformer with simulated non-contact AFM (nc-AFM) images to predict sample structures. Specifically, we used a transformer-based model capable of simultaneously predicting objects (atoms) and their relationships (bonds), ensuring accurate geometric characterization. To train this model, we utilized a high-throughput nc-AFM simulator, PPAFM, which provides high-resolution AFM images alongside molecular graph labels.

**Expected results**

- Accurate prediction of molecular graphs, including atom types and bond information, from AFM images.
- Enhanced workflow efficiency by transitioning to an end-to-end machine learning model.